\documentclass{aastex63}
\usepackage{multirow}
\usepackage{enumitem}

\begin{document}
\title{Constraining the Properties of GRB Accreting Magnetars with $R/I$ Evolutionary Effects Using \emph{Swift}/XRT Data}

\correspondingauthor{Lin Lan}
\email{lanlin@bao.ac.cn}

\correspondingauthor{He Gao}
\email{gaohe@bnu.edu.cn}

\author{Lin Lan}
\affiliation{CAS Key Laboratory of Space Astronomy and Technology, National Astronomical Observatories, Chinese Academy of Sciences, Beijing 100012, People's Republic of China; lanlin@bao.ac.cn}

\author{He Gao}
\affiliation{School of Physics and Astronomy, Beijing Normal University, Beijing 100875, People's Republic of China; gaohe@bnu.edu.cn}
\affiliation{Institute for Frontier in Astronomy and Astrophysics, Beijing Normal University, Beijing 102206, People’s Republic of China}

\author{Litao Zhao}
\affiliation{School of Mathematics and Science, Hebei GEO University, Shijiazhuang, People’s Republic of China}

\author{Shunke Ai}
\affiliation{Niels Bohr International Academy and DARK, Niels Bohr Institute, University of Copenhagen, Blegdamsvej 17, 2100, Copenhagen, Denmark}

\author{Jie Lin}
\affiliation{Department of Astronomy, Xiamen University, Xiamen, Fujian 361005, People's Republic of China}

\author{Long Li}
\affiliation{Department of Physics, School of Physics and Materials Science, Nanchang University, Nanchang
330031, People’s Republic of China}

\author{Lang Xie}
\affiliation{Department of Physics, School of Physics and Materials Science, Nanchang University, Nanchang
330031, People’s Republic of China}

\author{Li-Ping Xin}
\affiliation{CAS Key Laboratory of Space Astronomy and Technology, National Astronomical Observatories, Chinese Academy of Sciences, Beijing 100012, People's Republic of China; lanlin@bao.ac.cn}

\author{Jian-Yan Wei}
\affiliation{CAS Key Laboratory of Space Astronomy and Technology, National Astronomical Observatories, Chinese Academy of Sciences, Beijing 100012, People's Republic of China; lanlin@bao.ac.cn}

\begin{abstract}
A newly born millisecond magnetar has been proposed as one possible central engine of some long gamma-ray bursts (LGRBs) with X-ray plateau emission. In this work, we used a universal correlation between the initial spin period ($P_0$) and the surface magnetic field ($B_p$) of the newborn magnetar based on an LGRB sample in \cite{Lan2025} to explore the propeller properties of accreting magnetars with $R/I$ evolutionary effects. We found that the $B_p-P_0$ relation is approximately consistent with $B_p\propto P_{\rm eq}^{7/6}$. Here $P_{\rm eq}$ is the equilibrium spin period in the magnetic propeller model, where the electromagnetic dipole radiation and the propeller mechanism jointly modulate the spin evolution of a newborn magnetar. The $B_p-P_0$ relation indicates that $P_0$ constrained by X-ray plateau data may not be the true initial spin period of a newborn magnetar but had reached an equilibrium spin period via fallback accretion in the propeller model. The magnetar accretion rate in our LGRBs is in the range of $\dot{M}\sim10^{-5}-10^{-2}~M_{\odot}~\rm s^{-1}$ by incorporating $R/I$ evolutionary effects and using the transition relation between gravitational mass $M_g$ and baryonic mass $M_b$ in different equations of state and X-ray radiation efficiencies. Such accretion rates ensure that the accreting magnetars in our sample survive until reaching the equilibrium spin period, and the accretion rate is one order of magnitude lower compared to the statistical results in \cite{Stratta2018} and \cite{Linweili2020}, which used the constant $R/I/M_g$ scenario. The fallback rate of progenitor envelope materials onto the magnetar accretion disk for our LGRBs is compatible with the theoretical mass fallback rate of some low-metallicity progenitors. We suggested that adopting a constant $R/I/M_g$ scenario for modeling the propeller regime in accreting magnetars results in a higher mass accretion rate, which may impair our understanding of the physical nature of an accreting magnetar and its surroundings, and that the low-metallicity progenitors can provide enough material to satisfy the accretion requirements of the newborn accreting magnetar in LGRBs.
\end{abstract}
\keywords{gamma-ray burst: general - magnetars}

\section{Introduction}
As one of the most violent explosions in the Universe, Gamma-ray bursts (GRBs) have a huge amount of isotropic emission energy $E_{\gamma,\rm iso}$ from $\sim10^{46}$ to $\sim10^{55}~\rm erg$ \citep{Woosley2006a,Kumar2015,Zhang2018,Lan2023}. In general, they are thought to originate from massive-star collapse (for long GRBs (LGRBs)) or merger of binary compact stars (for short GRBs (SGRBs)). Within post-collapse or post-merger, two competing types of GRB central engine would be formed and power a relativistic outflow \citep{Paczynski1986,Eichler1989,Usov1992,Woosley1993,Thompson1994,Dai1998a,Dai1998b,MacFadyen1999,Zhang2001,Metzger2008,Zhang2011a}: one is a hyperaccreting stellar-mass black hole (BH) via neutrino-antineutrino annihilation \citep{Ruffert1997,Popham1999,Chen2007,Lei2009,Lei2013,Liu2017} or Blandford--Znajek mechanism \citep{Blandford1977,Lee2000,Li2000} to launch a relativistic outflow, and the other is a rapidly spinning, strongly magnetized neutron star (NS; also called a millisecond magnetar) via losing its rotational energy to power a relativistic outflow \citep{Usov1992,Thompson1994,Dai1998a,Dai1998b,Zhang2001,Metzger2008,Metzger2011,Bucciantini2012,Lv2014,Lv2015}. The observed $\gamma$-ray prompt emission and multiband afterglow emission can be explained by internal shocks/dissipated photosphere/magnetic dissipation models and synchrotron emission mechanism in the standard external shocks model, respectively \citep{Thompson1994,Rees1994,Meszaros1997,Sari1998,Kobayashi2000,Meszaros2002,Zhang2004,Rees2005,Peer2006,Zhang2011b,Gao2013,Gao2015,Kumar2015,Zhang2018}.

Theoretically, it is widely accepted that newborn millisecond magnetars are formed after the mergers of binary NSs or the core collapses of massive stars, accompanied by GRBs with strong X-ray plateau emission, and strong gravitational-wave (GW) radiation may be present during the X-ray plateau phase, due to a large ellipticity and a high spin \citep{Dai1998a,Dai1998b,Zhang2001,Corsi2009,Metzger2011,Fan2013a,Fan2013b,Metzger2014}. With the successful launch of the Neil Gehrels \emph{Swift} Observatory \citep{Gehrels2004}, the early-time afterglow emissions of GRBs have been revealed, and abundant X-ray observation data have been collected. A good fraction of the X-ray afterglows (LGRBs and SGRBs) pose a long-lasting plateau emission feature, which seems to coincide with the prediction of a newborn magnetar central engine \citep{Dai1998a,Dai1998b,Zhang2001,Metzger2011,Gompertz2013,Gompertz2014,Rowlinson2013,Rowlinson2014,Lv2014,Metzger2014,Lv2015}. The electromagnetic (EM) dipole emission of a newborn magnetar generates a Poynting flux that can undergo magnetic energy dissipation processes with high efficiency \citep{Zhang2011b} and inject additional energy into the GRB to induce transient X-ray plateau emission, and the transient X-ray plateau luminosity and temporal evolution are highly dependent on the spin-down behavior of the newborn magnetar.

In previous studies, the magnetar remnants of GRBs were usually treated as isolated NSs with a constant spin-down rate and therefore a constant level of EM emission. The physical properties and lifetime of the newly born magnetar could be inferred from the observed multiband afterglow morphological characteristics and plateau data \citep{Troja2007,Rowlinson2010,Rowlinson2013,Rowlinson2014,Fan2013a,Fan2013b,Gompertz2013,Gompertz2014,Lv2014,DallOsso2015,Lv2015,Lv2018,Lv2019,Gao2016,Gao2017a,Gao2017b,Lasky2016,Lasky2017,Ai2018,Ai2020,Lin2019,Zou2019,Zou2021b,Lan2020a,Lan2021,Linjie2020,Sarin2020,Zhao2020,Ai2021,Xie2022a,Xie2022b}. However, as the GRB's central engine, the newborn magnetar should be highly magnetized and rapidly rotating, due to the fact that it inherited prodigious angular momentum from its progenitor and underwent extreme magnetic field amplification during the progenitor's catastrophic process. Thus, it is natural to consider that a fraction of the progenitor fragment in post-collapse or post-merger could not escape from the central magnetar and could circulate into an accretion disk and interact with a nascent magnetar, which could have a strong influence on the spin evolution and transient EM emission of a newborn magnetar (namely, magnetic propeller model). To date, the propeller model has been widely used to explain some observations in GRBs and supernovae (SNe), such as the significant brightening of early afterglows in some GRBs \citep{Dai2012,Li2021,Yang2024}, the precursor emission and X-ray flares in LGRBs \citep{Bernardini2013,Gibson2018,Lan2018,Yu2024}, the extended emission and X-ray plateau emission in SGRBs \citep{Gompertz2014,Gibson2017,Lan2020b}, and some special SNe \citep{Piro2011,Lin2021}. Meanwhile, such an interacting magnetar--disk system tends to reach an equilibrium spin state, where the equilibrium spin period ($P_{\rm eq}$) and surface magnetic field strength ($B_p$) will follow a positive correlation for a given accretion rate, i.e., $B_p\propto P_{\rm eq}^{7/6}$ \citep{Piro2011}. Interestingly, such a correlation has been found in relevant statistical studies based on GRB and SN samples that are believed to be powered by the central magnetar \citep{Stratta2018,Linweili2020,Xie2022a}. Furthermore, \cite{Stratta2018} and \cite{Linweili2020} tentatively
investigated the mass accretion rate of newborn accreting magnetars in the propeller model based on their respective GRB samples, and the corresponding mass accretion rates are inferred as $\dot{M}\sim10^{-4}-10^{-1}~M_{\odot}~\rm s^{-1}$. However, the accretion rates derived from the propeller model are highly dependent on the $B_p$ and initial period ($P_0$) obtained through X-ray plateau data, as well as the gravitational mass ($M_g$), radius ($R$) and moment of inertia ($I$) of the newborn magnetar, and their constraints on accretion rates typically assume fixed values for the mass, radius, and moment of inertia of the newborn magnetar ($M_g\sim1.4~M_{\odot}$, $R\sim10~\rm km$, $I=0.35M_gR^2$), which simplified assumptions may lead to results that deviate from the true mass accretion rate of the newborn accreting magnetar. In particular, \cite{Lan2025} found that neglecting the $R/I$ evolutionary effects can lead to systematic overestimation or underestimation of magnetar physical parameters such as $B_p$ and $P_0$ from 20\% to 50\%. The $R/I$ evolutionary effects and different gravitational masses may significantly impact the accretion rate results obtained from the $B_p-P_0$ distribution for newborn accreting magnetars.

It is interesting to ask the following two questions: What is the true mass accretion rate of newborn accreting magnetar after incorporating $R/I$ evolutionary effects and different gravitational masses? What constraints can one pose on the progenitors in our LGRBs from the true mass accretion rates? In this paper, we systematically explore the propeller properties of accreting magnetars by utilizing an LGRB sample with an X-ray plateau in \cite{Lan2025}, and try to constrain the accretion rate of accreting magnetars in different equations of state (EoSs) and X-ray radiative efficiencies, which is crucial to helping us understand the nature and physical environment of a newborn accreting magnetar and its progenitor. Our paper is organized as follows: The sample selection and data reduction are presented in Section \ref{Sec:sample}. In Section \ref{Sec:model}, we briefly introduce the model of a magnetar propeller with fallback accretion. In Section \ref{Sec:B-P}, we show the $B-P$ statistical results in different EoSs and further constrain the accretion rate of accreting magnetars in different EoSs and X-ray radiative efficiencies. Finally, we summarize our conclusions and discussions in Section \ref{Sec:discussions}. Throughout the paper, a concordance cosmology (flat $\Lambda$CDM) with parameters $H_0 = 67.4$ km s$^{-1}$ Mpc $^{-1}$, $\Omega_M=0.315$, and $\Omega_{\Lambda}=0.685$ has been adopted according to the {\it Planck} results \citep{Planck2020}.

\section{Sample Selection and Data Reduction}
\label{Sec:sample}

Theoretically, the magnetar energy injection signature typically exhibits a shallow decay segment (or plateau), followed by a steeper decay segment in X-ray emission when it is spinning down. \cite{Lan2025} systematically searched for LGRBs with X-ray plateau emission in the available \emph{Swift} GRB catalog, which includes more than 1500 GRB outbursts between 2005 January and 2023 November. Finally, they found that 105 LGRBs were able to satisfy magnetar engine candidates, 43 of which have a measured redshift. We mainly used these 105 curated magnetar engine candidate samples to statistically study the propeller properties of LGRB remnants. It should be noted that the redshift measurement $z$ is vital to derive the intrinsic luminosity parameter, and we adopted $z=1$ for the GRBs without the redshift measurement in our sample. 

Once the sample has been determined, to investigate in depth the physical properties of newborn magnetars, one should next estimate the physical parameters of these magnetars. However, the magnetar parameters inferred from the observed X-ray plateau data are strongly dependent on the X-ray radiative efficiency $\eta_{\rm X}$ and the $R/I$ evolutionary effects. On the one hand, $\eta_{\rm X}$ is quite uncertain and could evolve over time. \cite{Xiao2019} claimed that $\eta_{\rm X}$ is largely dependent on the bulk saturation Lorentz factors ($\Gamma_{\rm sat}$) of magnetar wind and found that a typical value $\Gamma_{\rm sat}\sim 300$ in the GRBs corresponds to $\eta_{\rm X}$ possibly close to 0.01. \cite{Zhong2024} suggested that the radiative efficiency evolves with time during the magnetar spin-down process and estimated the time-averaged radiation efficiency as $\sim0.01$ for GRB 230307A. \cite{Rowlinson2014} suggested that the conversion efficiency of rotational energy from the magnetar wind to the observed X-ray plateau luminosity is $\leq0.2$. \cite{Gao2016} found the efficiency parameter to be within 0.5 by Monte Carlo simulation. It is worth noting that in the framework of magnetar spin-down, a low radiation efficiency, such as $\eta_{\rm X}<0.1$, would lead to a magnetar rotational speed that breaks through the Keplerian breakup limit, which may challenge the most current NS EoS models. Therefore, we took two radiation efficiency values to evaluate our results, i.e. $\eta_{\rm X}$ = 0.1 and 0.5. On the other hand, \cite{Lan2025} found that neglecting the evolutionary effects of $R/I$ can lead to systematic overestimation or underestimation of magnetar physical parameters such as $B_p$, $P_0$, and ellipticity ($\epsilon$) from 20\% to 50\%. Thus, when using the observed X-ray plateau data to analyze the properties of the newborn magnetar, we should simultaneously combine the EoS and $\eta_{\rm X}$ information as constraints to accurately estimate various parameters of the newborn magnetar.

We employed the Markov Chain Monte Carlo (MCMC) method from the emcee Python package \citep{Foreman-Mackey2013} to derive the best-fitting posterior magnetar parameters and their uncertainties in four EoSs [SLy \citep{Douchin2001}, WFF2 \citep{Wiringa1988}, ENG \citep{Engvik1996}, AP3 \citep{Akmal1997}] with baryonic mass $M_{b}=2.0~M_{\odot}$ and two X-ray radiation efficiency ($\eta_{\rm X}=0.1,~0.5$). The key physical parameters for defining the properties of a newborn magnetar are the initial spin period $P_0$, the dipole magnetic field strength $B_p$, and the deformed ellipticity $\varepsilon$. The prior $P_0$ was set within a uniform distribution over a broad range, and the prior $B_p$ and $\varepsilon$ were set within a log-uniform distribution over a broad range, 
specifically $P_0 \in [0.3,40]$ ms, log $(B_p/\rm G)\in [13,16]$, and log$~\varepsilon \in [-6,-2]$. Based on the value of $\chi^2$/degrees of freedom, we determined the best-fitting model parameters for each GRB in different EoSs and X-ray radiation efficiencies. For more details on constraining magnetar parameters by taking into account the $R/I$ evolutionary effects, please refer to our latest paper in \cite{Lan2025}. Ultimately, we collected magnetar parameters ($B_p$ and $P_0$) in various combination situations by considering constraints from different EoSs and $\eta_{\rm X}$.

\section{Magnetar Propeller Accretion Model}
\label{Sec:model}

The propeller model was first theoretically proposed by \cite{Illarionov1975} to elucidate the dynamical processes of how highly magnetized NSs can lose angular momentum and regulate accretion when their magnetic fields interact with inflowing materials from the accretion disk. If a NS has a strong magnetic field ($B_{p}\gtrsim10^{12}~\rm G$) and is rapidly rotating, as expected for the NSs in GRBs, it is often classified as a millisecond magnetar. In this case, the local materials of accretion disk could be ejected owing to centrifugal force rather than being accreted onto the NS. Finally, the interaction between a magnetar and its surrounding accretion disk has an effect on its spin evolution and hence the accretion outflows. This propeller accretion model has been widely discussed in the context of GRBs that are likely powered by magnetars \citep{Dai2012,Bernardini2013,Gompertz2014,Gibson2017,Gibson2018,Lan2018,Lan2020b,Li2021,Linweili2020,Yang2024,Yu2024}.

Within the propeller accretion model, the interaction between a magnetar and its surrounding accretion disk can be modeled by defining the relative positions of Alfv\'{e}n radius (the radius at which the dynamics of the gas within the disk is strongly influenced by the magnetic field, $r_{\rm m}$), corotation radius (the radius at which material in the disk orbits at the same rate as the magnetar surface, $r_{\rm c}$), and light cylinder radius (the radius at which the magnetic field lines rotate at the speed of light in order to maintain rigid rotation with the NS surface, $r_{\rm lc}$). The magnetic dipole field of the central magnetar is given by $B = \mu /r^3$, where $\mu = B_{p} R^3$ is the magnetic dipole moment for an NS with surface dipole field $B_{p}$ and radius $R$. The magnetic pressure at a given radius $r$ can be expressed as
\begin{equation}
P_{\rm mag}=\frac{B^2}{2 \mu_0}=\frac{\mu^2}{2 \mu_0 r^6},
\end{equation}
where $\mu_0$ is the permeability in vacuum. There is also an inward ram pressure on the accreted material that falls back from the accretion disk, which competes with the $P_{\rm mag}$. For the case of spherically symmetric accretion, the ram pressure can be given by
\begin{equation}
P_{\rm ram}=\frac{\dot{M}}{8\pi}\bigg{(}\frac{2GM_g}{r^5}\bigg{)}^{1/2},
\end{equation}
where $G$ is the gravitational constant, $M_g$ and $\dot{M}$ are the gravitational mass and the accretion rate of the central magnetar, respectively. Equating these two pressures gives the radius at which the fallback material is strongly affected by the dipole magnetic field, known as the Alfv\'{e}n radius $r_{\rm m}$, which could be written as 
\begin{equation}
r_{\rm m} = \mu^{4/7}(GM_g)^{-1/7}\dot{M}^{-2/7},
\label{eq:r_m}
\end{equation}
$r_{\rm m}$ is one of the three key radii that determine the spin evolution behavior of the central magnetar. The second key radius is the corotation radius $r_{\rm c}$, which is the radius at which the material orbits at the same rate as the stellar surface. Given that inflowing materials rotate at the local Keplerian angular velocity, i.e., $\Omega_{\rm K} = (GM_g/r^3)^{1/2}$, their corotation angular velocity with
the central magnetar occurs at a radius of
\begin{equation}
r_{\rm c} = (GM_g/\Omega^2)^{1/3},
\label{eq:r_c}
\end{equation}
where $\Omega = {2\pi}/{P}$ and $P$ are the angular velocity and spin period of the central magnetar, respectively. Furthermore, another critical radius is the radius of the light cylinder $r_{\rm lc}$, where the magnetic field lines will rotate at the speed of light to maintain a rigid rotation with the magnetar surface. Thus, $r_{\rm lc}$ is defined as 
\begin{equation}
r_{lc} = c/\Omega,
\label{eq:r_lc}
\end{equation}
The dynamical evolution process and the accretion outflow transport pathway of the magnetar--disk system are fundamentally dictated by the spatial hierarchy among $r_{\rm m}$, $r_{\rm c}$, and $r_{\rm lc}$. Specifically speaking, when $r_{\rm m}<r_{\rm c}<r_{\rm lc}$, the disk materials in the inner radius revolve faster than the magnetar, and the magnetic field will first slow the disk materials within $r_{\rm c}$. Once the angular momentum of the materials is lost, the disk materials can no longer keep Keplerian rotation at the orbit and tend to accrete to the surface of the NS along the magnetic field lines, resulting in the spin-up of the magnetar, i.e., the so-called accretion regime. Conversely, if $r_{\rm c}<r_{\rm m}\ll r_{\rm lc}$, the magnetic field is spinning faster than the inner disk materials, and materials already within $r_{\rm c}$ accrete to the surface of the magnetar, while the materials within the range of $r_{\rm c}$ and $r_{\rm m}$ will be accelerated to a super-Keplerian velocity and propelled away from the system by the centrifugal force. The angular momentum of the magnetar is transferred to the inner disk materials and causes the spin-down of the magnetar, i.e., the so-called propeller regime. If the propeller power is not strong enough for the materials, it cannot reach the potential well. Then, the materials will return to the disk without any emission signals to be detected.

Notably, the relative positions of $r_{\rm c}$ and $r_{\rm m}$ always change dynamically in the accretion regime and the propeller regime. In the accretion regime, the magnetar is spun up by gaining angular momentum from accretion, and the increased spin velocity will cause $r_{\rm c}$ to shrink. Then, as the accretion rate decreases, the Alfv\'{e}n radius $r_{\rm m}$ will expand until $r_{\rm c}<r_{\rm m}$, which will switch on the propeller regime. In the propeller regime, the magnetar is spun down by transferring angular momentum to the inner disk materials, and the decreased spin velocity will in turn cause $r_{\rm c}$ to expand. Once $r_{\rm c}>r_{\rm m}$, the system will switch on the accretion regime again. As we describe above, there would be several critical values for the accretion rate of such a magnetar--disk system. The magnetar spin-down rate due to the loss of angular momentum from a magnetized wind is enhanced when the Alfv\'{e}n radius resides inside the light cylinder. This condition ($r_{\rm m} < r_{\rm lc}$) is satisfied above a critical accretion rate of
\begin{eqnarray}
\dot{M}_\mathrm{lc} = \left(\frac{c}{2\pi}\right)^{-7/2}(GM_g)^{-1/2}B_p^2R^6P^{-7/2}~(r_{\rm m} = r_{\rm lc}),
\label{eq:Mdotlc}
\end{eqnarray}
As $r_{\rm m}$ continues to shrink, the enhancement of the spin-down rate reaches saturation when $r_{\rm m}$ is pushed all the way down to the NS surface, a condition that requires an even higher accretion rate of
\begin{eqnarray}
\dot{M}_\mathrm{NS} = (GM_g)^{-1/2}B_p^2R^{5/2}~(r_{\rm m} = R),
\label{eq:Mdotns}
\end{eqnarray}
Obviously, $\dot{M}_\mathrm{NS}$ depends only on the NS EoS and is independent of the rotation period. When $R_{\rm NS}<r_{\rm m}<r_{\rm lc}$, whether the materials accrete freely onto the magnetar or whether the system may enter the more complicated propeller regime depends on the location of $r_{\rm c}$ relative to $r_{\rm m}$. Ultimately, such a magnetar--disk system tends to evolve toward $r_{\rm c} \simeq r_{\rm m}$ if the spin evolution of the magnetar is dominated by the interaction with its accretion disk. When $r_{\rm c}$ equals $r_{\rm m}$, the accreting magnetar would reach an equilibrium spin period and accretion rate, which can be expressed as \citep{Piro2011}
\begin{eqnarray}
\dot{M}_\mathrm{eq} = (2\pi)^{7/3}(GM_g)^{-5/3}B_{p}^{2}R^{6}P_{\rm eq}^{-7/3},
\label{eq:Mdoteq}
P_\mathrm{eq} = 2\pi(GM_g)^{-5/7}B_{p}^{6/7}R^{18/7}\dot{M}_{\rm eq}^{-3/7},
\label{eq:Peq}
\end{eqnarray}
Distinctly, the equilibrium spin period is independent of the initial spin period of the newborn magnetar but correlates with the magnetic field strength, fallback accretion rate, gravitational mass, and radius of the central magnetar. The gravitational mass and radius of NSs remain intrinsically tied to the poorly constrained EoS of matter at the nuclear density. Once the EoS and baryonic mass are given, one can infer the gravitational mass and radius of the NS by using the universal baryonic-to-gravitational mass conversion relation, combined with a numerical solution to the NS field equations. Therefore, the mass accretion rate can be quantitatively estimated via statistical analysis of the $B_p–P_0$ distribution in a population of accreting magnetars, where the equilibrium spin condition links the relation $\dot{M}\propto B_{p}^2P_{\rm eq}^{-7/3}$. Furthermore, assuming a constant mass accretion rate, the evolutionary timescale, before the newborn magnetar reaches such a spin equilibrium, can be estimated by \citep{Metzger2018}
\begin{eqnarray}
t_\mathrm{ev}\simeq\frac{2\pi I/P_\mathrm{eq}}{\dot{M}(GM_gr_\mathrm{m})^{1/2}} 
= (GM_g)^{2/7}IB_{p}^{-8/7}R^{-24/7}\dot{M}^{-3/7},
\label{eq:Tev}
\end{eqnarray}
where $I=0.35M_gR^2$ is the moment of inertia. Obviously, the larger mass accretion rate and the stronger magnetic field correspond to a shorter timescale $t_{\rm ev}$. For an accreting magnetar that has reached equilibrium period rotation, our estimate of $t_\mathrm{ev}$ can be considered as the lower limit for the accretion timescale ($t_\mathrm{acc}$). 

In addition, the interaction between a magnetar and its surrounding accretion disk has an effect on its spin evolution. The energy reservoir of a newly born magnetar
is the total rotational energy, which could be estimated as
\begin{eqnarray}
E_{\rm rot} = \frac{1}{2} I \Omega^{2},
\label{eq:Erot}
\end{eqnarray}
Typically, the rotation energy boundary is $E_{\rm rot} \simeq 3\times 10^{52}$ erg for an NS of gravitation mass 1.4$M_{\odot}$ rotating near the breakup period of $P \simeq 1$ ms, but reservoirs up to $E_{\rm rot} \simeq 10^{53}$ erg are possible for more massive NSs ($M_{g} \gtrsim 2M_{\odot}$) with a stiff EoS \citep{Metzger2015}. In the spin-down process, we consider that the rotational energy of the newborn magnetar is mainly lost to a polar Poynting-flux-dominated wind at a rate 
\begin{eqnarray}
\dot{E}_{\rm sd} = \left(\frac{f_{\Phi}}{f_{\Phi, \rm lc}}\right)^{2}\frac{\mu^{2}\Omega^{4}}{c^{3}},
\label{eq:Edot}
\end{eqnarray}
here we have assumed that the magnetic dipole moment and rotational axes are aligned. The $f_{\Phi} = \int_{0}^{\theta_{\rm lc}}\sin \theta d\theta \simeq \theta_{\rm lc}^{2}/2$ is the fraction of the magnetar surface threaded by open magnetic flux, and $\theta_{\rm lc}$ is the latitude from the pole of the last closed field line \citep{Bucciantini2006}. For an isolated accreting magnetar, the closed field lines are typically assumed to extend to the light cylinder $r_{\rm lc}$, corresponding to $\theta_{\rm lc} \simeq \sin^{-1}(R/r_{\rm lc})^{1/2} \simeq (R/r_{\rm lc})^{1/2}$ \citep{Contopoulos1999,Spitkovsky2006}, thus providing a minimum open flux of $f_{\Phi,\rm lc} = R_{\rm NS}/2r_{\rm lc}$. In this limit ($f_{\Phi} = f_{\Phi,\rm lc}$), one recovers from equation (\ref{eq:Edot}) the normal magnetic dipole spin-down rate $\dot{E}_{\rm sd} \propto \Omega^{4}$ of the force-free magnetar wind. By comparison, for an accreting magnetar with $r_{\rm m} \lesssim r_{\rm lc}$, closed field lines are truncated where the accretion disk begins, in which case $\theta_{\rm lc} \simeq (R/r_{\rm m})^{1/2}$ and $\dot{E}_{\rm sd}$ is thus enhanced over the standard magnetic dipole rate by a factor of $(r_{\rm m}/r_{\rm lc})^{1/2}>1$ \citep{Parfrey2016}. The maximum spin-down rate is achieved for the maximum open flux $f_{\Phi,\rm max} \simeq 1$ at $r_{\rm m} = R$, i.e., the limit of a split monopole spin-down. Combining above results, the magnetar wind energy-loss rate of an accreting magnetar is given by
\begin{eqnarray} \dot{E}_{\rm sd} &\simeq& \left\{
\begin{array}{lr} 
\frac{\mu^{2}\Omega^{4}}{c^{3}}\frac{r_{\rm lc}^{2}}{R^{2}}, & \dot{M} \gtrsim \dot{M}_{\rm NS} \\
\frac{\mu^{2}\Omega^{4}}{c^{3}}\frac{r_{\rm lc}^{2}}{r_{\rm m}^{2}},~~~~& \dot{M}_{\rm lc} \lesssim \dot{M} \lesssim \dot{M}_{\rm NS} \\
\frac{\mu^{2}\Omega^{4}}{c^{3}}, &
\dot{M} \lesssim \dot{M}_{\rm lc}\\
\end{array}
\right. , \nonumber \\
 &\simeq& \left\{
\begin{array}{lr}  \frac{4\pi^2}{c}B_{p}^{2}R^4P^{-2}, & \dot{M} \gtrsim \dot{M}_{\rm NS} \\
\ \frac{4\pi^2}{c}(GM_g)^{2/7}(B_{p}R^3)^{6/7}P^{-2}\dot{M}^{4/7},~~~~& \dot{M}_{\rm lc} \lesssim \dot{M} \lesssim \dot{M}_{\rm NS} \\
\frac{16\pi^4}{c^3}B_{p}^{2}R^6P^{-4}, &
\dot{M} \lesssim \dot{M}_{\rm lc}\\
\end{array}
\label{eq:Edotcases}
\right. ,
\end{eqnarray}
Combining equations (\ref{eq:Mdotlc}), (\ref{eq:Mdotns}), (\ref{eq:Mdoteq}), and (\ref{eq:Edotcases}), one can give the contours of various fixed spin-down powers, which can help us to infer the physical properties and energy mechanisms of the accretion magnetar.

\section{$B-P$ Statistical Properties and Accreting Magnetar Properties}
\label{Sec:B-P}

\begin{figure*}
\centering
\includegraphics  [angle=0,scale=0.30]   {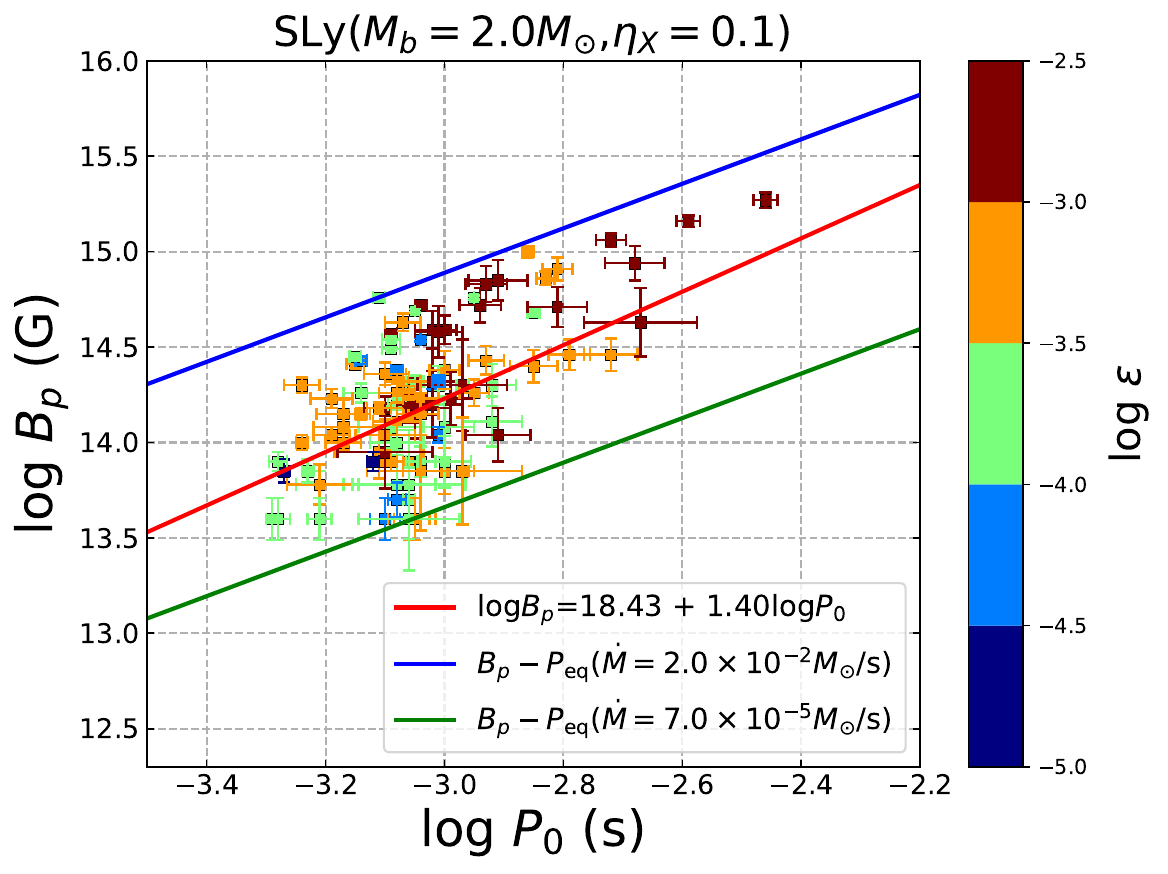}
\includegraphics  [angle=0,scale=0.30]   {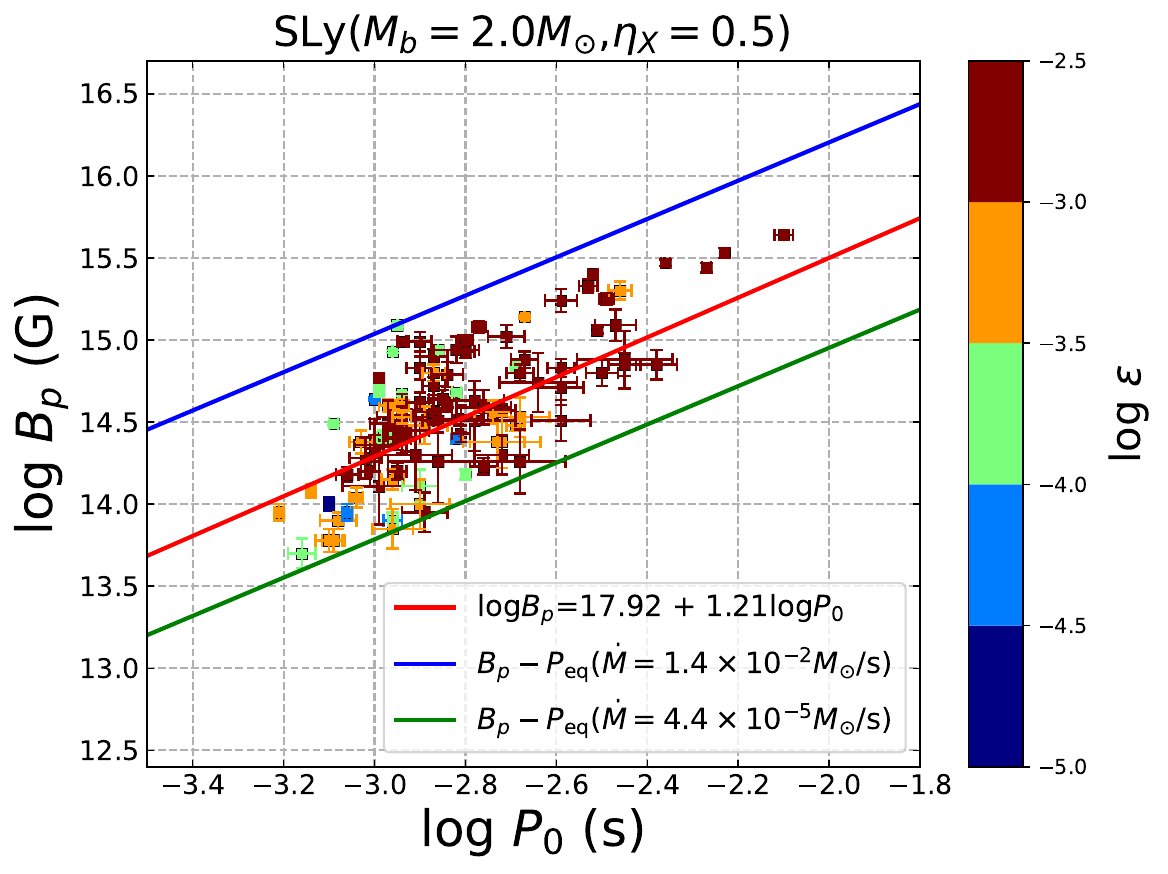}
\includegraphics  [angle=0,scale=0.30]   {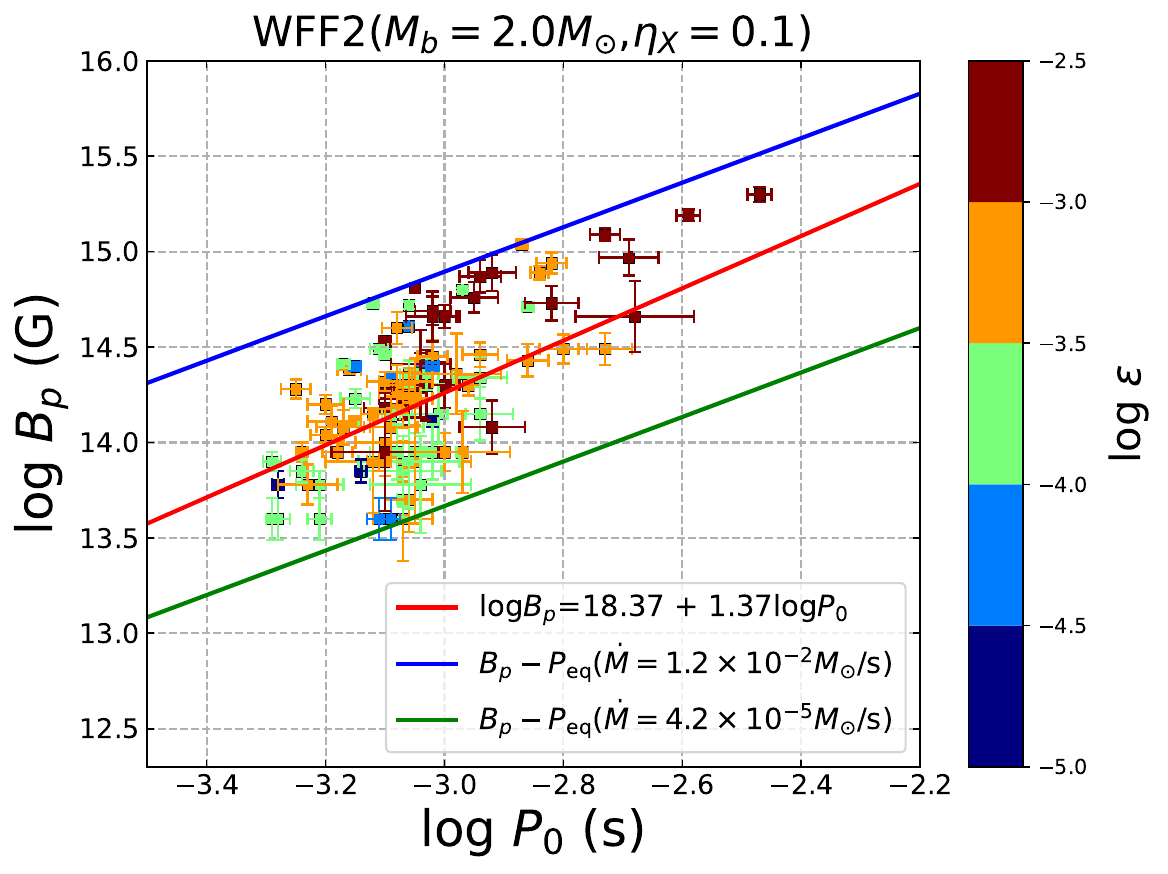}
\includegraphics  [angle=0,scale=0.30]   {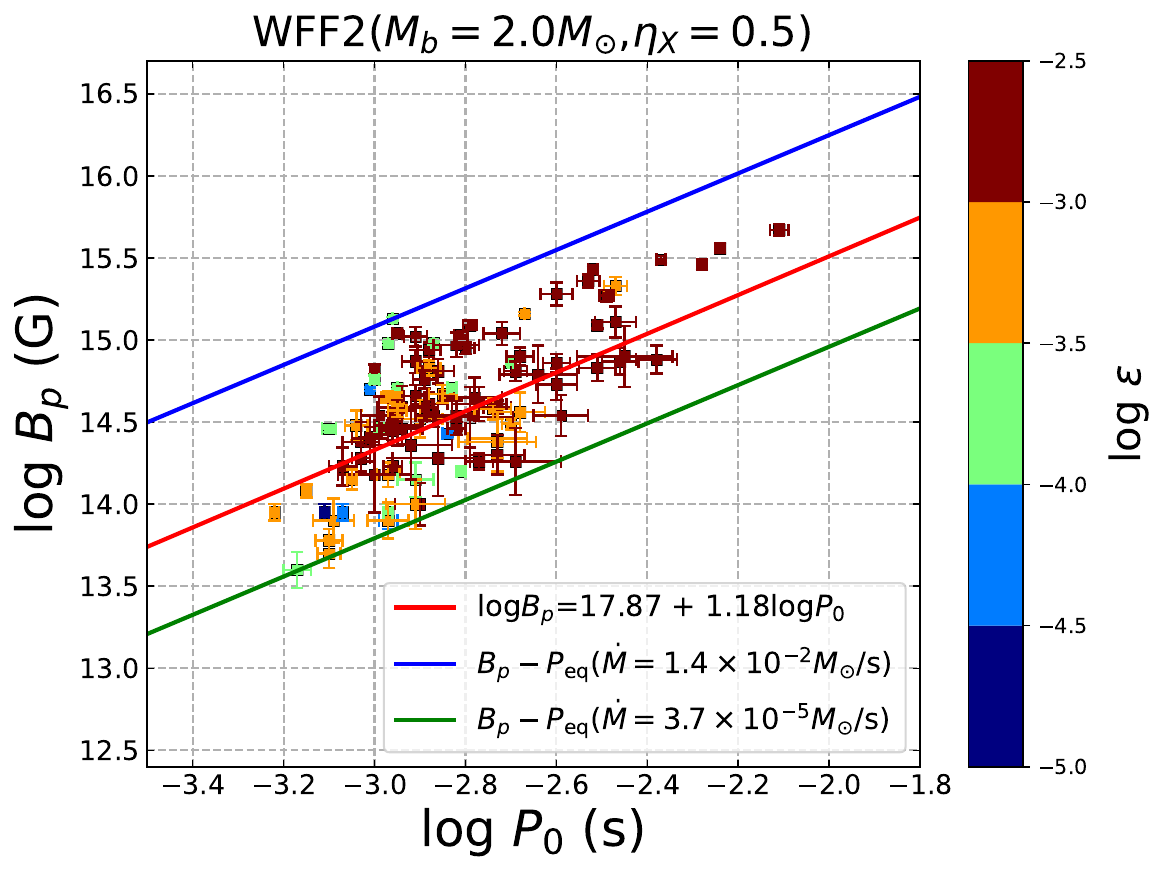}
\includegraphics  [angle=0,scale=0.30]   {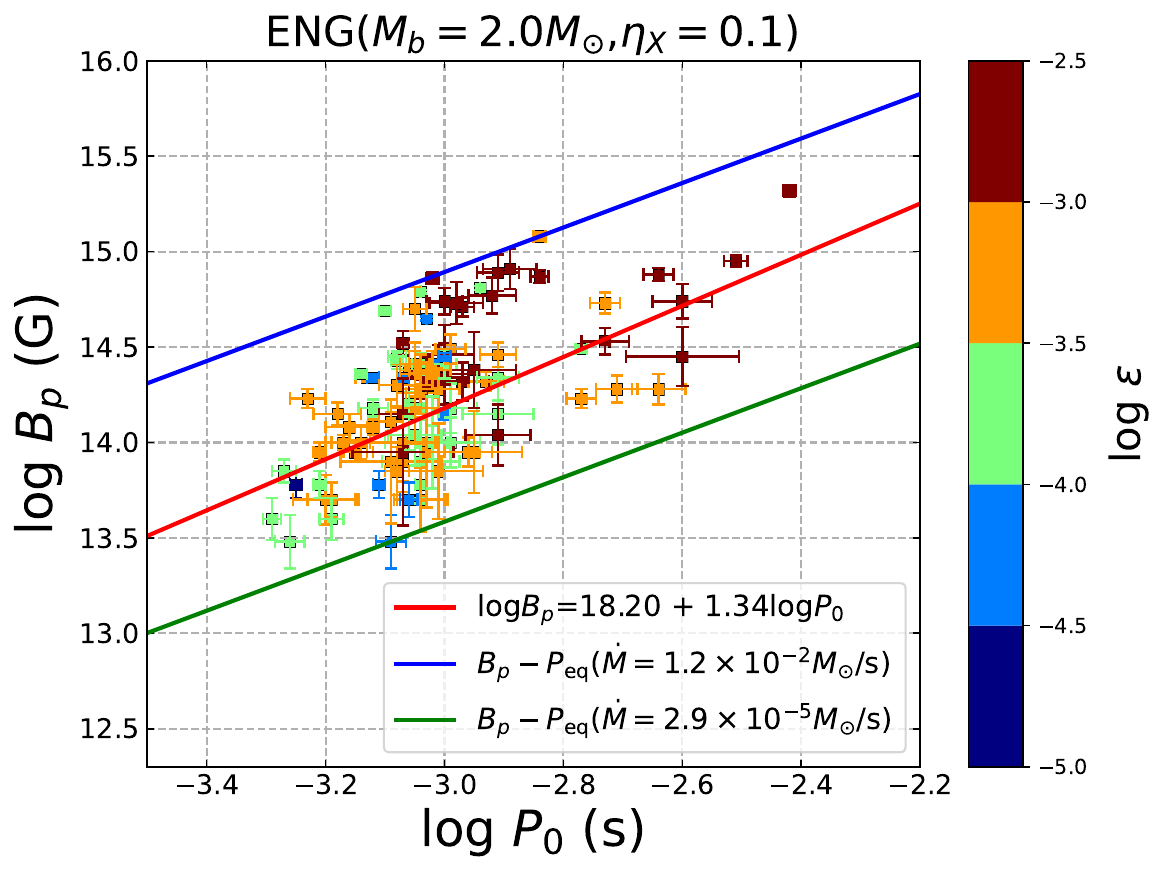}
\includegraphics  [angle=0,scale=0.30]   {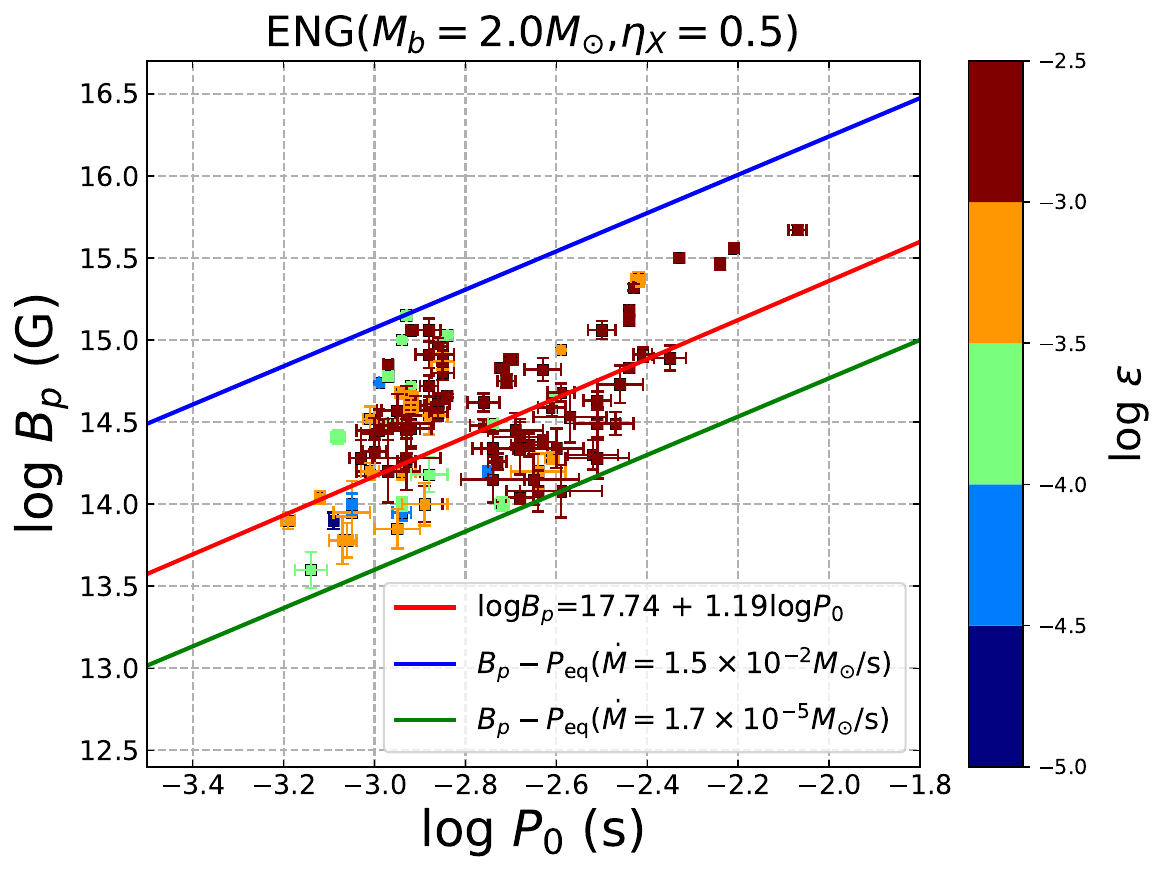}
\includegraphics  [angle=0,scale=0.30]   {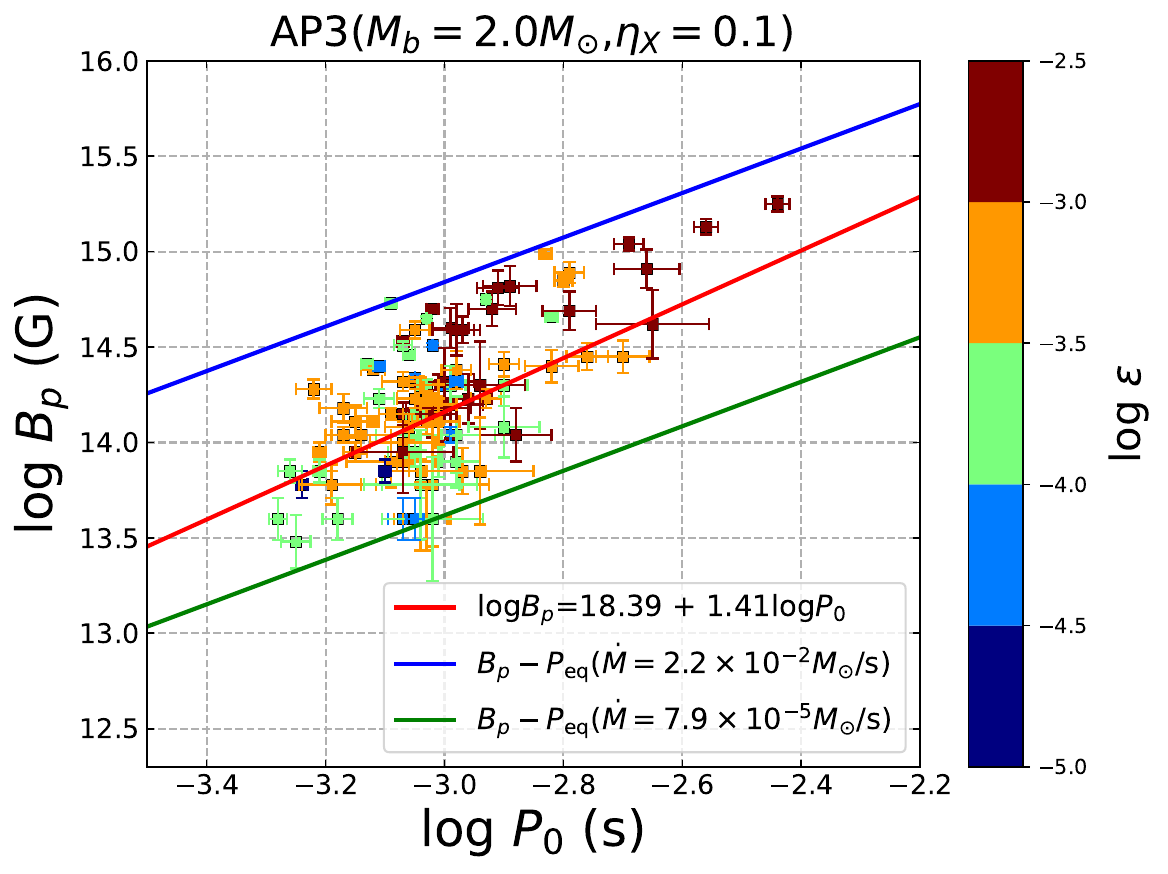}
\includegraphics  [angle=0,scale=0.30]   {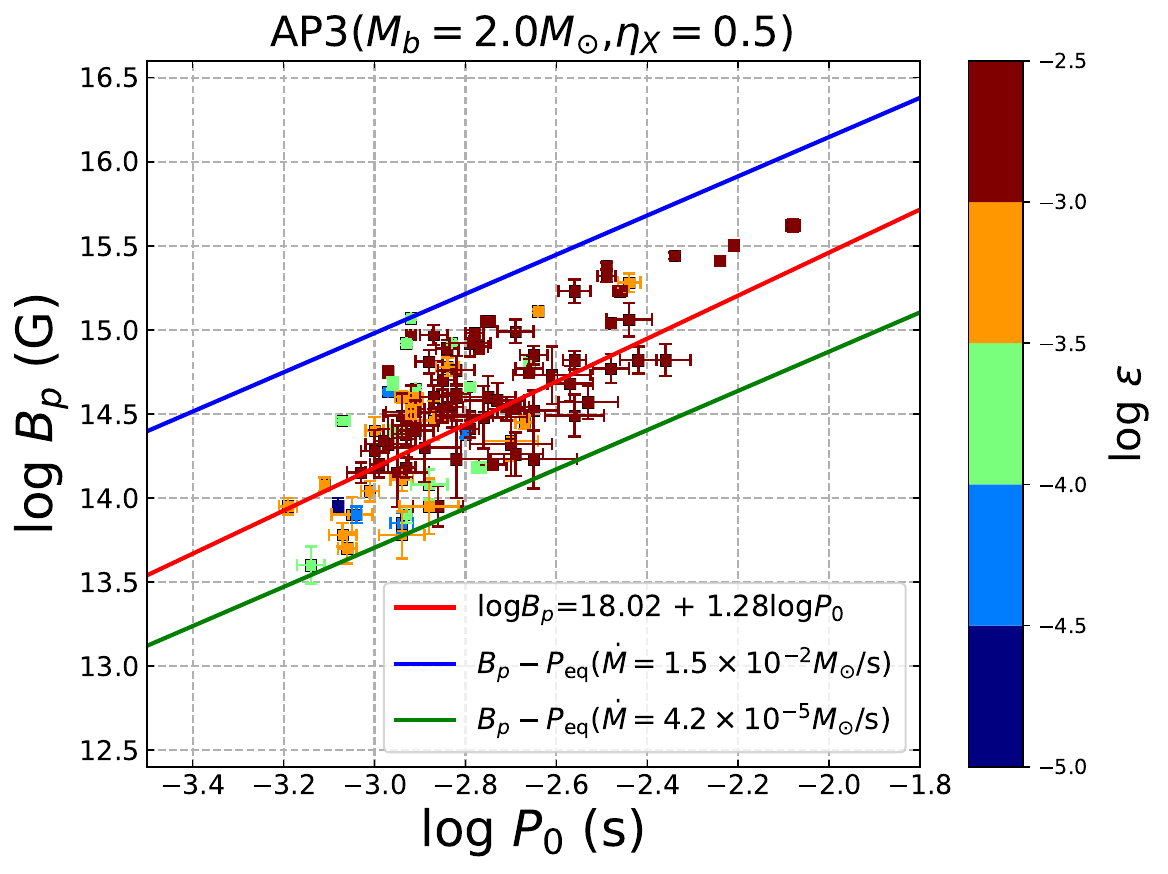}
\caption{The distributions between the $B_p$ and $P_0$ for our LGRB sample in four EoSs with $M_b=2.0M_{\odot}$ and $\eta_{\rm X}=0.1,~0.5$. The different colors of these circles correspond to different $\epsilon$ values. The red solid lines show the best-fitting results, and green solid lines and blue solid lines represent the expected $B_p-P_{\rm eq}$ correlations at the lower and upper limits of the accretion rate derived from our LGRB sample, respectively.}
\label{fig:Bp-P0}
\end{figure*}

With the above MCMC method and the magnetar propeller accretion model, we can accurately estimate the physical parameters of the newborn magnetar and perform the $B_p-P_0$ statistical analysis to constrain the accretion rate of accreting magnetars in different EoSs and $\eta_{\rm X}$ based on the 105 LGRBs with X-ray plateau emission that are thought to be fed by the central magnetar. As shown in Figure \ref{fig:Bp-P0}, Putting all the combinations of (EoS, $\eta_{\rm X}$) together, we can find that the newborn magnetars of LGRBs in our sample with $P_0\sim1~\rm ms$ usually possess $B_p\sim10^{13.5}-10^{15}~\rm G$, while those with $P_0\gtrsim3~\rm ms$ are accompanied by a strong magnetic field of $B_p\sim10^{14.5}-10^{15.5}~\rm G$. Furthermore, we systematically investigated possible correlations between $B_p$ and $P_0$ of the newborn magnetar by applying a least-squares regression algorithm to all combinations of (EoS, $\eta_{\rm X}$).

In Figure \ref{fig:Bp-P0}, we show a series of $B_p-P_0$ scatter distribution diagrams for all combinations of (EoS, $\eta_{\rm X}$). Interestingly, we found that there are generally strong correlations between $B_p$ and $P_0$ for all combinations of (EoS, $\eta_{\rm X}$). Most GRBs in our sample fall into the $3\sigma$ deviation region of the best-fit power-law model. The $B_p/P_0$ center values of the Gaussian fit and the $B_p-P_0$ best-fit correlation results for various combinatorial scenarios are reported in Table \ref{table-1}. The strong correlations of $B_p-P_0$ appeared to be universal for all combinations of (EoS, $\eta_{\rm X}$), which may indicate that the magnetic field of the newborn magnetar is coupled to the initial spin period. More interestingly, if we put all EoSs together, the overall best-fit correlations of $B_p-P_0$ can be approximately described as $B_p\propto P_0^{1.30\pm0.16}$, with the 1$\sigma$ deviation included. Such a correlation seems to hint that the initial spin period $P_0$ estimated from X-ray plateau observations could deviate from that of the magnetar at birth but strongly links to the equilibrium spin period when the accreting magnetar has reached equilibrium owing to its interactions with the surrounding accretion disk, known as the magnetar propeller mechanism. In the context of the magnetar propeller mechanism, the magnetic field and spin period of accreting magnetars satisfy power-law relation $B_p\propto P_{\rm eq}^{7/6}$ for a given mass accretion rate. According to equation (\ref{eq:Peq}), we can also use the statistical distribution $B_p-P_0$ in a population of accreting magnetars to constrain the mass accretion rate in the given gravitational mass and radius of the newborn magnetar, which are highly dependent on the EoS and baryon mass of the NS and can be inferred by using the universal baryonic-to-gravitational mass conversion relation, combined with a numerical solution to the NS field equations.      

In NS problems, two masses are widely discussed, i.e. the baryonic mass $M_b$ and the gravitational mass $M_g$. The former ($M_b$) refers to the total mass of all the baryons (protons, neutrons, etc.) that make up an NS when they are not gravitationally bound and is theoretically relevant because it is directly connected to the initial mass of the iron core of the progenitor massive star, which undergoes gravitational collapse to form the NS. On the other hand, the latter ($M_g$) refers to the mass measured directly from observations and is smaller than $M_b$ owing to the subtraction of the nuclear binding energy. Studying the conversion relationship between $M_b$ and $M_g$ is crucial to set interesting lower limits on the maximum mass ($M_{\rm TOV}$) of a nonrotating NS for ruling out some soft EoS models and can accurately constrain the mass accretion rate of nascent magnetars at different EoSs and baryon masses. \cite{Gao2020} investigated the transformational relations between $M_b$ and $M_g$ for a nonrotating NS or a rotating NS and found that the transformation of the baryonic mass to gravitational mass is universally approximated using the EoS-independent quadratic formula to all selected EoSs
\begin{equation}
M_b=M_g+A\times M_g^2,
\label{eq:Mb-Mg}
\end{equation}
where $A$ is usually adopted as a constant number that depends on the EoS and spin period. Putting together the ($M_b$, $M_g$) results for all adopted EoSs, \cite{Gao2020} found that the overall best-fit $A$ value is 0.080 for a nonrotating NS only and 0.073 when different spin periods are considered. In this work, in order to precisely estimate the mass accretion rate of newborn magnetars at different EoSs and baryon masses, we need to use the statistical distribution $B_p-P_0$ in a population of accreting magnetars and to know the gravitational mass and radius of the NS in different EoSs and spin periods. Based on the best-fit Gaussian distribution center values $P_0$ for our sample of newborn magnetars under various scenarios, combined with the $M_b-M_g$ transition relation and public code RNS \citep{Stergioulas1995} to solve the field equations for rotating NSs, we can infer the corresponding gravitational masses and radii of newborn magnetars in various combinations of (EoS, $M_b$, $\eta_{\rm X}$) for different spin periods $P_0$. The details are as follows: (1) according to Table \ref{table-1} to get the central values $P_0$ of the spin period distribution in various (EoS, $M_b$, $\eta_{\rm X}$) combinations; (2) by numerically solving the field equations for rotating NSs using the public RNS code \citep{Stergioulas1995}, we can obtain the Keplerian periods $P_k$ and the initial radii at $P_0$ for various (EoS, $M_b$, $\eta_{\rm X}$) combinations; (3) to estimate the $P_k$ and $P_0$ relations in various (EoS, $M_b$, $\eta_{\rm X}$) combinations, and to determine the $A$ values by searching for the best-fit results for the $M_b-M_g$ conversion of rotating NSs in Table 2 of \cite{Gao2020}, and then use the universal relation $M_b=M_g+A\times M_g^2$ to obtain the corresponding $M_g$ in various (EoS, $M_b$, $\eta_{\rm X}$) combinations; (4) finally, the ranges of mass accretion rate for newborn magnetar in various (EoS, $M_b$, $\eta_{\rm X}$) scenarios are accurately estimated from Equation (\ref{eq:Peq}) and the $B_p-P_0$ distribution in our GRB sample. Table \ref{table-2} reports the characteristic parameters and the corresponding mass accretion rate for various (EoS, $\eta_{\rm X}$) combinations under the $M_b=2.0M_{\odot}$ scenario.

According to the results of Figure \ref{fig:Bp-P0} and Table \ref{table-2}, in the context of the magnetar propeller mechanism, we can find that all our LGRB data in the $B_p-P_0$ plane are encompassed by two lines corresponding to the range of mass accretion rates $10^{-5}~M_{\odot}~{\rm s^{-1}}<\dot{M}<10^{-2}~M_{\odot}~\rm s^{-1}$ in combining the results of (EoS, $\eta_{\rm X}$) together. The results are inconsistent with the statistical results in \cite{Stratta2018} and \cite{Linweili2020} that are in the range $10^{-4}~M_{\odot}~{\rm s^{-1}}<\dot{M}<10^{-1}~M_{\odot}~\rm s^{-1}$, which used the constant $R$, $I$ and $M_{g}$ ($M_{g}=1.4M_{\odot}$, $R=12~\rm km$, $I=0.35M_gR^2$) for their GRB sample, and the results for our constrained accretion rate are one order of magnitude lower compared to their results. A lower accretion rate appears to be more physically justified, since under a high accretion rate ($0.1~M_{\odot}~\rm s^{-1}$), the mass accreted onto the surface of the magnetar could exceed $1M_{\odot}$ before reaching the equilibrium spin period. This would likely cause the magnetar to collapse into a BH before attaining spin equilibrium, rendering the magnetar propeller mechanism ineffective. The main reason for this discrepancy should be that we have corrected the magnetar parameters $B_p$ and $P_0$ by systematically incorporating $R/I$ evolutionary effects and then using the $M_b-M_g$ transition relation and the $B_p-P_0$ statistical distributions of different EoSs and X-ray radiative efficiencies in combination with the propeller model. In other words, the propeller properties of accreting magnetars are obtained by using the constant $R/I$ scenario in previous studies that will present results with a higher mass accretion rate, which may impair our understanding of the physical nature of an accreting magnetar and its surroundings, and will further diminish the reliability of inverse inference about the physical properties of its progenitor. The precise constraint on the mass accretion rate utilizing evolved $R/I$ will be crucial to understanding the physical nature of an accreting magnetar and its progenitor. 

\begin{figure*}
\centering
\includegraphics  [angle=0,scale=0.40]   {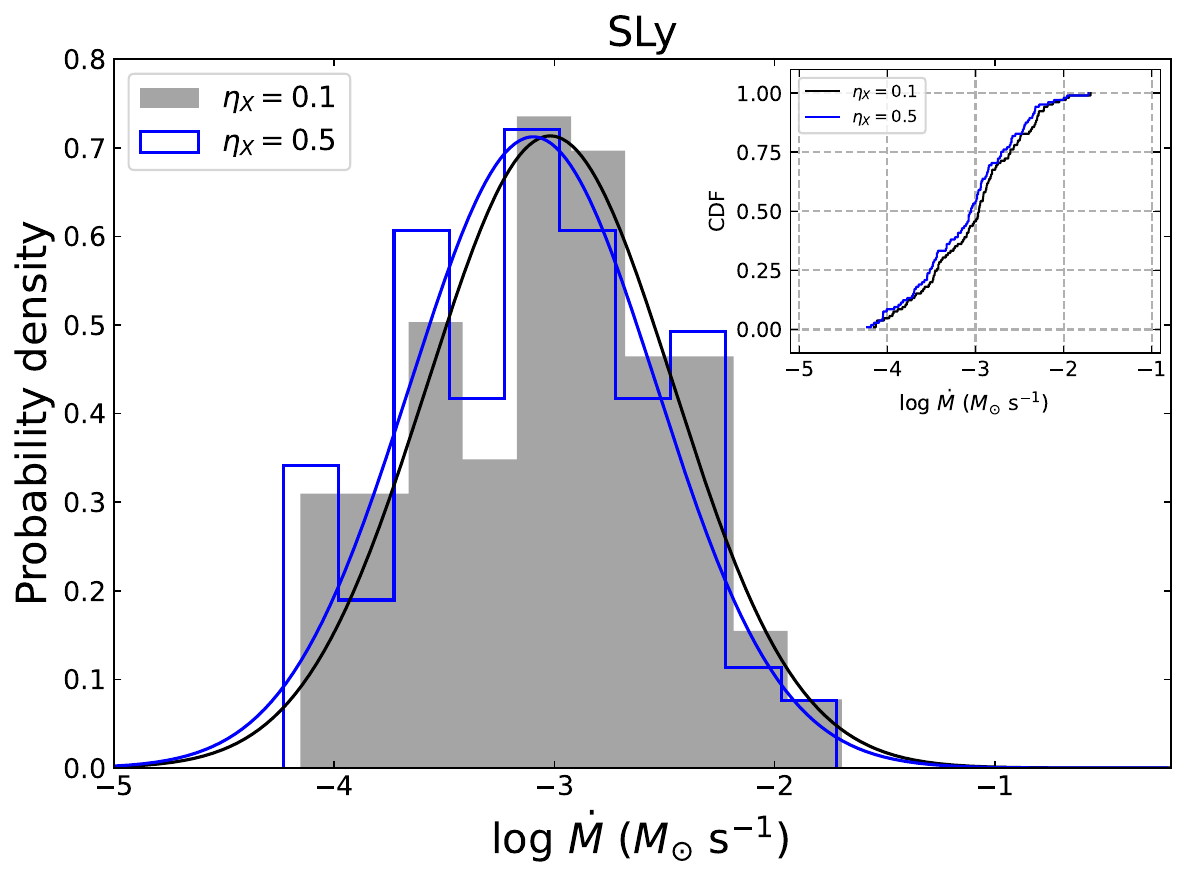}
\includegraphics  [angle=0,scale=0.40]   {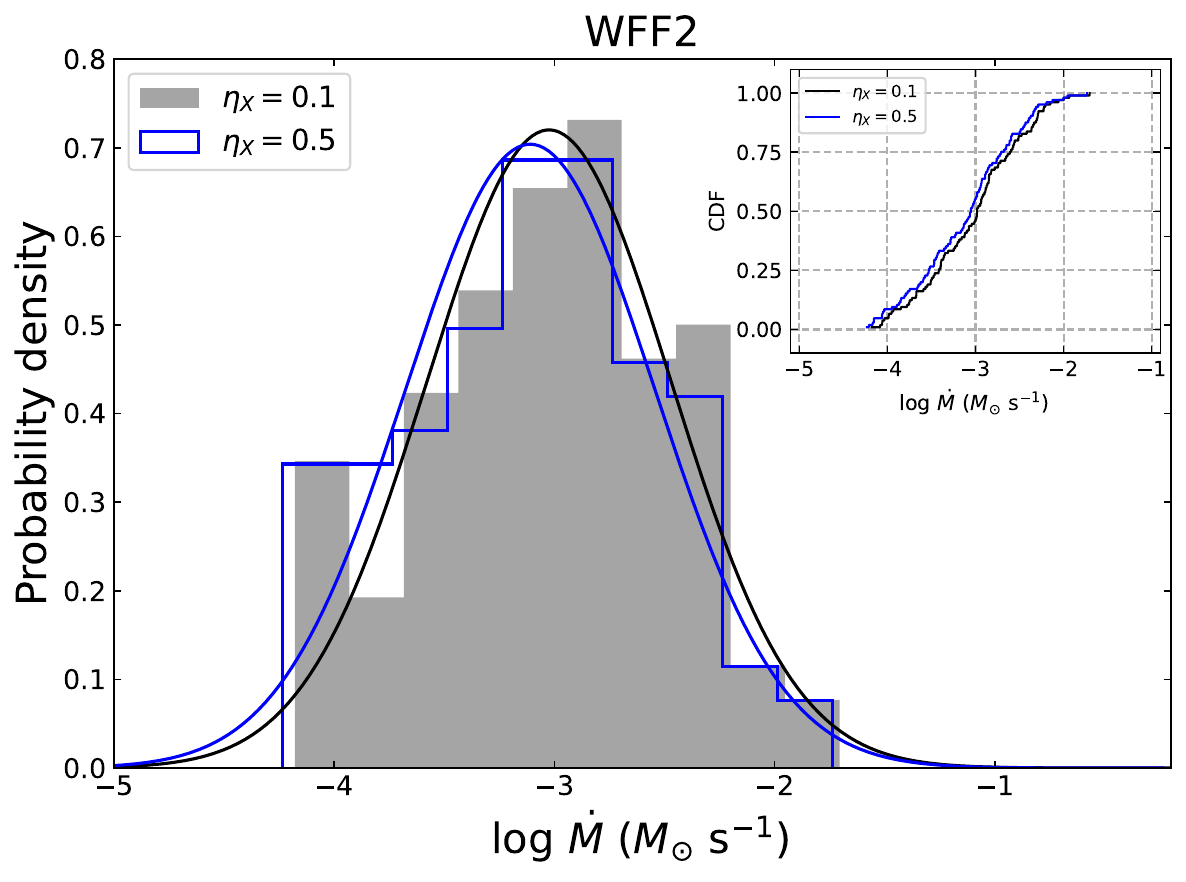}
\includegraphics  [angle=0,scale=0.40]   {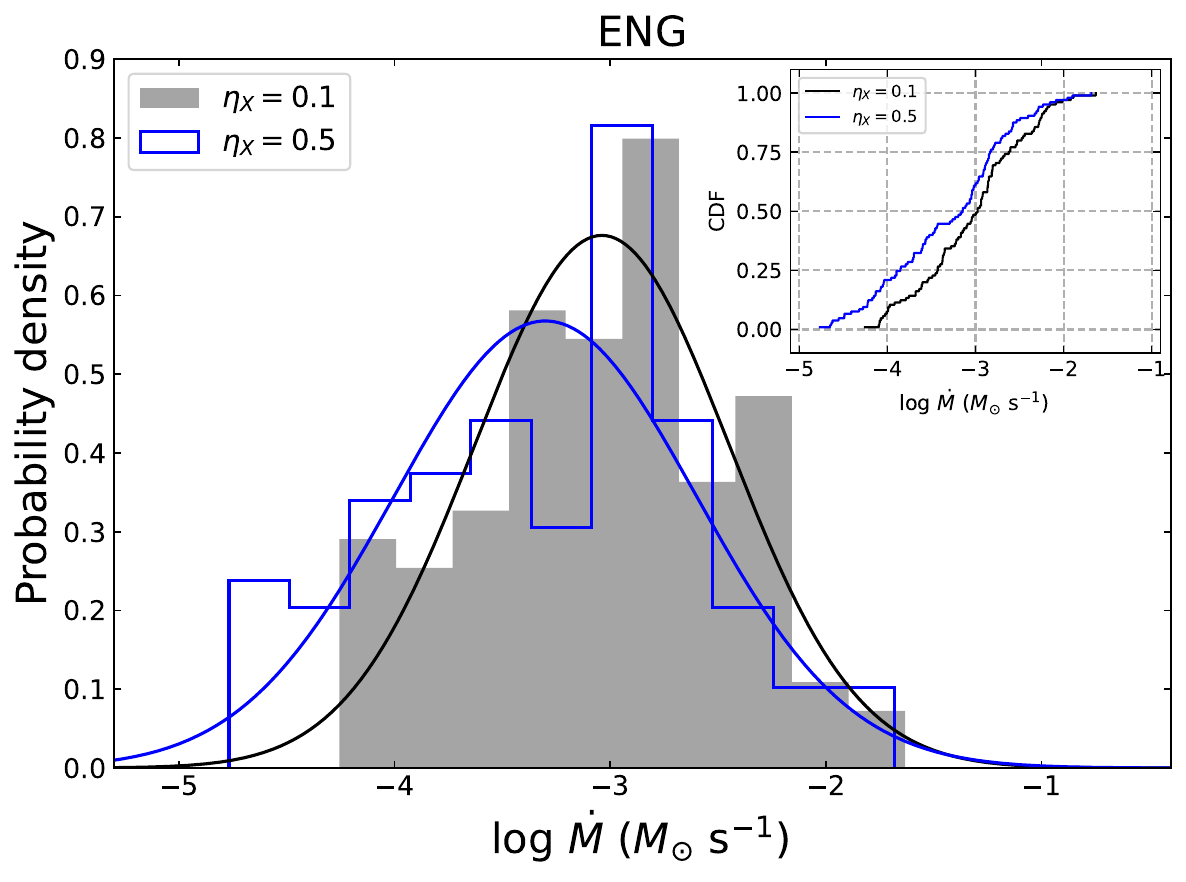}
\includegraphics  [angle=0,scale=0.40]   {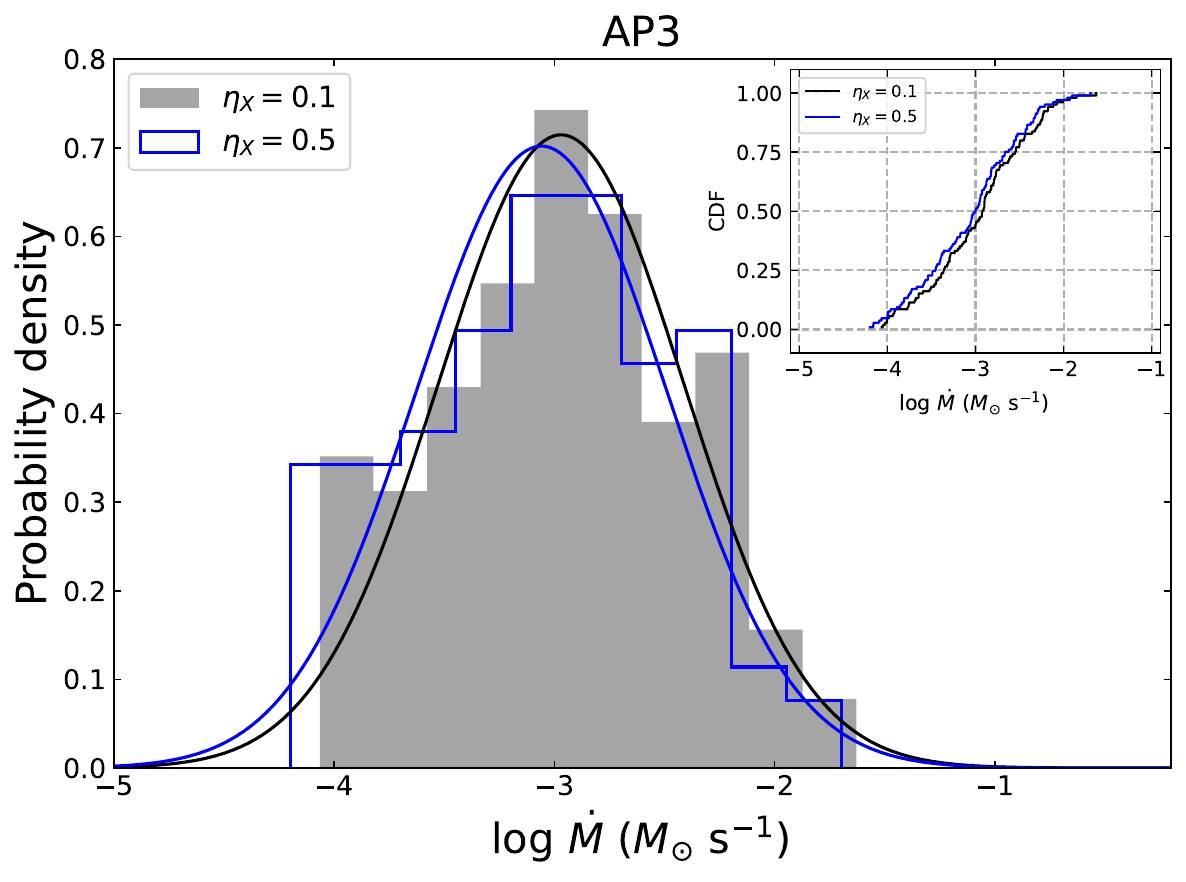}
\includegraphics  [angle=0,scale=0.40]   {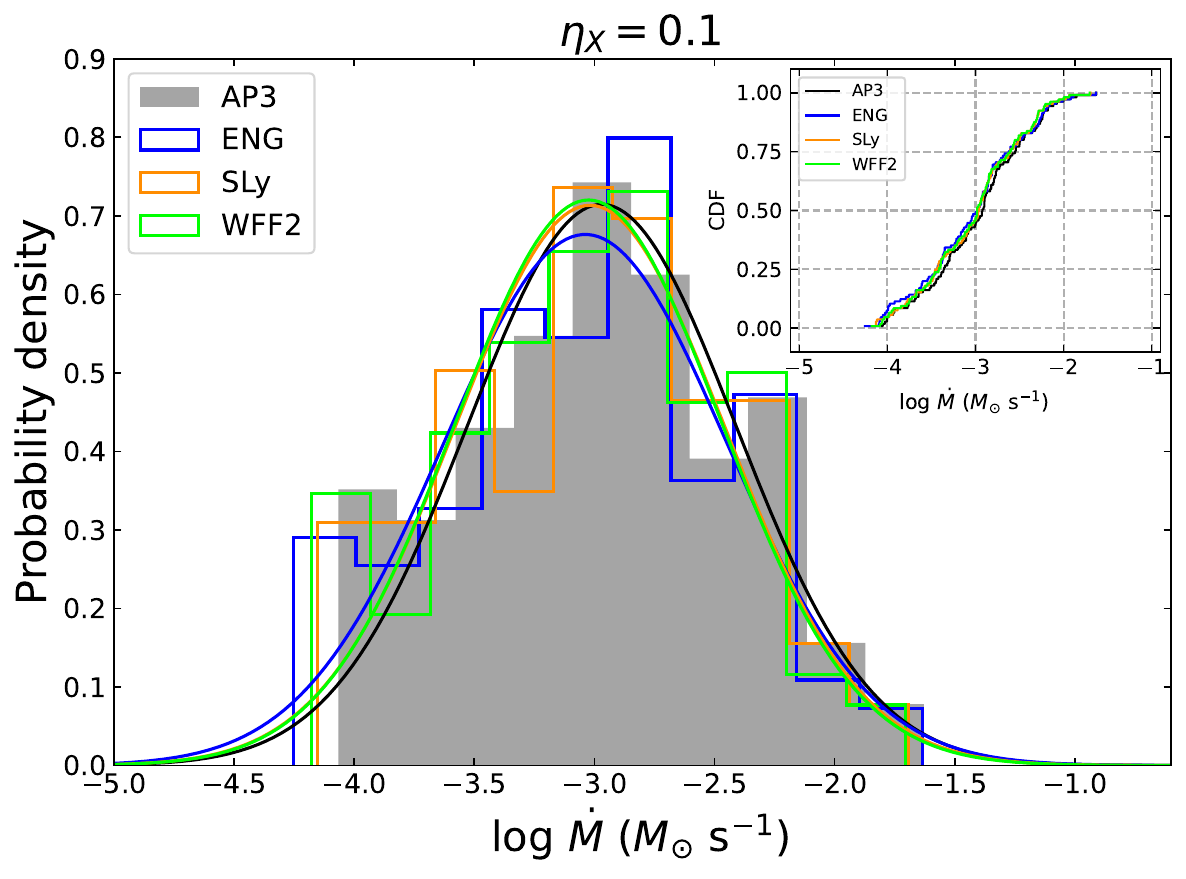}
\includegraphics  [angle=0,scale=0.40]   {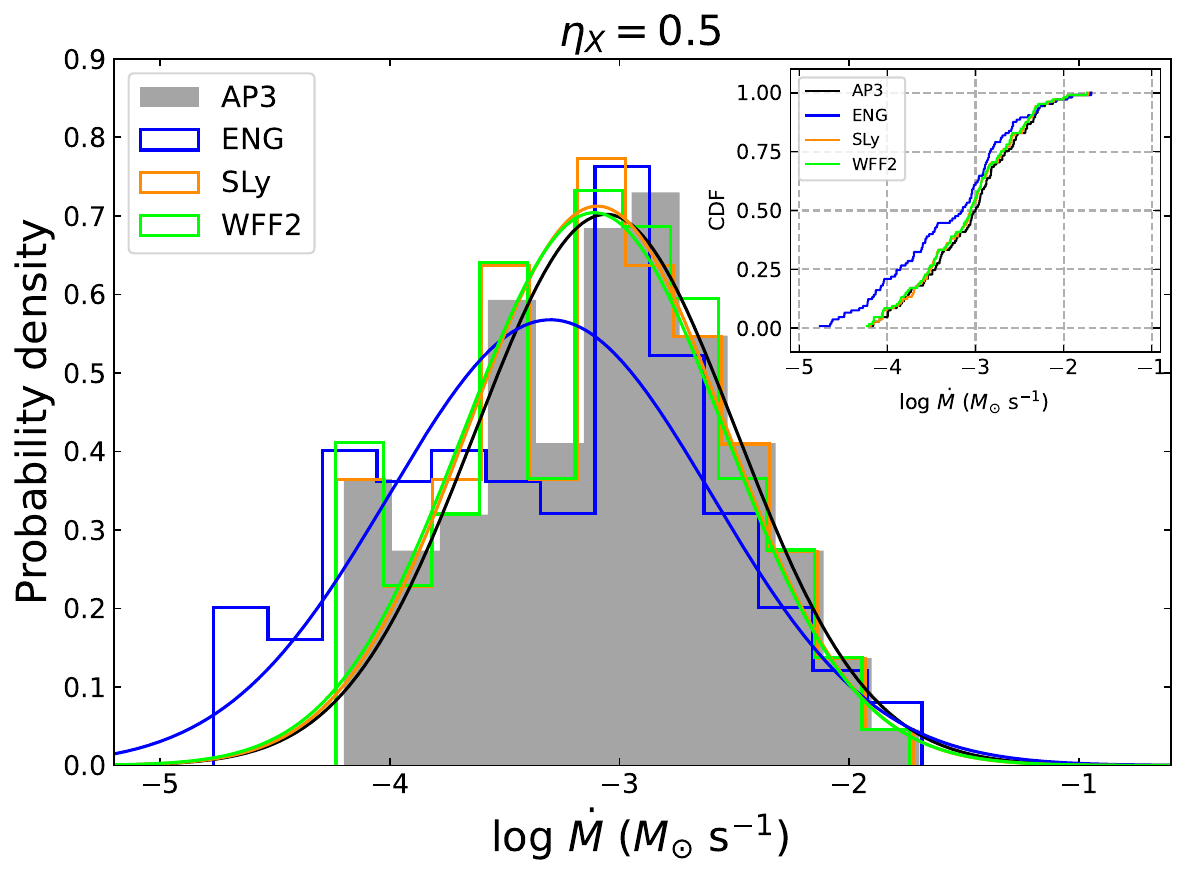}
\caption{Comparisons of the derived magnetar accretion rate parameter $\dot{M}$ histograms in different EoSs and $\eta_X$ for our LGRB sample. The solid lines of different colors are the best Gaussian fits. The cumulative distributions corresponding to $\dot{M}$ in different EoSs and $\eta_X$ are also displayed in the inset.}
\label{fig:Mdot distribution}
\end{figure*}

Furthermore, we performed the relevant statistical analysis for the mass accretion rate in our GRB sample. In Figure \ref{fig:Mdot distribution}, we show the probability density distributions and the cumulative distributions of mass accretion rate in different NS EoSs and radiative efficiencies. The log-normal distributions can be described as
$\log\dot{M}^{(\rm SLy,\eta_X=0.1)}/{M_{\odot}~\rm s^{-1}}=-3.02\pm0.56$, $\log\dot{M}^{(\rm SLy,\eta_X=0.5)}/{M_{\odot}~\rm s^{-1}}=-3.10\pm0.56$, $\log\dot{M}^{(\rm WFF2,\eta_X=0.1)}/{M_{\odot}~\rm s^{-1}}=-3.03\pm0.55$, $\log\dot{M}^{(\rm WFF2,\eta_X=0.5)}/{M_{\odot}~\rm s^{-1}}=-3.11\pm0.57$, $\log\dot{M}^{(\rm ENG,\eta_X=0.1)}/{M_{\odot}~\rm s^{-1}}=-3.04\pm0.59$, $\log\dot{M}^{(\rm ENG,\eta_X=0.5)}/{M_{\odot}~\rm s^{-1}}=-3.30\pm0.70$, $\log\dot{M}^{(\rm AP3,\eta_X=0.1)}/{M_{\odot}~\rm s^{-1}}=-2.97\pm0.56$, $\log\dot{M}^{(\rm AP3,\eta_X=0.5)}/{M_{\odot}~\rm s^{-1}}=-3.06\pm0.57$. To compare whether there are significant differences in the constrained mass accretion rates for different EoSs and radiative efficiencies, we used the Kolmogorov--Smirnov (K-S) algorithm to test the deviations of data set for different EoSs and radiative efficiencies. Putting $\eta_{\rm X}$ results for our adopted EoSs together, the overall $p$-value of the K-S test for the derived mass accretion rates between different NS EoSs and different radiative efficiencies can be approximately expressed as $p_{\rm KS}<7.3\times10^{-1}$, and $p_{\rm KS}<5.3\times10^{-2}$, respectively. The K-S test results indicate that the null hypothesis that the sets of mass accretion rate in different EoSs and radiative efficiencies scenarios come from the same population cannot be rejected. Combining the results of Figure \ref{fig:Mdot distribution} and the K-S test, we can find that there are no significant differences in the constrained mass accretion rates for different EoSs, and the constraints on magnitude of mass accretion rate seem to be independent of the EoS stiffness. However, there are systematic differences in the constrained mass accretion rates for different radiative efficiencies, and with higher X-ray radiative efficiency, the constrained mass accretion rates tend to become larger.

\begin{figure*}
\centering
\includegraphics  [angle=0,scale=0.29]   {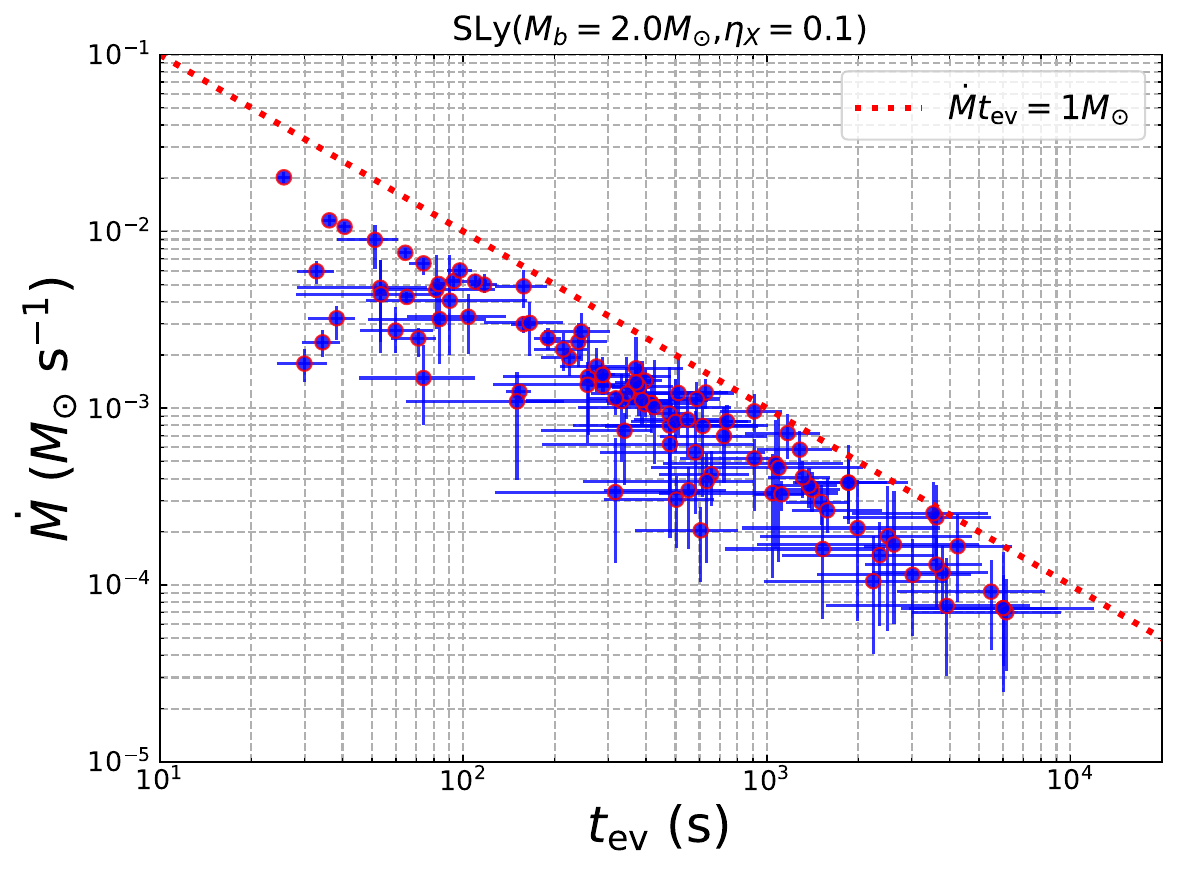}
\includegraphics  [angle=0,scale=0.29]   {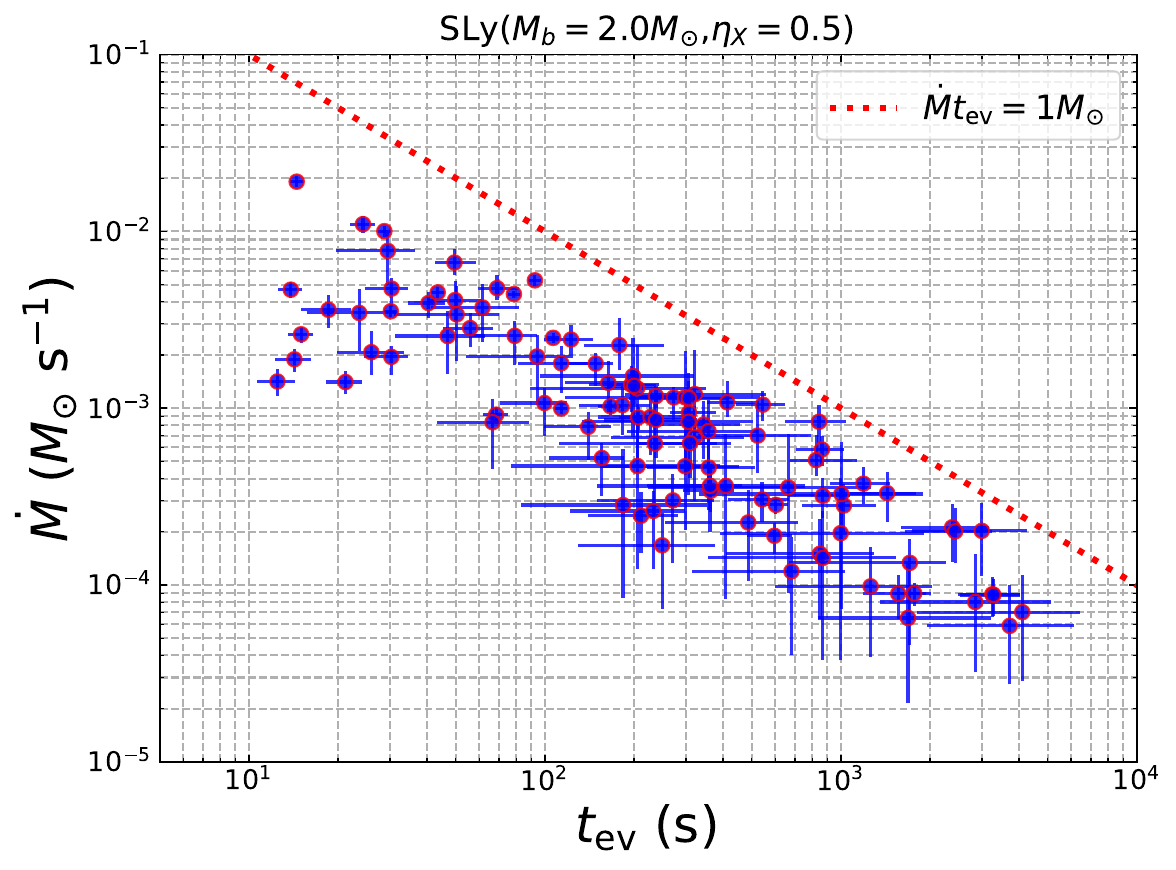}
\includegraphics  [angle=0,scale=0.29]   {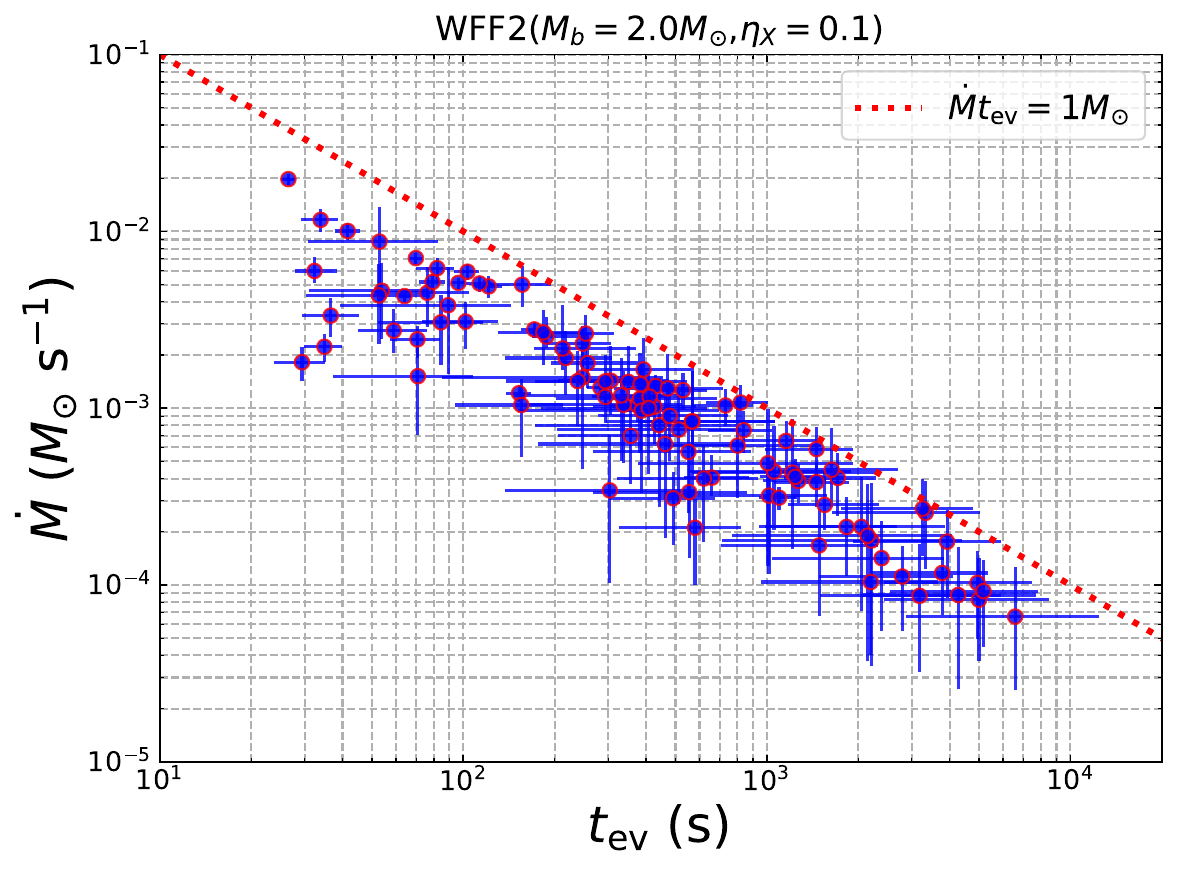}
\includegraphics  [angle=0,scale=0.29]   {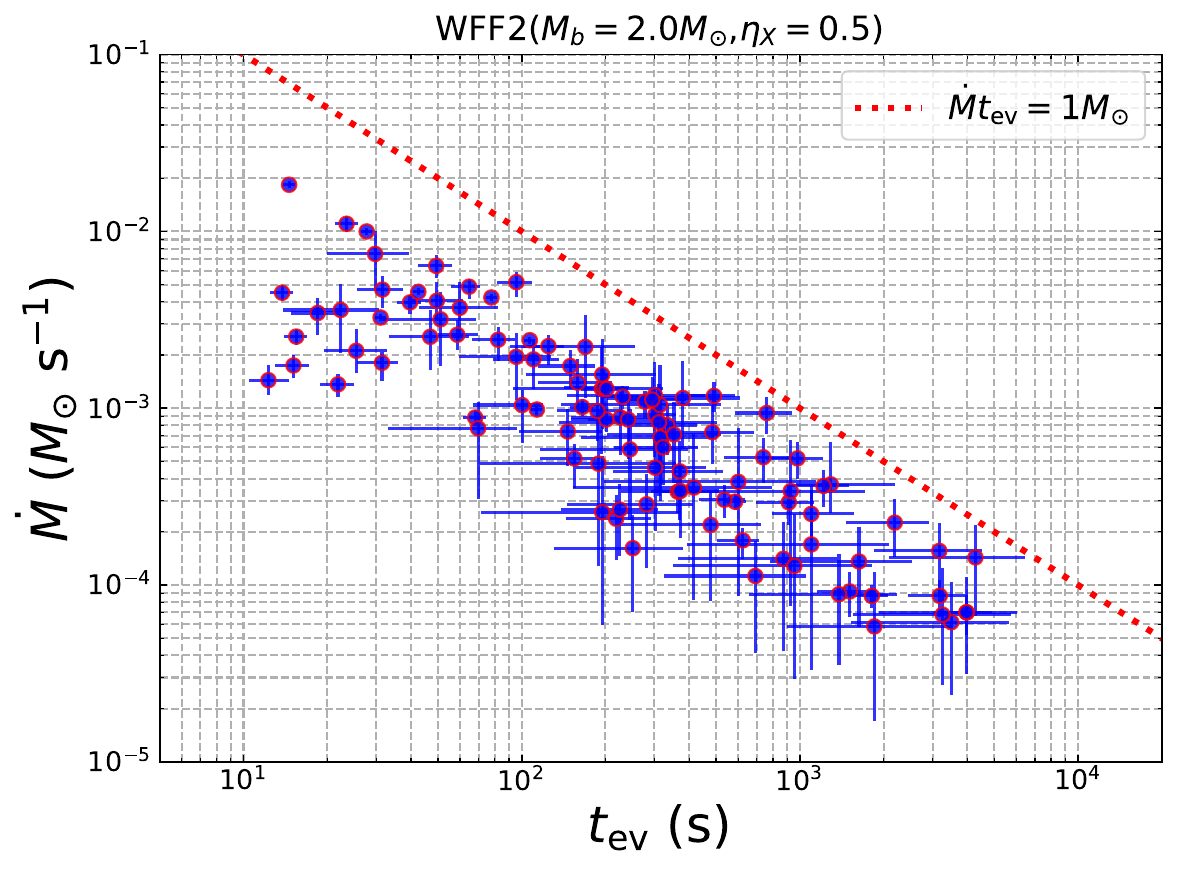}
\includegraphics  [angle=0,scale=0.29]   {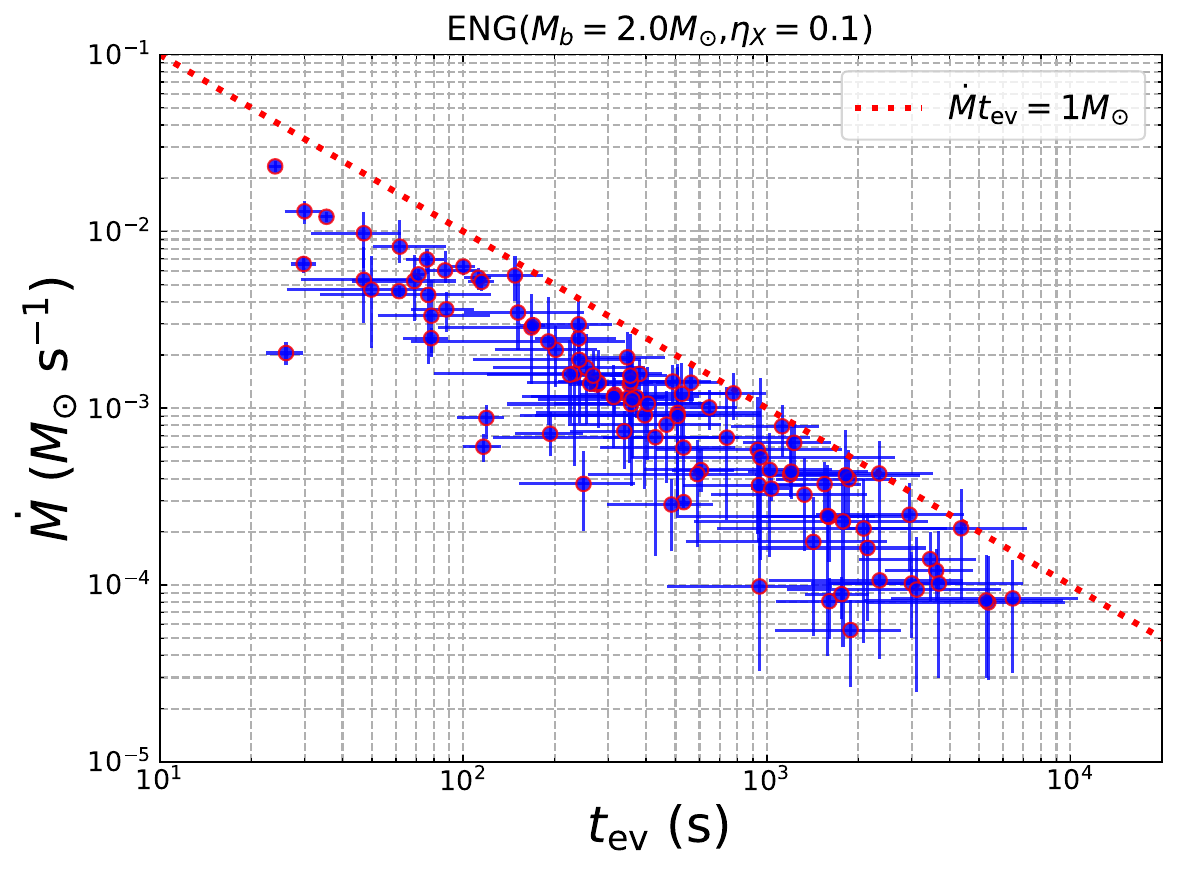}
\includegraphics  [angle=0,scale=0.29]   {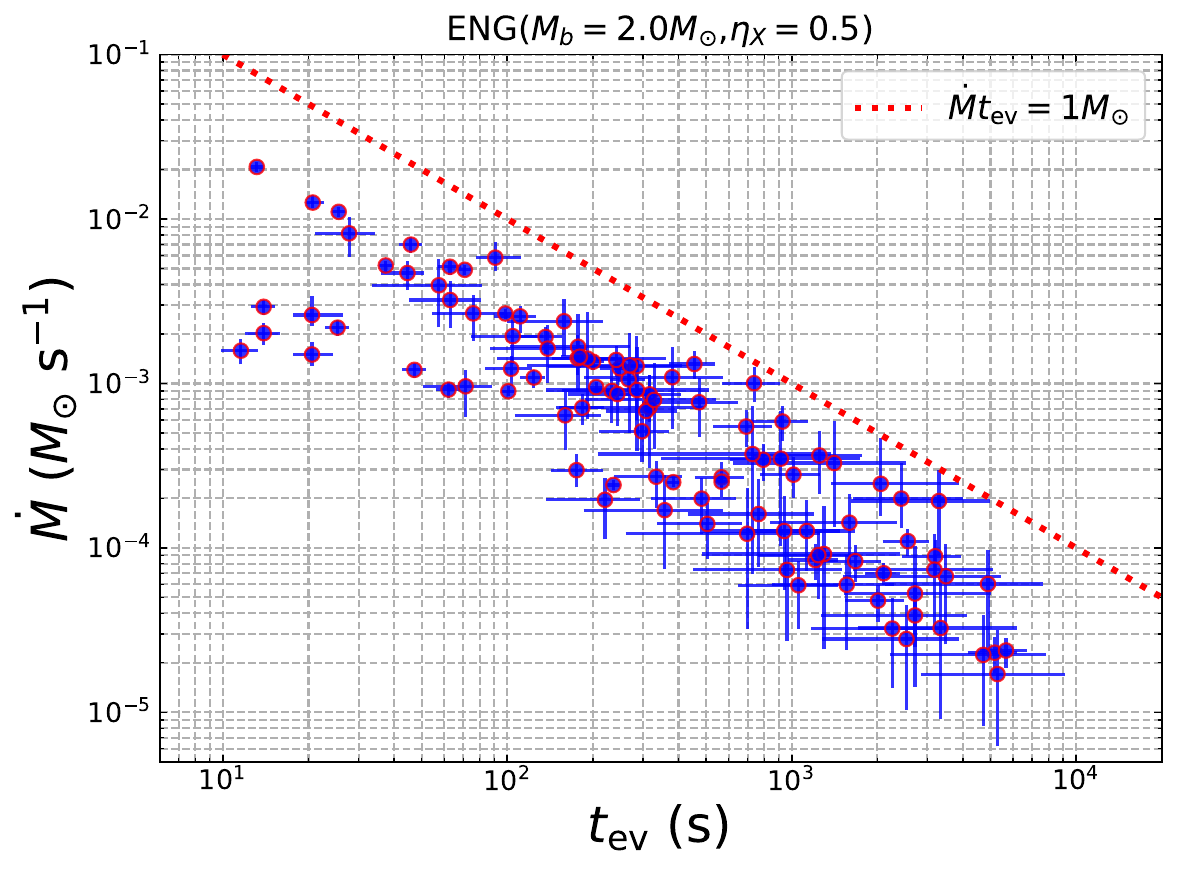}
\includegraphics  [angle=0,scale=0.29]   {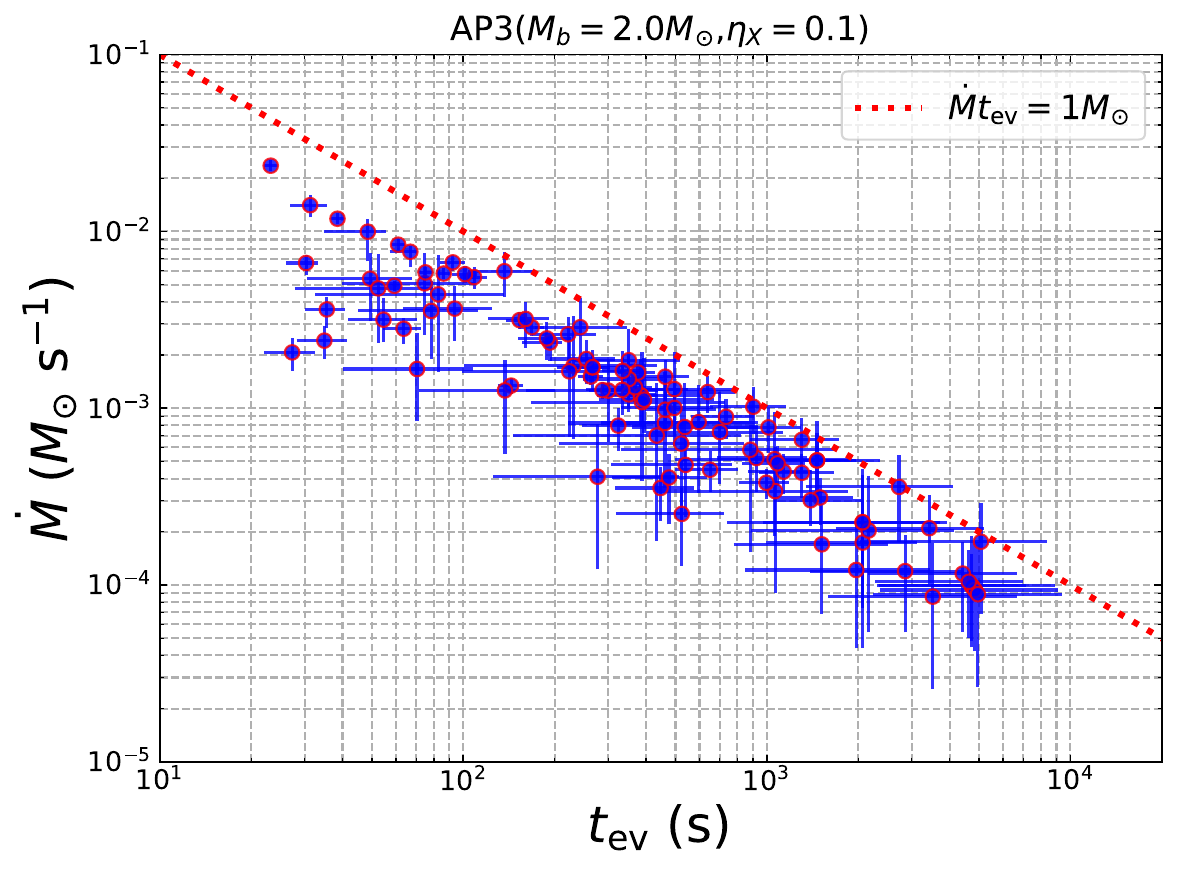}
\includegraphics  [angle=0,scale=0.29]   {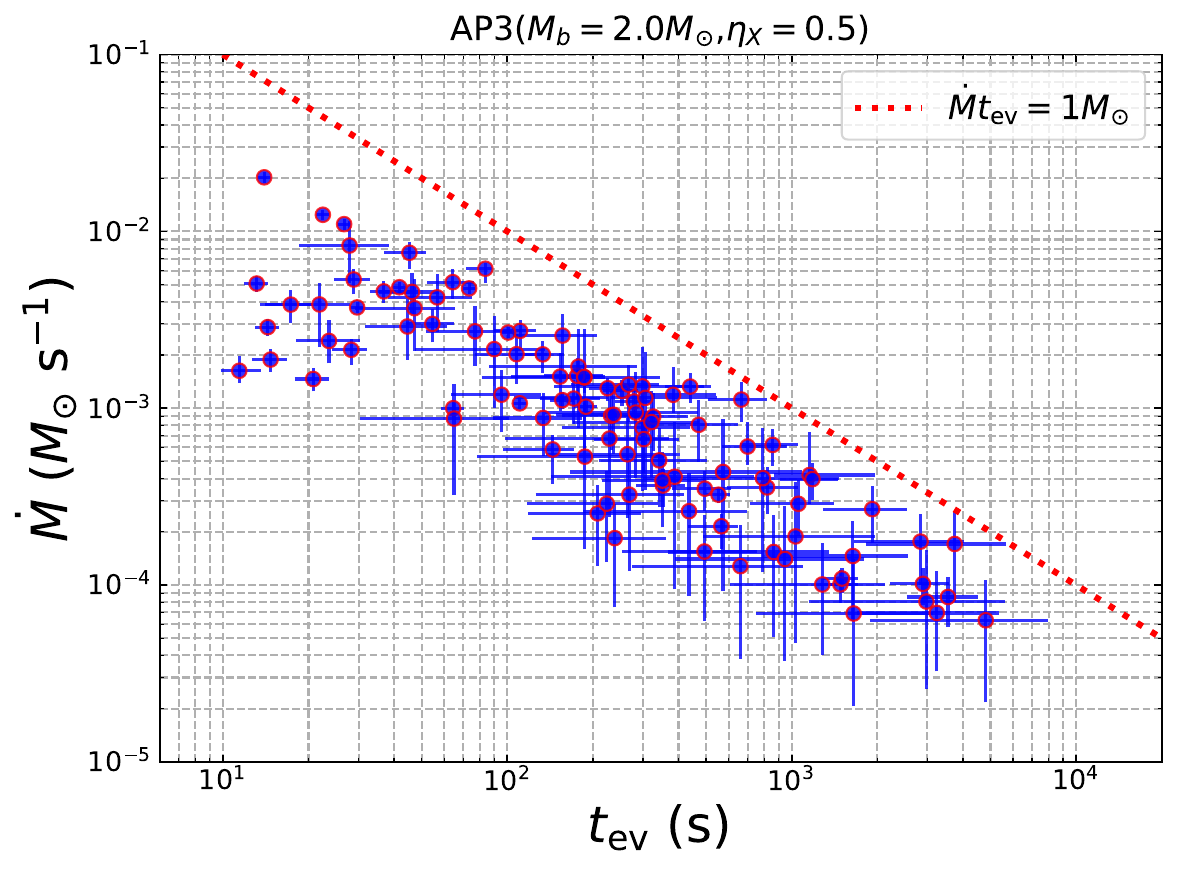}
\caption{The accretion rate $\dot{M}$ vs. the lower limit of accretion timescale $t_{\rm ev}$ in different EoSs and $\eta_X$ for our LGRB sample. The red dotted lines correspond to $\dot{M}t_{\rm ev}=1M_{\odot}$.}
\label{fig:Mdot-Tev}
\end{figure*}

It is important to note that the newborn magnetar is likely to collapse into a BH if it accretes a significant amount of material and exceeds the maximum gravitational mass ($M_{\rm max}$) of a stable NS. One interesting question is, can the newborn accreting magnetars in our GRB sample have collapsed into a BH before reaching the equilibrium spin period as a result of accreting a significant amount of material and exceeding the maximum gravitational mass of an NS? The collapse process would directly affect the sustainability of the magnetar central engine and the jet energy injection mechanism. Under certain approximations, we can estimate the accretion mass of newborn magnetar in our GRB sample once they reach the equilibrium spin period. According to Equation (\ref{eq:Tev}), we first estimated the evolutionary timescales $t_{\rm ev}$ for these accreting magnetars to reach an equilibrium spin period in various (EoS, $\eta_{\rm X}$) combination scenarios. In Figure \ref{fig:Mdot-Tev}, we plotted a series of $\dot{M}-t_{\rm ev}$ scatter diagrams for different EoSs and radiative efficiencies based on our GRB sample. We find that the accretion masses of all newborn accreting magnetars and the majority of newborn accreting magnetars in our GRB sample are less than 1$M_{\odot}$ and 0.5$M_{\odot}$ for various (EoS, $\eta_{\rm X}$) combination scenarios, respectively. Such statistical results seem to imply that the majority of newborn accreting magnetars in our GRB sample will not exceed the maximum gravitational mass of an NS with a stiff EoS before reaching the accretion equilibrium spin period and will not collapse into a BH before reaching the accretion equilibrium spin period for an NS with a stiff EoS. In order to test whether these accreting magnetars in our GRB sample can actually survive the equilibrium spin period for our selected EoSs, we need to further estimate the maximum gravitational mass $M_{\rm max}$ and the total gravitational mass ($M_{\rm tot}$) of accreting NSs for our selected EoSs. 

\begin{figure*}
\centering
\includegraphics  [angle=0,scale=0.40]   {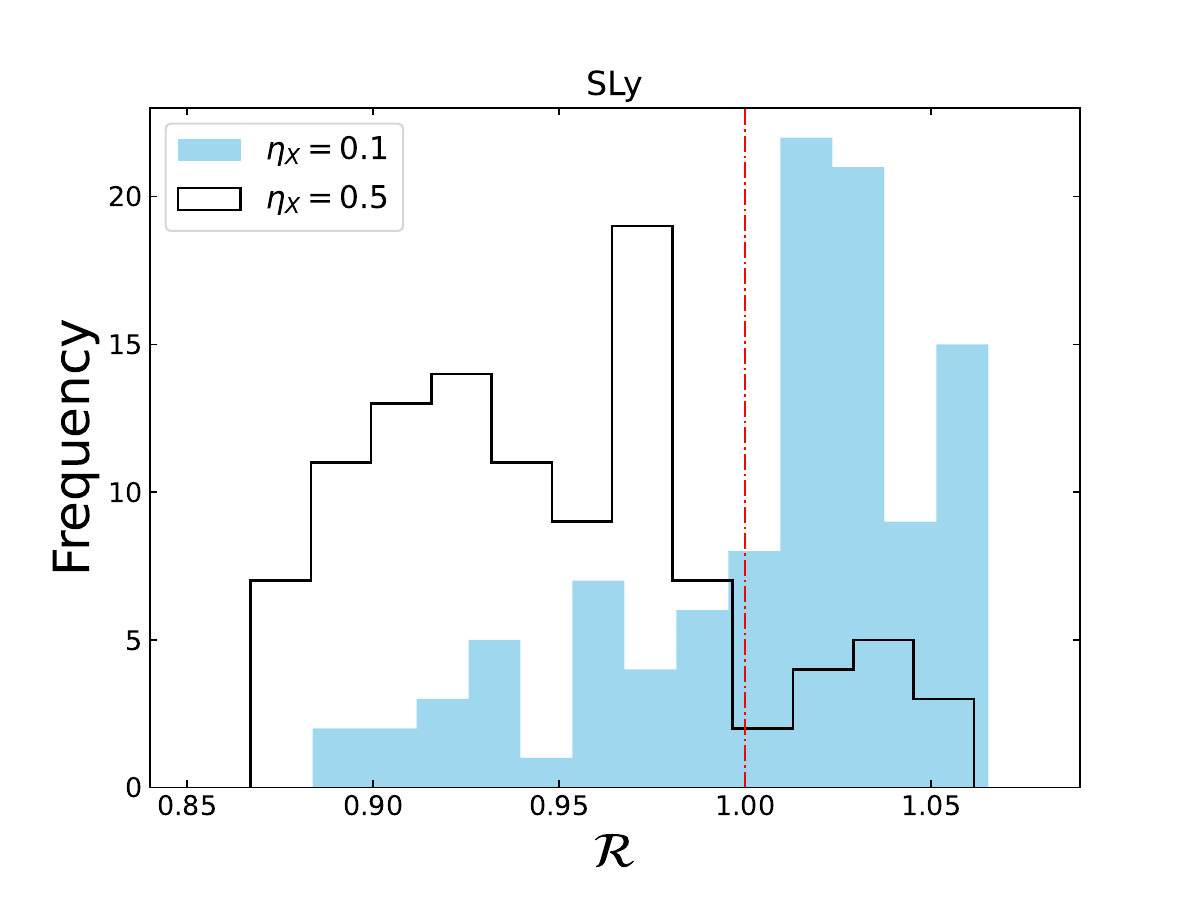}
\includegraphics  [angle=0,scale=0.40]   {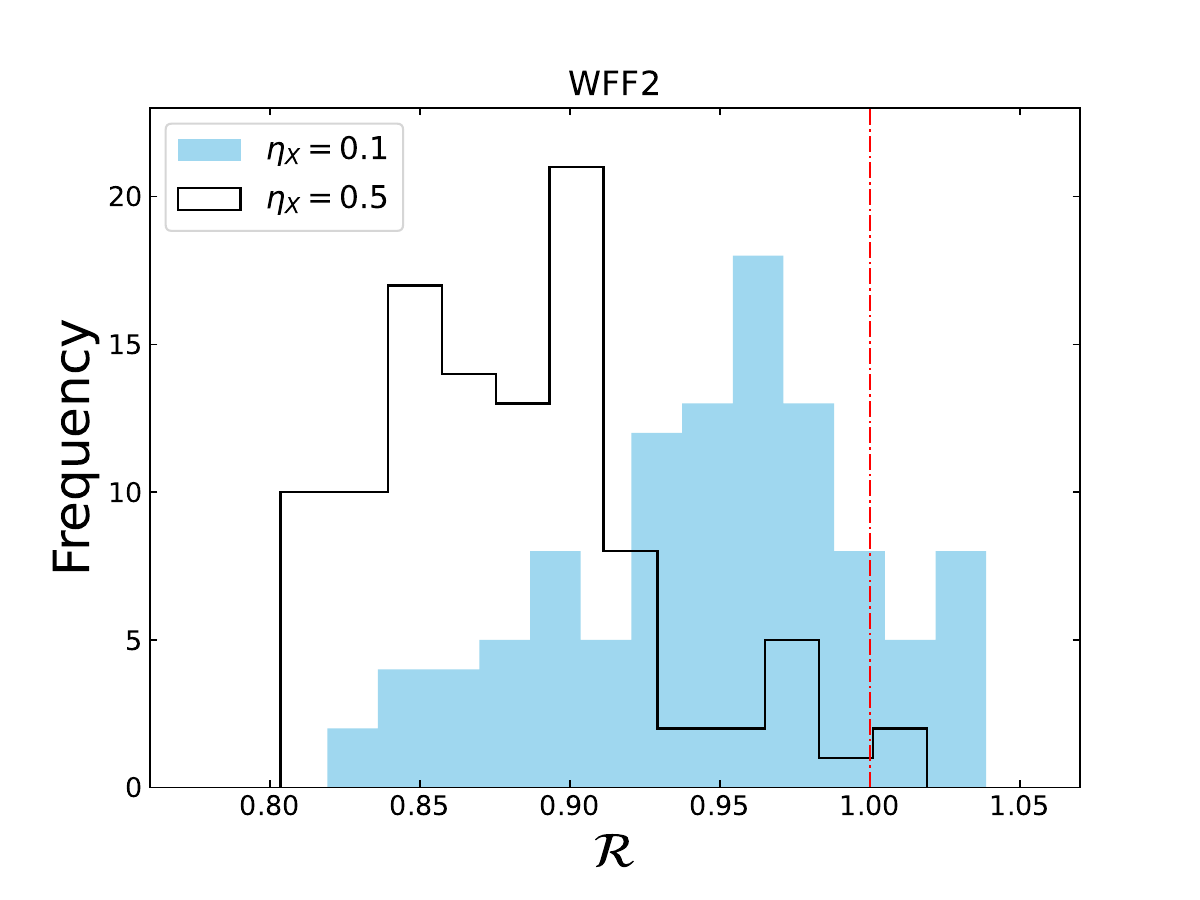}\\
\includegraphics  [angle=0,scale=0.40]   {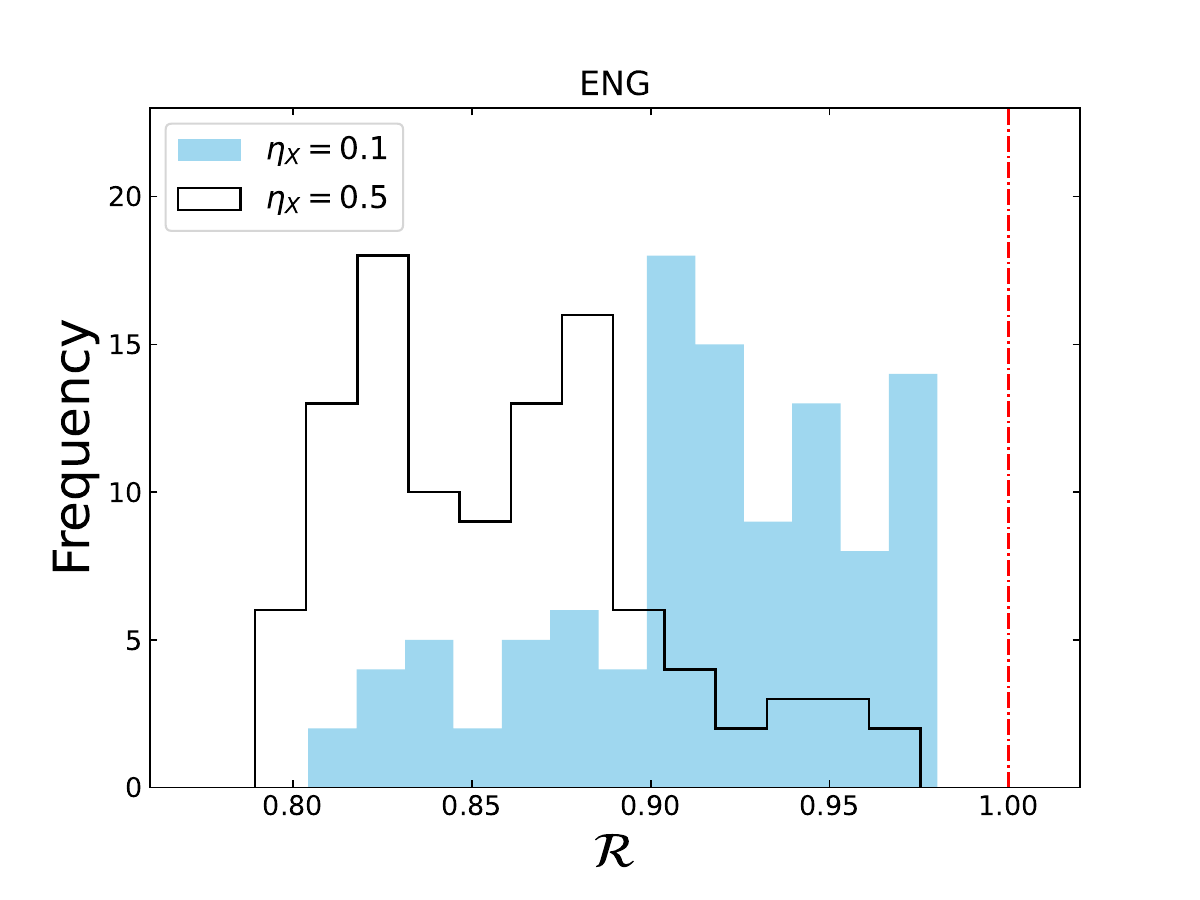}
\includegraphics  [angle=0,scale=0.40]   {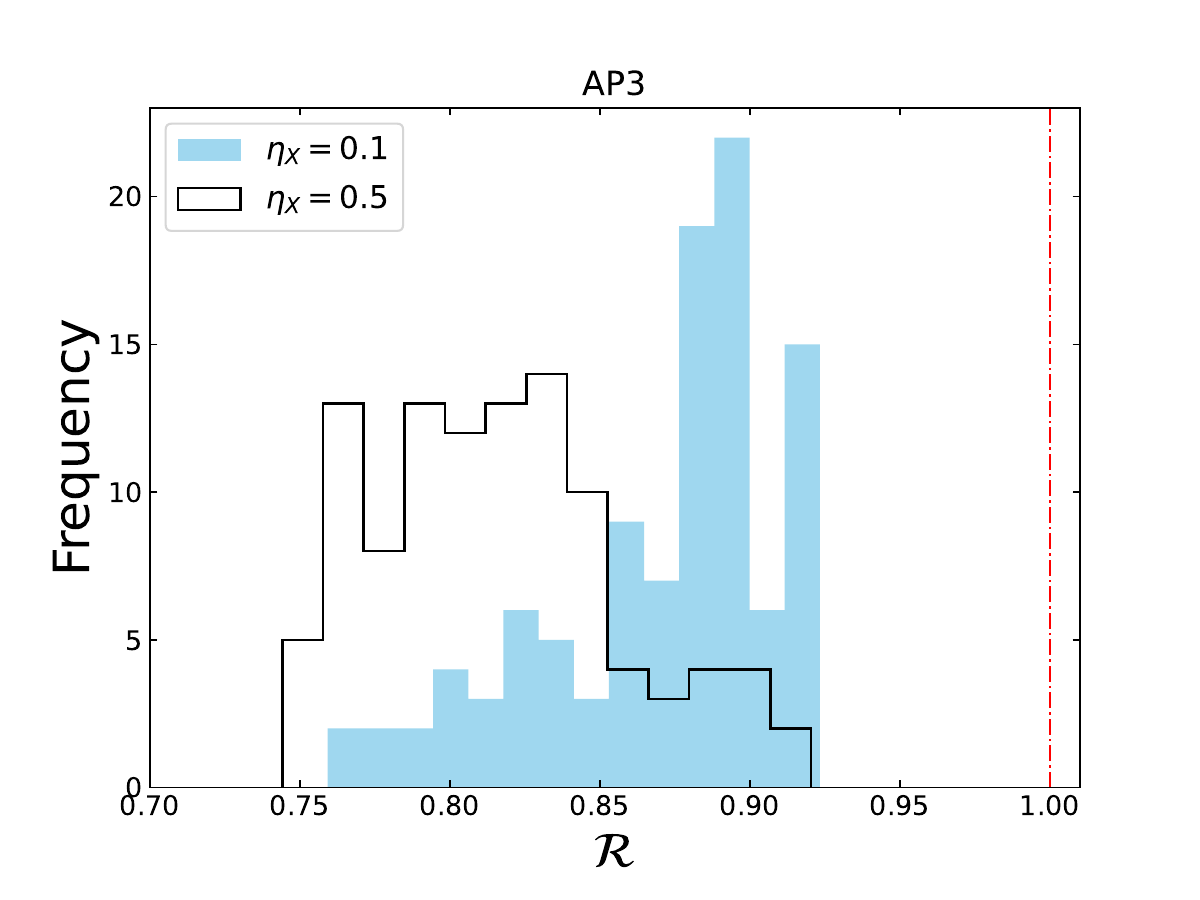}\\
\caption{The distributions of the mass ratio parameter ${\cal R}=M_{\rm tot}/M_{\rm max}$ in different EoSs and $\eta_X$ for our LGRB sample. The red vertical dashed--dotted lines correspond to ${\cal R}=1$, which marks the boundary where an accreting magnetar collapses into a BH.}
\label{fig:Mtot-Mmax}
\end{figure*}

For a given EoS, the maximum gravitational mass $M_{\rm max}$ of a rotating NS depends on the rotation period and the maximum mass of a nonrotating NS. It can be expressed as \citep{Lyford2003,Lasky2014}
\begin{eqnarray}
M_{\rm max} = M_{\rm TOV}\left(1+\alpha P^{\beta}\right)
\label{eq:M_max}
\end{eqnarray}
where $\alpha$ and $\beta$ depend on the NS EoS, and the values of $\alpha$ and $\beta$ for our selected EoSs are presented in Table \ref{table-2}. For an accreting magnetar, the total gravitational mass energy when it reaches the accretion equilibrium spin period can be expressed as
\begin{eqnarray}
M_{\rm tot} = M_{g}+M_{\rm acc}
\label{eq:M_tot}
\end{eqnarray}
where $M_{\rm acc}\simeq \dot{M}_{\rm eq}t_{\rm ev}$ approximately represents the mass accreted onto the magnetar surface before reaching accretion equilibrium in the interacting magnetar--disk system. In order to identify whether the accreting magnetar has collapsed into a BH before reaching accretion equilibrium, we define a dimensionless mass ratio parameter ${\cal R}=M_{\rm tot}/M_{\rm max}$. If ${\cal R}>1$ is satisfied for the newborn accreting magnetar, we can assume that it has collapsed into a BH before it reaches the accretion equilibrium spin period. If ${\cal R}<1$ is satisfied for the newborn accreting magnetar, we can assume that it remains a stable NS when it reaches the accretion equilibrium spin period and will not collapse into a BH. Combining Equations (\ref{eq:M_max}) and (\ref{eq:M_tot}), we can derive the distributions of mass ratio ${\cal R}$ for our GRB accreting magnetar sample in various (EoS, $\eta_{\rm X}$) combination scenarios. In Figure \ref{fig:Mtot-Mmax}, we show the distributions of mass ratio ${\cal R}$ for our GRB accreting magnetar sample in different EoSs and radiative efficiencies. We find that no matter whether the radiative efficiency $\eta_{\rm X}$ is 0.1 or 0.5, most of the accreting magnetars in our GRB sample can actually survive the equilibrium spin period for two soft SLy and WFF2 EoSs ($M_{\rm TOV}=2.05~M_{\odot}$ and $2.20~M_{\odot}$), and all the accreting magnetars in our GRB sample can actually survive the equilibrium spin period for two stiff EoSs ($M_{\rm TOV}=2.24~M_{\odot}$ and $2.39~M_{\odot}$). In other words, for various (EoS, $\eta_{\rm X}$) combination scenarios, the majority of the nascent accreting magnetars in our GRB sample are able to survive until they reach the equilibrium spin period and will not collapse into BHs because their total mass after accreting materials to the magnetar surface does not exceed its maximum gravitational mass. 

Further, we can try to infer the physical nature of the progenitors of accreting magnetars in our GRB sample under certain approximations. As the central engine of LGRBs with X-ray plateau emission, the newborn magnetar should be born in SN explosion and be highly magnetized and rapidly rotating. Thus, it is natural to consider that a fraction of the progenitor fragment in post-collapse could fall back to circulate into an accretion disk around the newborn magnetar with an accretion rate greatly exceeding the Eddington limit ($\dot{M}_{\rm Edd}$). The highly super-Eddington accretion disk is expected to generate a geometrically very thick, advection-dominated accretion disk \citep{Beloborodov1998}, which likely drives large-scale outflows within the time range of our interest \citep[$\gtrsim1000$ s since SN explosion;][]{Dexter2013}. Assuming that the accretion rate at the outer radius of the disk is equal to the mass fallback rate \citep{Michel1988,Metzger2018,Xu2019}, we can derive the following formula:
\begin{eqnarray}
\dot{M}_\mathrm{d,out}=\dot{M}_\mathrm{fb}=\frac{2M_\mathrm{fb}}{3t_\mathrm{fb}}\left(1+t/t_\mathrm{fb}\right)^{-5/3},
\label{eq:Mdot_fb}
\end{eqnarray} 
where $M_{\rm fb}$ is the total fallback mass available for the accretion disk, $t_{\rm fb}=\epsilon t_{\nu}$ is the fallback timescale, and $t_{\nu}$ and $\epsilon$ are the viscous timescale and the ratio between the fallback timescale and the viscous timescale, respectively. Due to the fact that the fallback accretion process is often accompanied by mass outflow, only a fraction ($\eta_1$) of the accretion rate would reach the inner disk radius, which is the true accretion rate onto the magnetar surface and can be expressed as
\begin{eqnarray}
\dot{M}_\mathrm{d,in}=\eta_1 \dot{M}_\mathrm{d,out},
\label{eq:Mdot_in}
\end{eqnarray}
where $\eta_1$ is a dimensionless efficiency value with $0<\eta_1<1$. Considering the effects of advection-dominated process and mass outflow, \cite{Mushtukov2019} found $\eta_1\gtrsim0.6$ or $\eta_1\gtrsim0.4$ when the disk outflow is powered by half or all of the viscously dissipated energy, respectively. Their numerical simulations also show that $\eta_1$ will tend to approach the minimum as the initial accretion rate increases from $1$ to $\sim1000~\dot{M}_{\rm Edd}$, which is significantly lower than our derived accretion rate for a newborn magnetar in various (EoS, $\eta_{\rm X}$) combination scenarios. It is therefore a reasonable approximation to take $\eta_1=0.8$ after ignoring the possible effect of chemical composition of the disk. In the context of the magnetar propeller model, the evolution of the accretion disk mass can be expressed as
\begin{eqnarray}
\dot{M}_\mathrm{d}=\dot{M}_\mathrm{fb}-\dot{M}_\mathrm{prop}-\dot{M}_\mathrm{acc}
\label{eq:Mdot}
\end{eqnarray}
where $\dot{M}_{\rm prop}$ is the mass lost through the propeller mechanism and $\dot{M}_{\rm acc}\simeq \dot{M}_{\rm d,in}$ is the mass lost through the accretion onto the magnetar surface. Based on the description of \cite{Gibson2017}, $\dot{M}_{\rm prop}$ and $\dot{M}_{\rm acc}$ can be defined as follows:
\begin{eqnarray}
\dot M_{\rm{prop}}=\eta_2\left(\frac{M_{\rm{d}}}{t_\nu}\right),
\label{eq:Mdot_prop}
\end{eqnarray}
\begin{eqnarray}
\dot{M}_{\rm{acc}}=(1-\eta_2)\left(\frac{M_{\rm{d}}}{t_\nu}\right),
\label{eq:Mdot_acc}
\end{eqnarray}
where $\eta_2$ is the efficiency of the magnetar propeller mechanism. \cite{Gibson2018} and \cite{Yu2024} used the magnetar propeller accretion model to fit the X-ray data in their respective LGRB samples and found that the magnetar propeller mechanism is highly efficient, with the propeller efficiency reaching $\sim0.8$ for most LGRBs in their sample. Therefore, it is a reasonable approximation to take $\eta_2=0.8$ for our LGRB sample. When the material in the accretion disk is depleted, the propeller mechanism will no longer be effective. This indicates that all materials falling back to the disk are ultimately either accreted or thrown by the propeller after ignoring the possible effect of chemical composition of the disk. If we take the equilibrium spin period timescale $t_{\rm ev}$ as the lower limit of the accretion timescale $t_{\rm acc}$, based on equations (\ref{eq:Mdoteq}), (\ref{eq:Mdot_fb}), (\ref{eq:Mdot_in}), (\ref{eq:Mdot}), (\ref{eq:Mdot_prop}) and (\ref{eq:Mdot_acc}), one can express the lower limit of the fallback time scale $t_{\rm fb}$ as
\begin{eqnarray}
t_{\rm fb}\simeq4t_{\rm ev}\simeq4(GM_g)^{2/7}IB_{p}^{-8/7}R^{-24/7}\dot{M}^{-3/7},
\label{eq:tfall}
\end{eqnarray}
Based on the above equations, we can estimate that the fallback rate of progenitor envelope materials onto the magnetar accretion disk in our LGRB sample can reach a range of $\dot{M}_{\rm fb}\simeq2\dot{M}_{\rm acc}\sim 10^{-5}-10^{-2}~M_{\odot}~\rm s^{-1}$ at fallback times ranging from $t_{\rm fb}\sim20-10^4~\rm s$ for various (EoS, $\eta_{\rm X}$) combinations. Such an accretion timescale seems consistent with the fallback timescale derived for Wolf-Rayet (W-R) star, i.e. $\sim10^2-10^5$ s\footnote{The radii ($r_\mathrm{e}$) of the envelopes of W-R stars are $\sim10^{10}-10^{12}$ cm \citep{Koesterke1995}. The freefall timescale of the extended envelopes can be estimated by $t_\mathrm{ff}\sim(r_\mathrm{e}^3/GM)^{1/2}$, i.e. $10^2\lesssim t_\mathrm{ff}\lesssim10^5$ s.}. For a shorter timescale, the fallback materials could come from the core of the progenitors, which suggests an origin of compact progenitor stars for some LGRBs \citep{Campana2006,Woosley2006b}. Next, we try to test whether the progenitor of the LGRBs in our sample can provide enough envelope materials to satisfy the accretion of a newborn magnetar.

Consider an axisymmetric rotating star at the onset of core collapse. For simplicity, let us ignore the pressure forces and take the trajectory of each particle to correspond to freefall. Assuming that the progenitor star in our LGRBs is symmetric about its rotation axis and mirror-symmetric across the equator, when a particle (Particle 1) falls back from its initial position to the equatorial plane, it will collide with another particle (Particle 2) with a velocity of $-v_{\rm z}$ in the equatorial plane. The positions of the two particles are mirror-symmetric across the equator; see Figure 1 in \cite{Kumar2008}. The time it takes for a particle at the initial position ($r$, $\theta$, $\phi$) with angular velocity $\Omega(r, \theta)$ to fall back to the equatorial plane of the progenitor star at the end of its evolution trajectory is approximately \citep{Kumar2008,Zhao2025} 
\begin{eqnarray}
t_{\rm eq} \approx t_s(r) + {\pi\over 2^{3/2}\Omega_{\rm k}} \left[ 1 + {3\over4}\left({\Omega\sin\theta\over\Omega_{\rm k}}\right)^2\right],
\label{eq:teq}
\end{eqnarray}
where $\Omega_{\rm k}$ is the local Kepler angular velocity of the particle and $t_s(r)$ is the sound propagation time from the center of the progenitor star to the initial position $r$ of the particle, which is roughly the time it takes (from the start of collapse at the center) for gas at $r$ to realize the loss of pressure support and begin its fall toward the center of the progenitor star.

We assume that the difference between the polar and equatorial radii of an equal-collapse-time surface is $\delta r$. For a fixed $t_{\rm eq}$, the radial difference between particles that start at $\theta=0$ and those that start at $\theta=\pi/2$ is \citep{Kumar2008,Zhao2025}
\begin{eqnarray}
\delta r \approx {3\pi\over 2^{7/2}} {[\Omega(r)/\Omega_{\rm k}(r)]^2\over \Omega_{\rm k}(r) d(t_s + \pi 2^{-3/2}\Omega_{\rm k}^{-1})/dr},
\label{eq:delr1}
\end{eqnarray}
Introducing the notation
\begin{eqnarray}
H_t^{-1} \equiv \left| {d\over dr}\ln[t_s + \pi 2^{-3/2}\Omega_{k}^{-1}] \right|,
\label{eq:Ht}
\end{eqnarray}
combining equations (\ref{eq:delr1}) and (\ref{eq:Ht}), one can obtain
\begin{eqnarray}
\delta r \sim {H_t\over 2}\left[{\Omega(r)\over\Omega_{\rm k}(r)}\right]^2,
\label{eq:delr2}
\end{eqnarray}

In the above equations, we have assumed $t_s\sim \Omega_{\rm k}^{-1}$, which is valid outside a relatively small core region. Since $\delta r \ll r$, one can approximate the surface in equal collapse time as a sphere, and the mass fallback rate of materials on the equatorial plane of the progenitor star should be equal to the mass-loss rate of the progenitor star per unit time \citep{Kumar2008,Zhao2025} 
\begin{eqnarray}
\dot M_{\rm eq,fb}(t_{\rm eq}) \approx {d M(r)\over dr} {dr\over d t_{\rm eq}} \approx {4\pi r^2\rho(r) H_t\over t_{\rm eq}} \approx {4\pi r^2\rho(r)\over t_{\rm eq} \left| d\ln\Omega_{\rm k}/dr \right|},
\label{eq:Mfb}
\end{eqnarray}

The formula shows that $M_{\rm eq,fb}$ is insensitive to the rotation profile in the progenitor star. However, the fallback radius where the stellar matter circularizes has a strong dependence on the angular velocity $\Omega$. After fallback to the equatorial plane of a progenitor star, the velocity of the materials can be decomposed into an azimuth $\phi$ component around the equatorial plane and a radial $r$ component toward the center. The velocity components in the two directions satisfy $\left|v_r\right|>v_{\phi}$, where the sign of $v_r$ is negative and independent of the initial position of the particle. This will cause the initial disk formed during the core collapse process to shrink rapidly toward the center, ending when it reaches a radius defined as the fallback radius $r_{\rm fb}$ 
\begin{eqnarray}
r_{\rm fb} \approx r \left({\Omega\over\Omega_k}\right)^2,
\label{eq:rfb}
\end{eqnarray}

At the fallback radius, the specific angular momentum of the inflowing materials is equal to that of the local circular orbit. For our selected LGRBs with X-ray plateau emission, their progenitor stars should have a core-envelope structure, as is common in stellar models. At the end of the progenitor star’s life, most of the mass in its core part collapses into a rapidly spinning NS, and a fraction of the progenitor fragment in post-collapse could circulate into an accretion disk and interact with the newborn magnetar. Such an interacting magnetar--disk system will undergo the accretion regime and propeller regime to accrete material onto the magnetar's surface and to throw material away from the system, which would have a strong influence on the spin evolution and transient EM emission of the newborn magnetar. It should be noted that the shrink timescale for the initial accretion disk is much shorter than $t_{\rm eq}$. Thus, we can equate the fallback rate of materials onto the accretion disk with the fallback rate of materials onto the equatorial plane, i.e., $\dot M_{\rm eq,fb}=\dot M_{\rm fb}$. Assuming that the formation time of the newborn NS is $t_{\rm NS}$, and given the density and spin profiles of the progenitor, we can obtain the fallback rate $\dot M_{\rm fb}(t)$ of the star’s material on the accretion disk for $t=t_{\rm eq}-t_{\rm NS}$.

\begin{figure*}
\centering
\includegraphics  [angle=0,scale=0.29]   {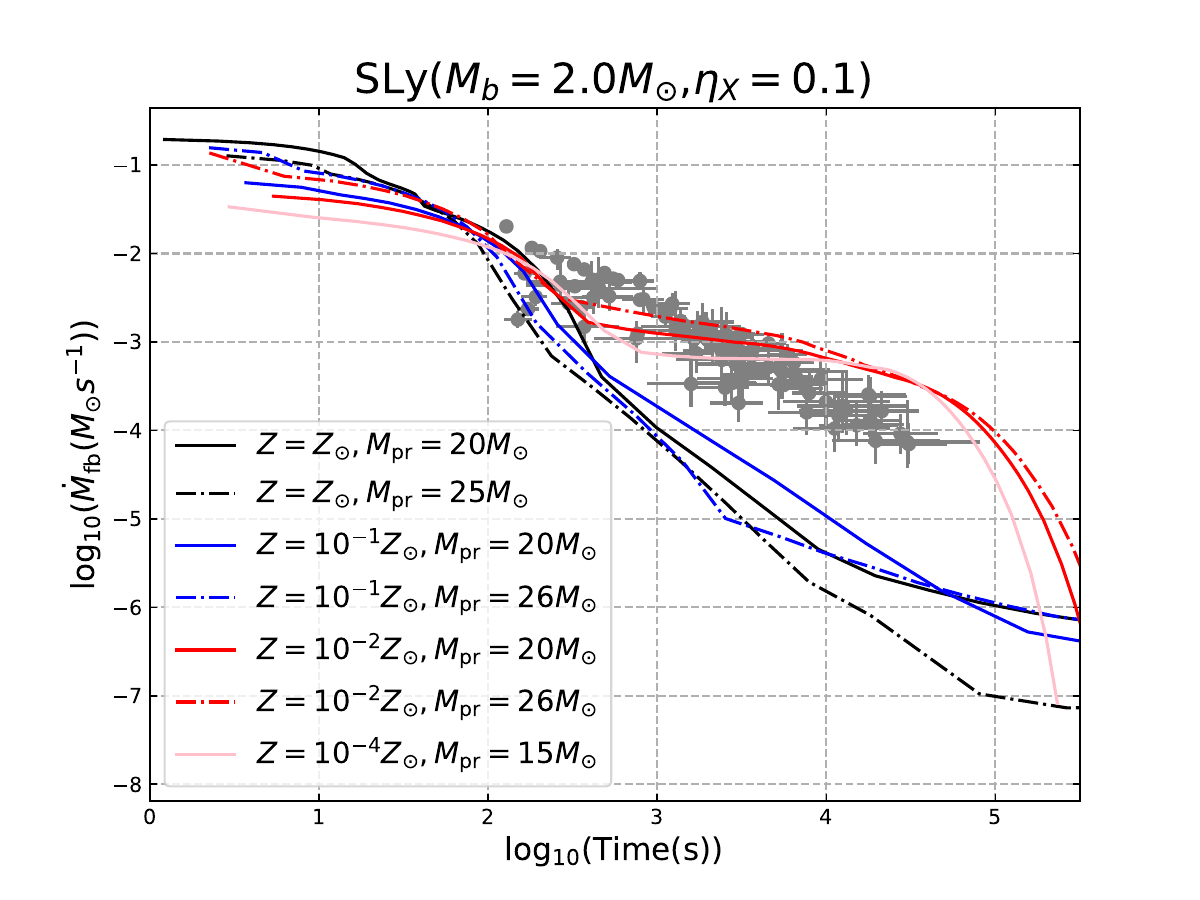}
\includegraphics  [angle=0,scale=0.29]   {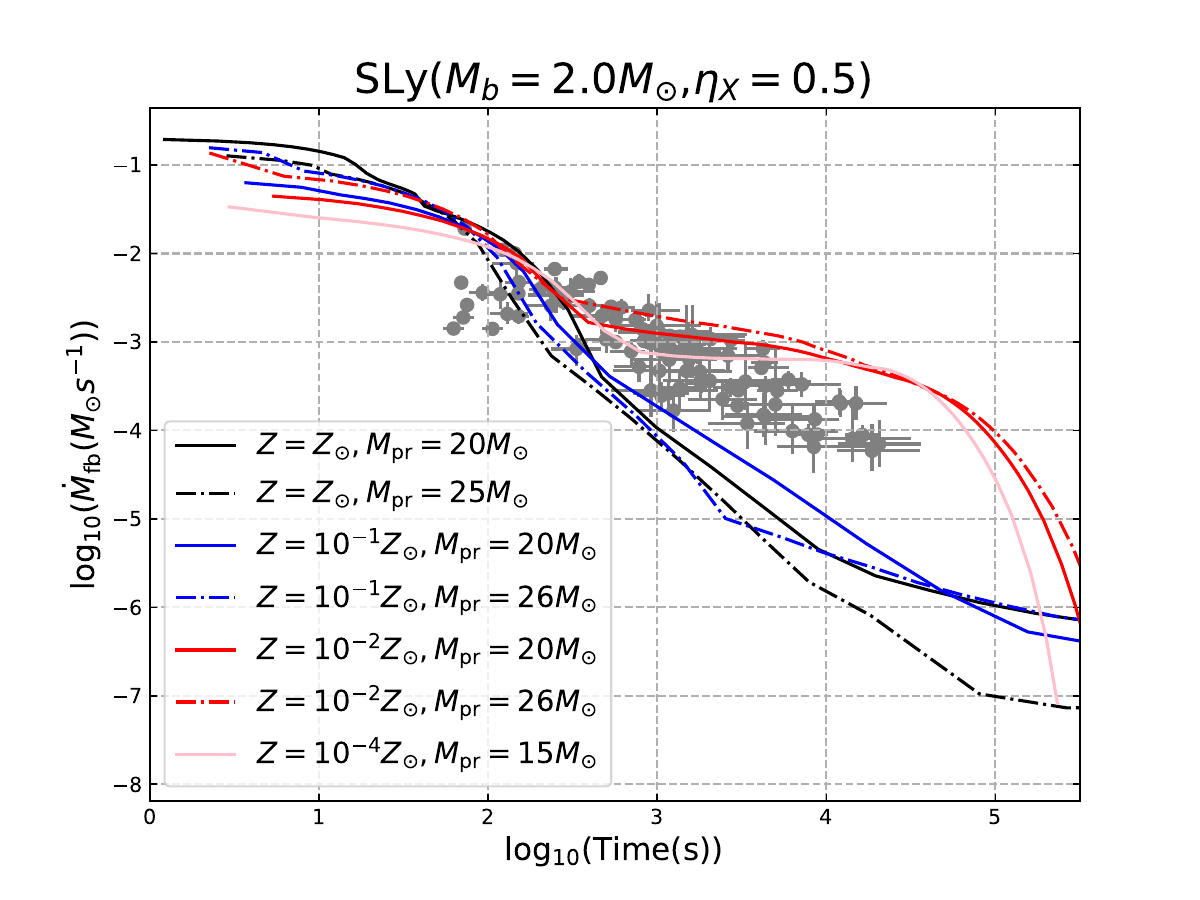}
\includegraphics  [angle=0,scale=0.29]   {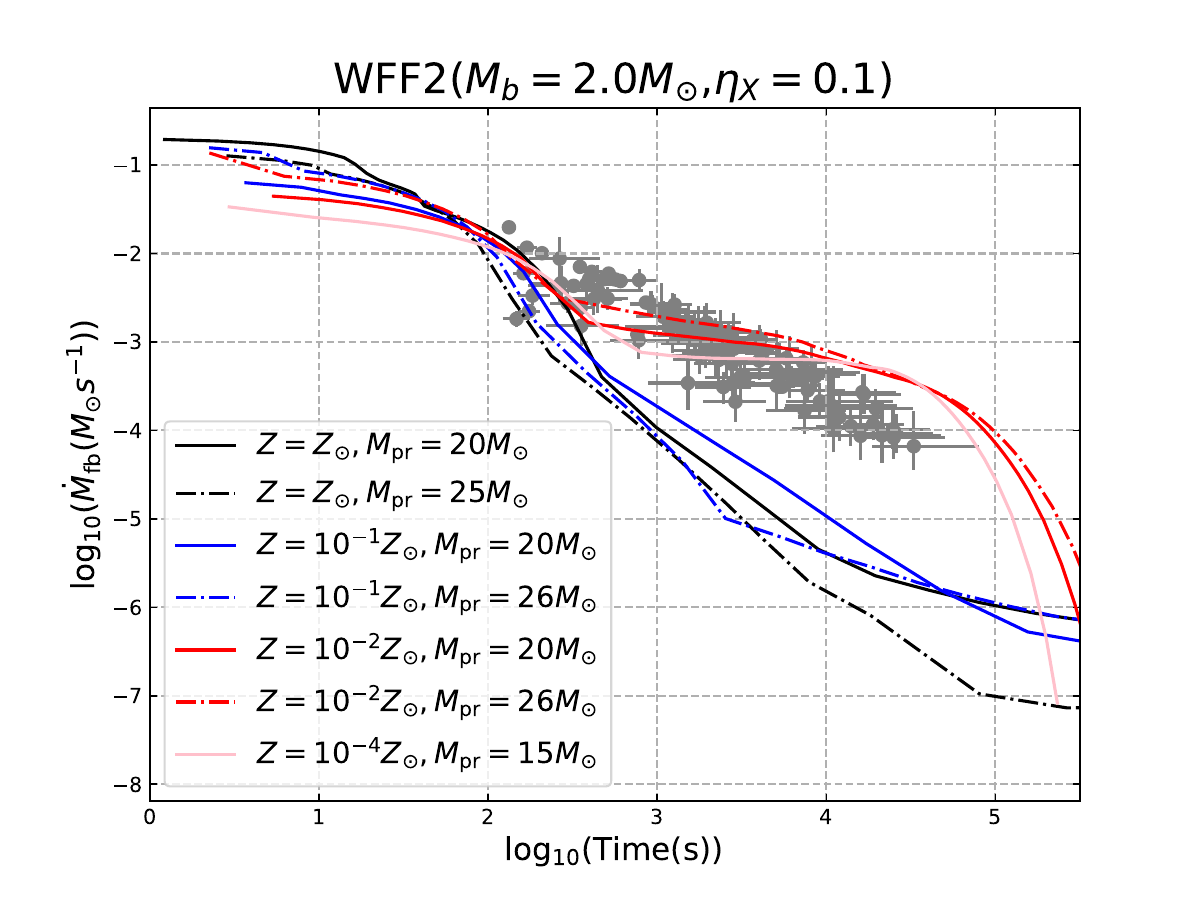}
\includegraphics  [angle=0,scale=0.29]   {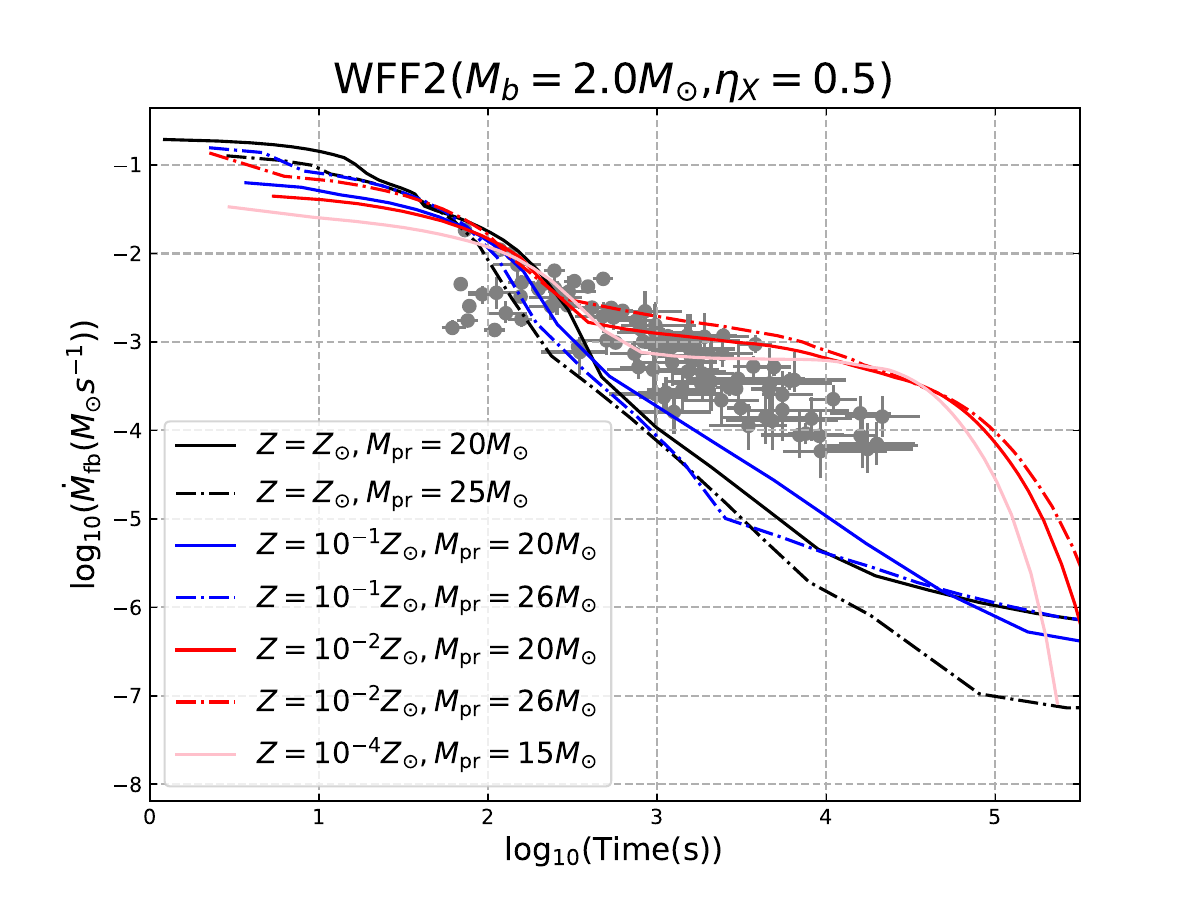}
\includegraphics  [angle=0,scale=0.29]   {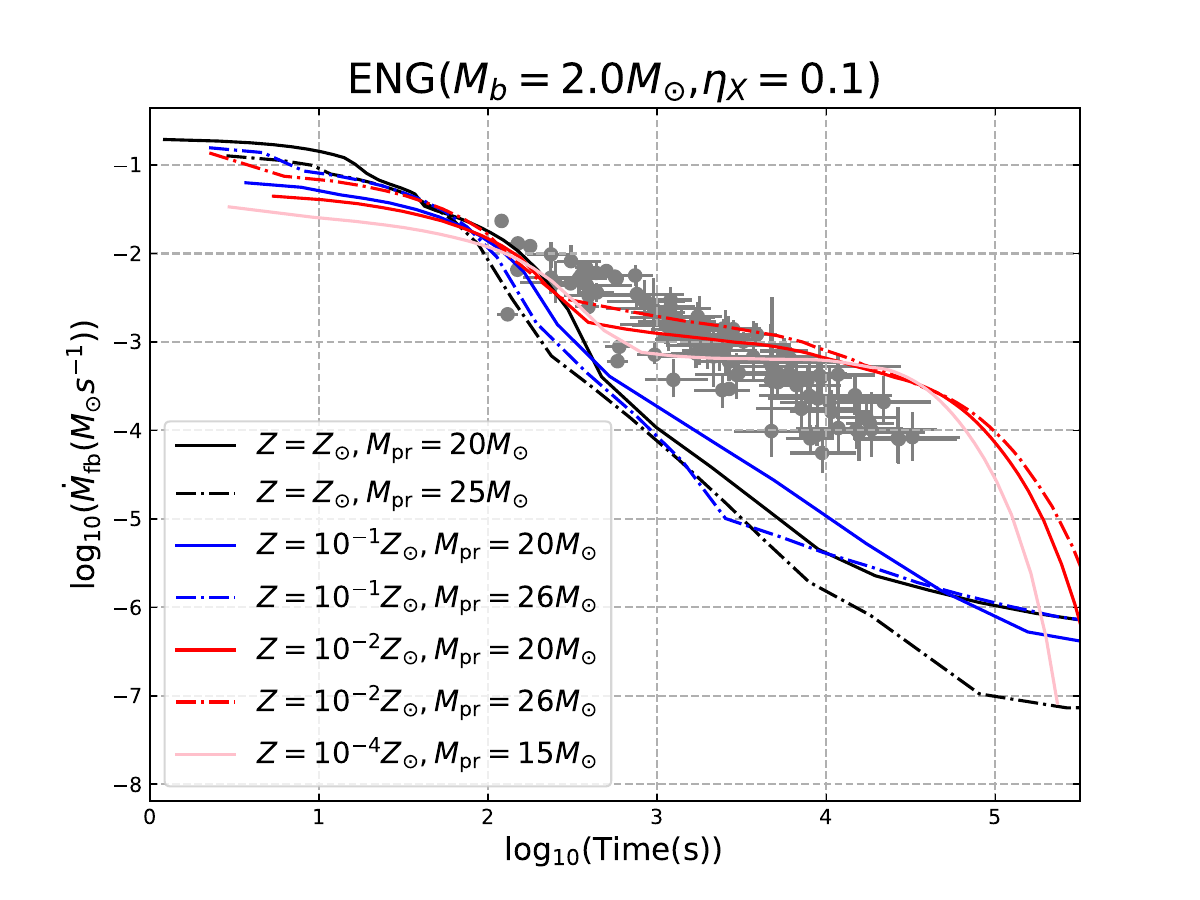}
\includegraphics  [angle=0,scale=0.29]   {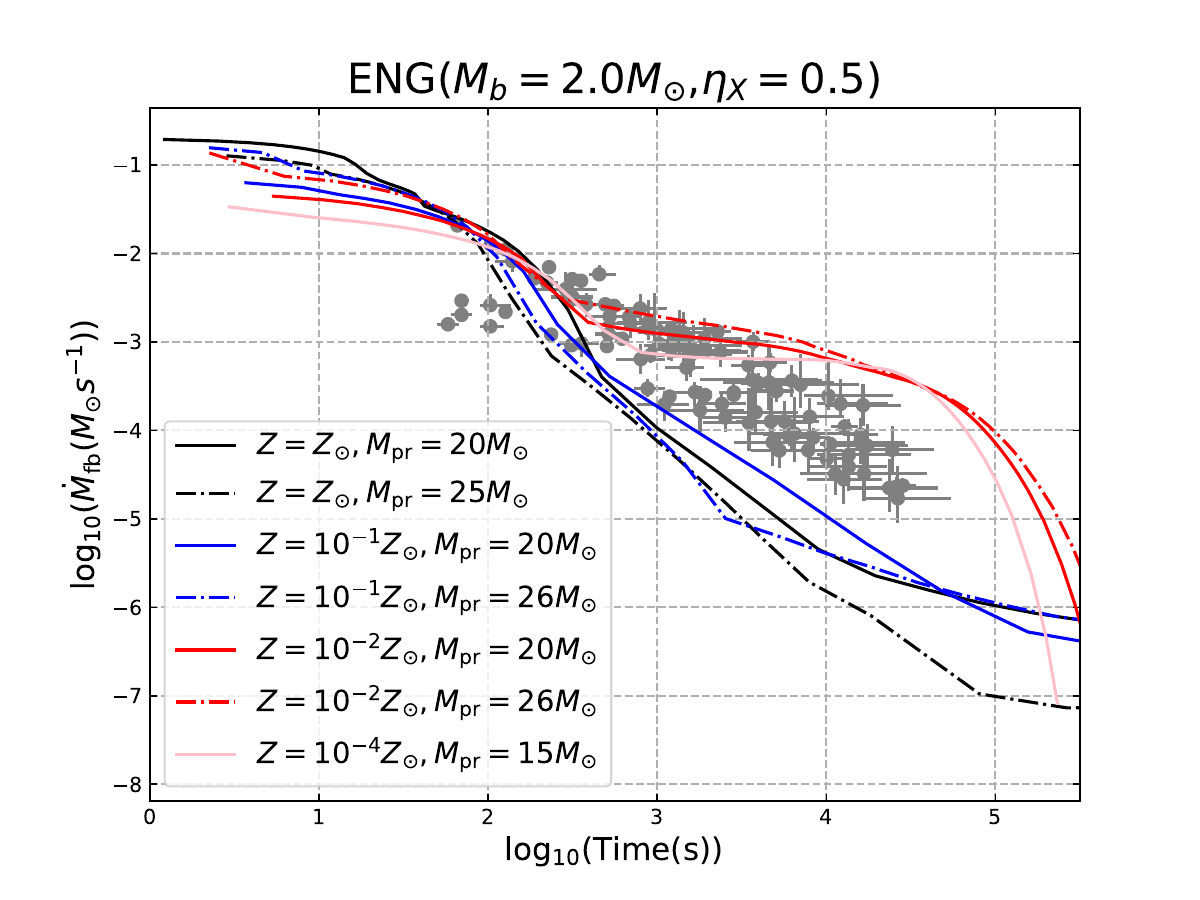}
\includegraphics  [angle=0,scale=0.29]   {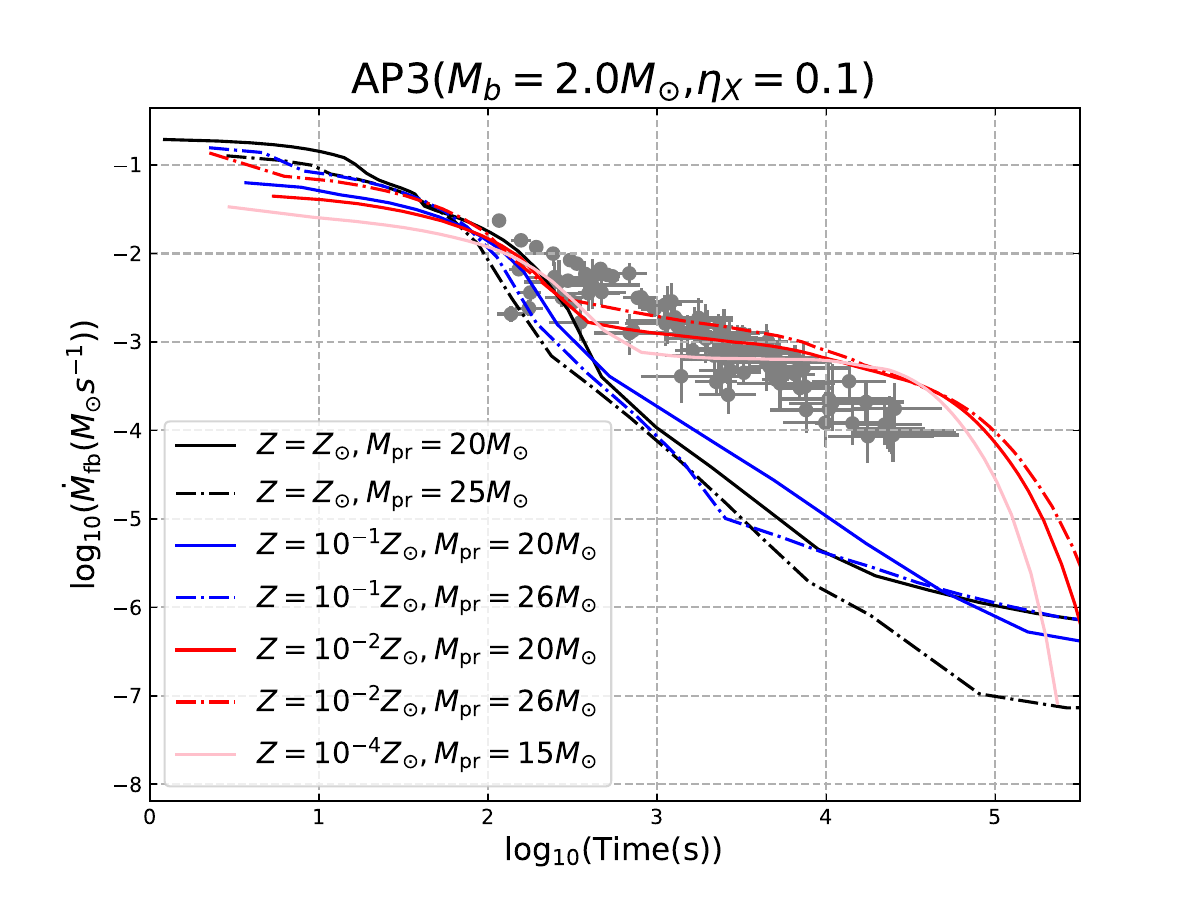}
\includegraphics  [angle=0,scale=0.29]   {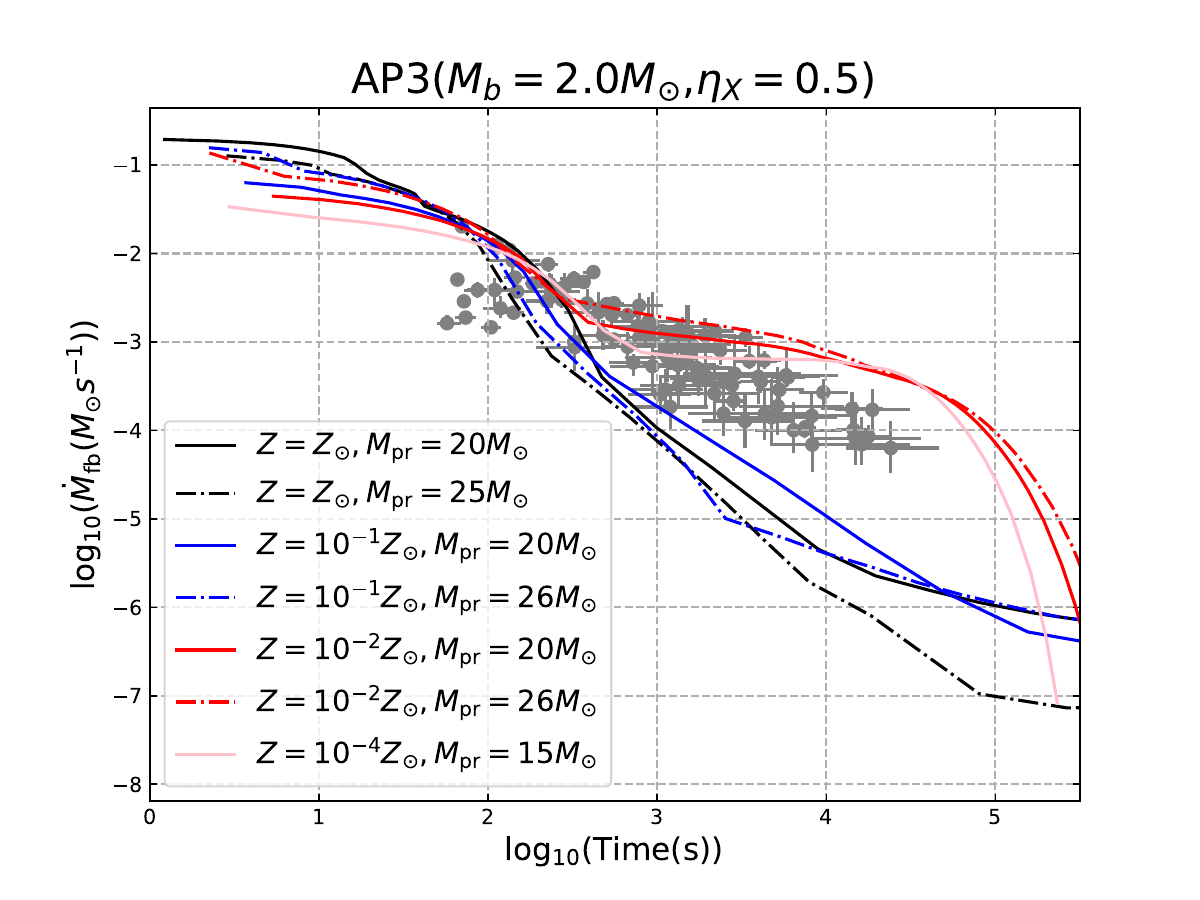}
\caption{The evolution of the fallback rate of progenitor materials with different masses and metallicities onto the accretion disk for different EoSs and $\eta_{\rm X}$. Lines of different colors and shapes represent different progenitor star masses and metallicities. The gray circles represent the fallback rate in our LGRB sample.}
\label{fig:Mdot-Tfb}
\end{figure*}

Based on the above equations, once we know the density profile and spin profile of the progenitor star, we can numerically solve for the fallback accretion rate at any given time. \cite{Liu2018} provided the density profiles of the progenitor star with different metallicities and masses, and \cite{Kumar2008} gave the spin profile of the progenitor star. We thus calculated the evolution of the fallback rate for various progenitor materials with different metallicities and masses on the accretion disk. To ensure that the residual remnant is an NS, we selected only progenitor stars with masses below $25M_{\odot}$ for the numerical solution. As shown in Figure \ref{fig:Mdot-Tfb}, we can find that the fallback rate of progenitor envelope materials onto the magnetar accretion disk for our LGRB sample in various (EoS, $\eta_{\rm X}$) combination scenarios is compatible with the theoretical mass fallback rate of some low-metallicity massive progenitor stars, such as those with ($Z\lesssim10^{-2}Z_{\odot}$, $M\gtrsim26M_{\odot}$), ($Z\lesssim10^{-2}Z_{\odot}$, $M\gtrsim20M_{\odot}$), ($Z\lesssim10^{-4}Z_{\odot}$, $M\gtrsim15M_{\odot}$), and so on. These results suggest that the low-metallicity progenitors can provide enough material to satisfy the accretion requirements of the newborn accreting magnetar in our LGRB sample for various (EoS, $\eta_{\rm X}$) combination scenarios. The solar-metallicity stars might not be so available to serve as the progenitors for our LGRB sample.

\begin{figure*}
\centering
\includegraphics  [angle=0,scale=0.29]   {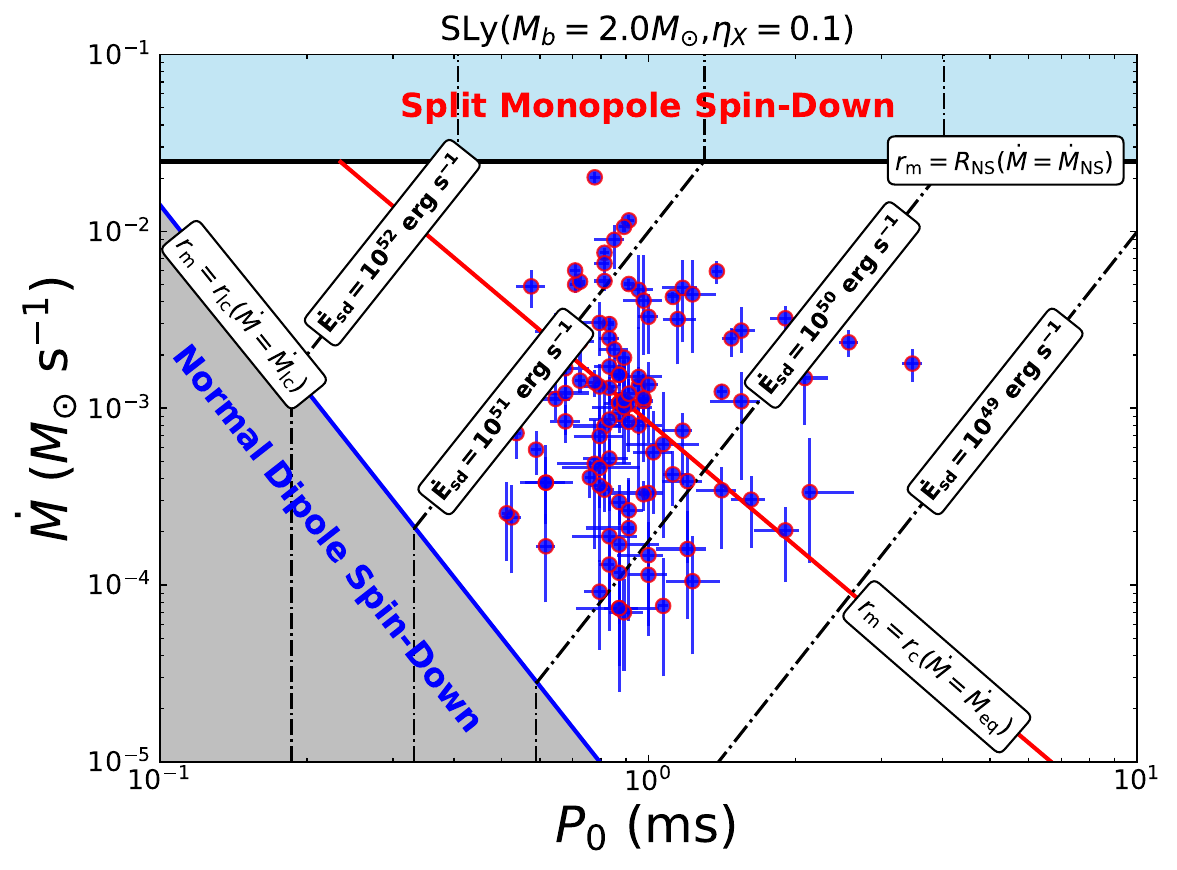}
\includegraphics  [angle=0,scale=0.29]   {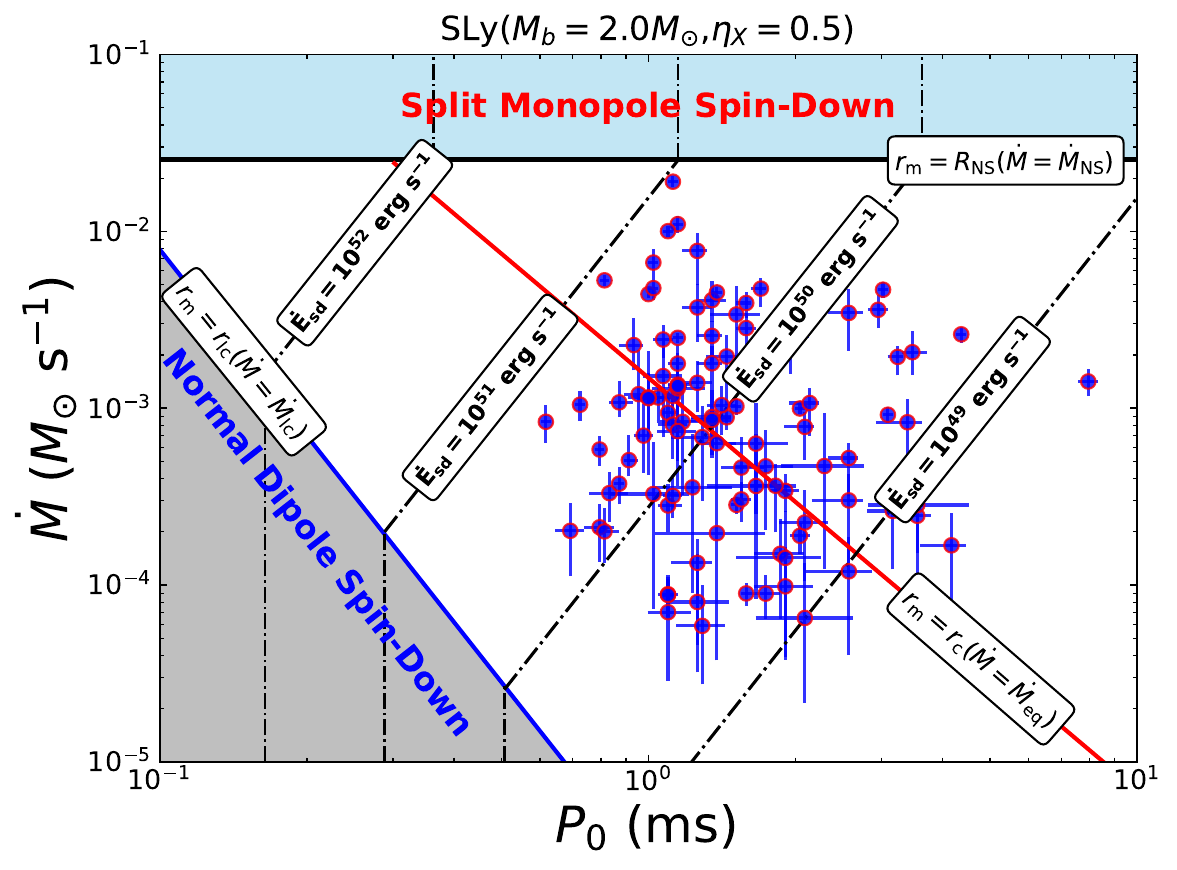}
\includegraphics  [angle=0,scale=0.29]   {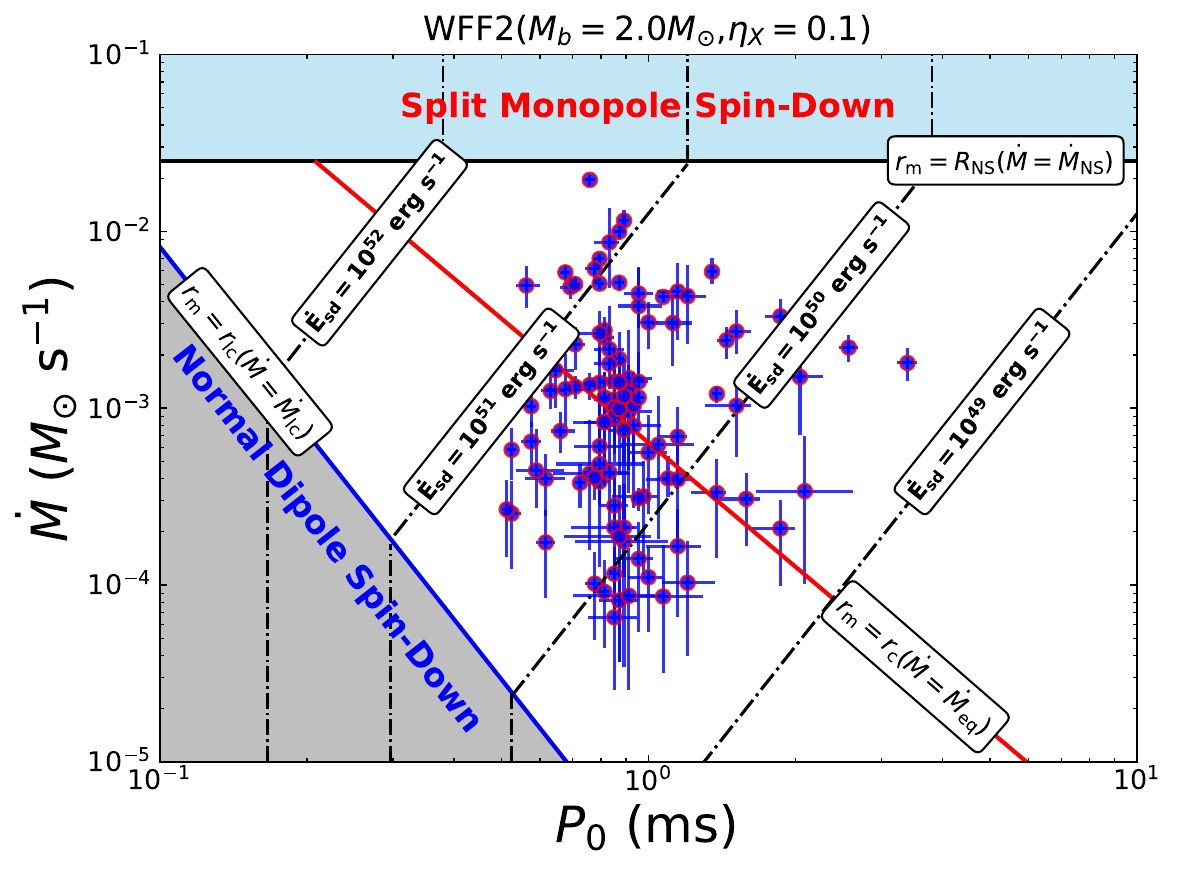}
\includegraphics  [angle=0,scale=0.29]   {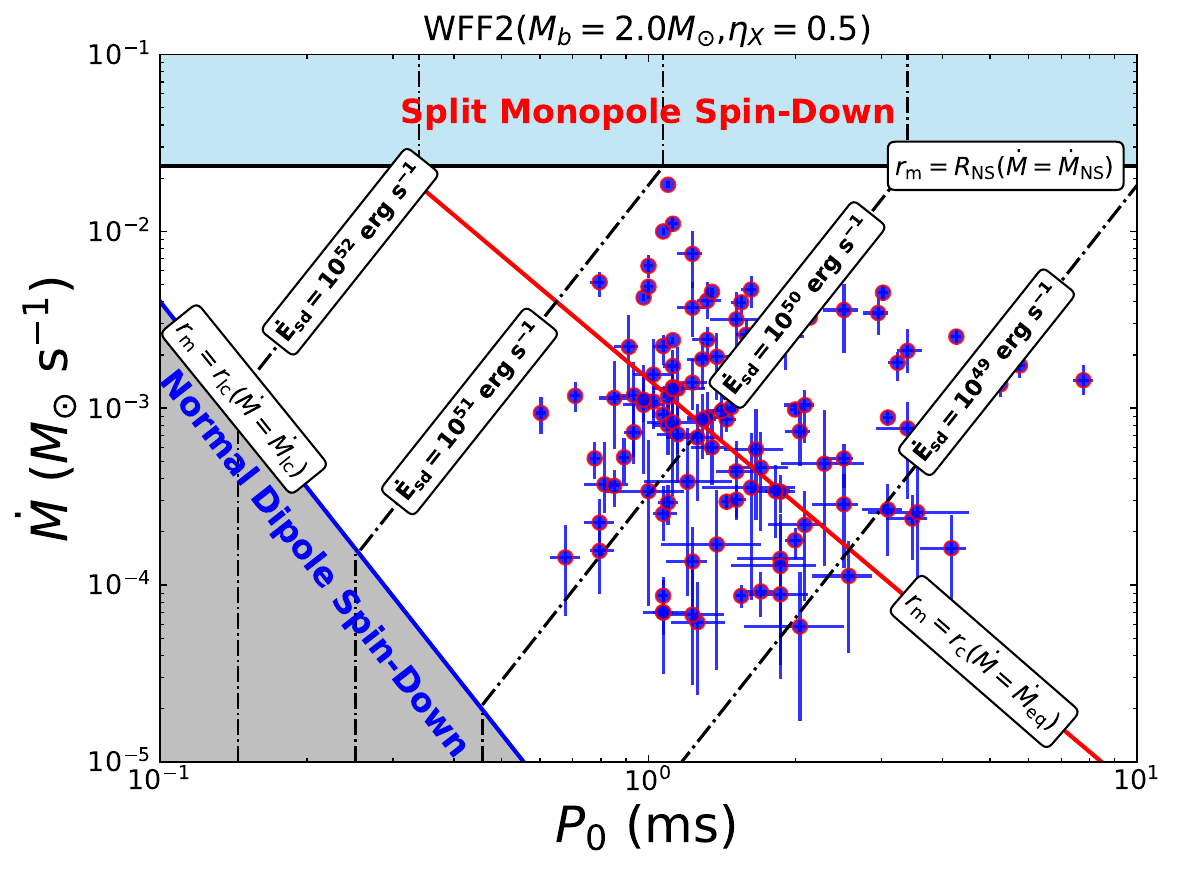}
\includegraphics  [angle=0,scale=0.29]   {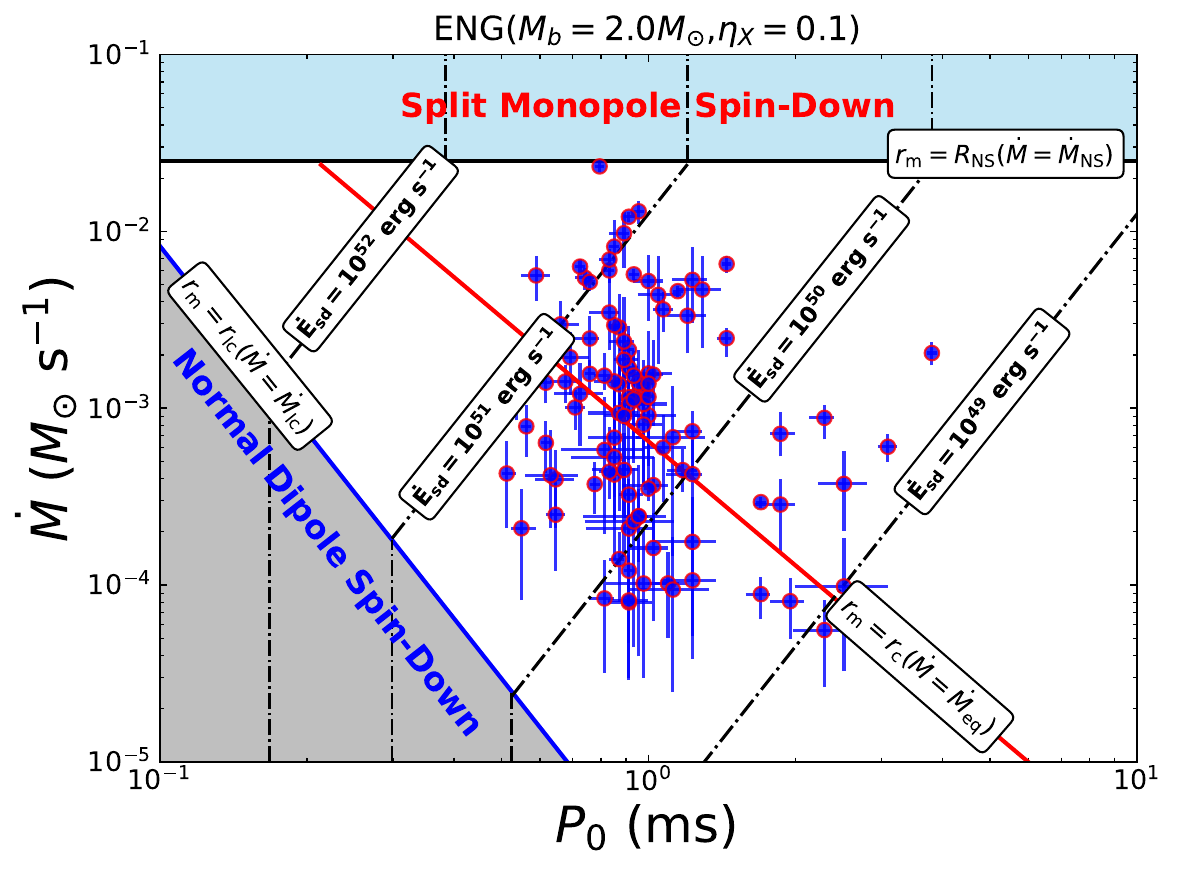}
\includegraphics  [angle=0,scale=0.29]   {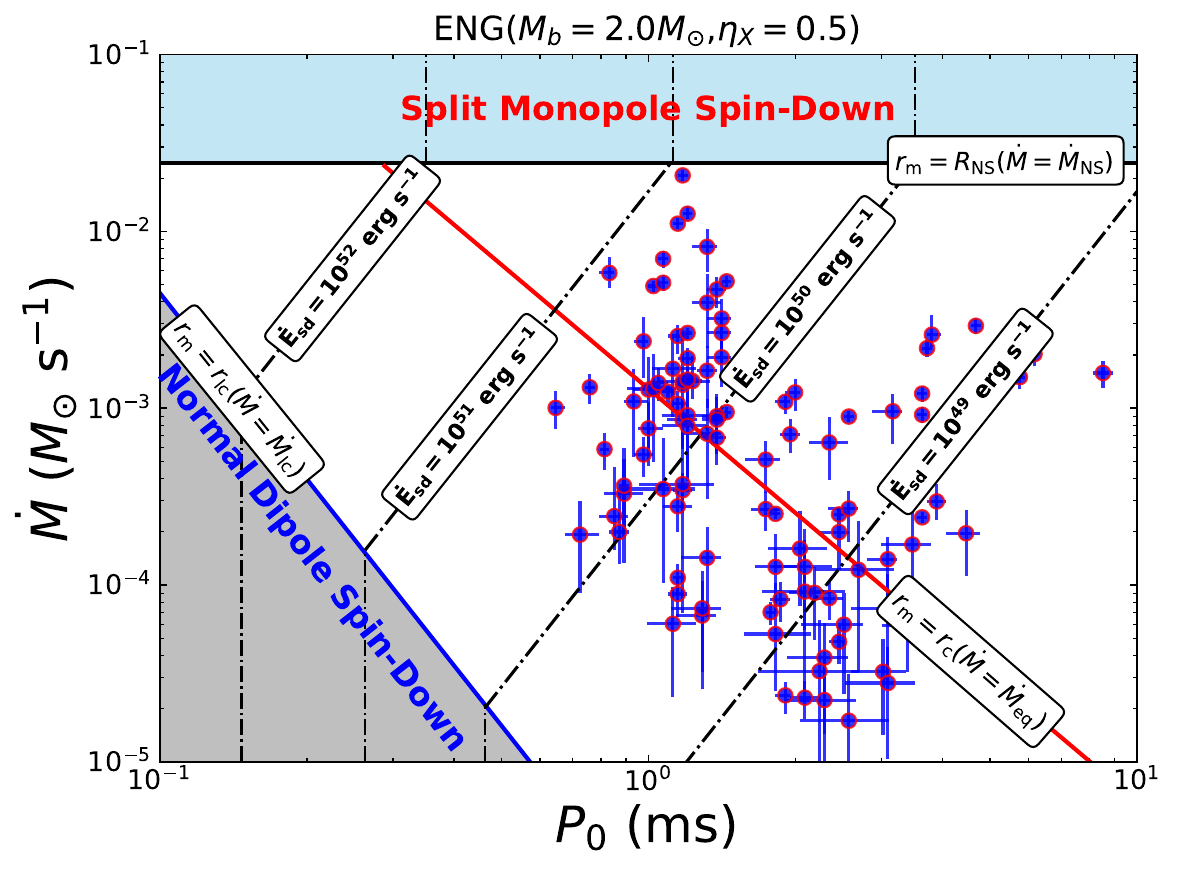}
\includegraphics  [angle=0,scale=0.29]   {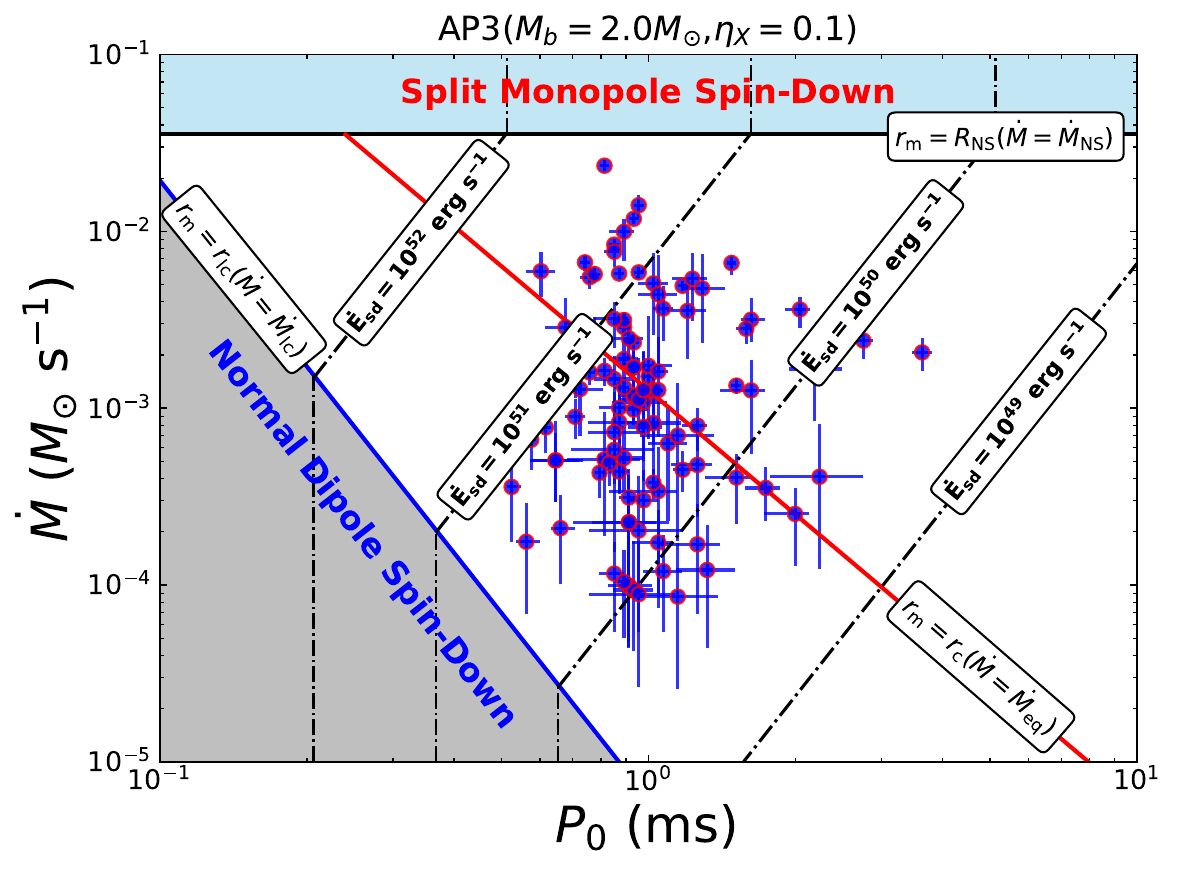}
\includegraphics  [angle=0,scale=0.29]   {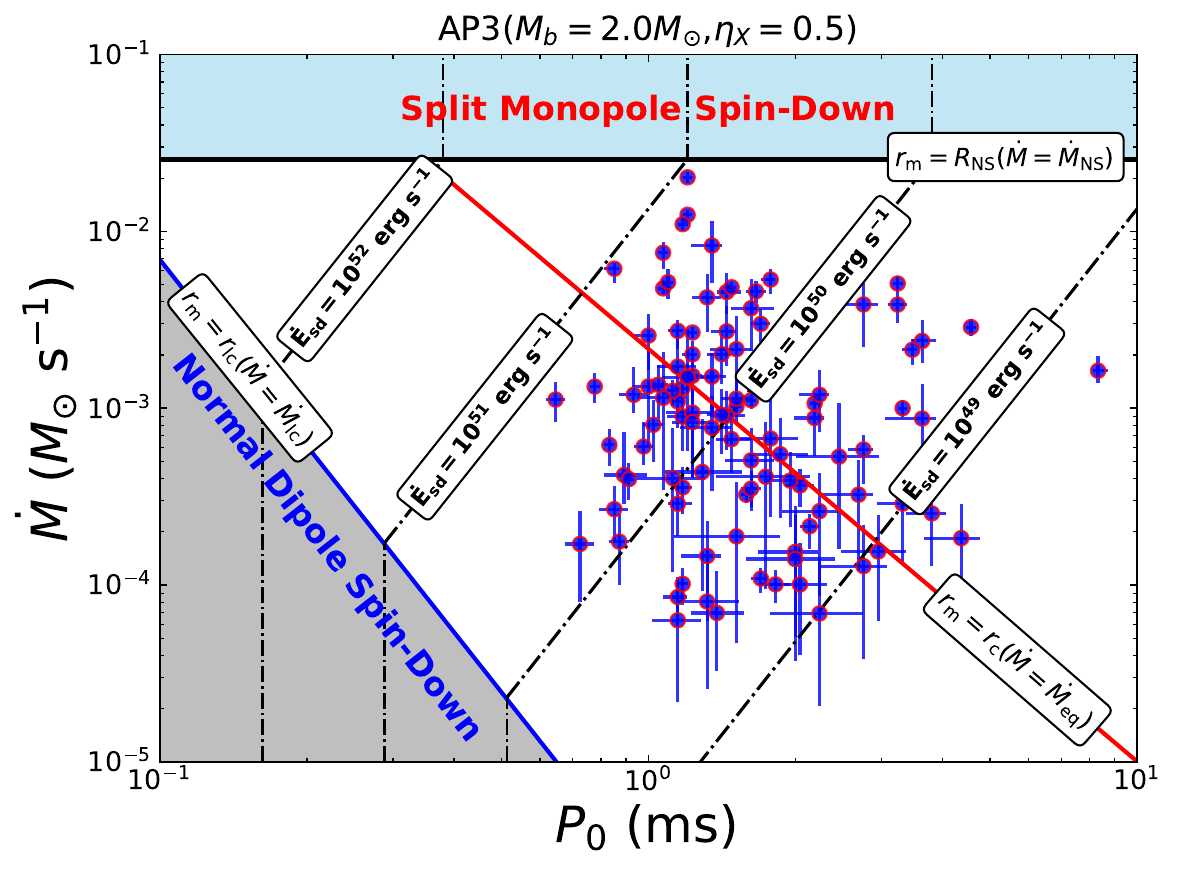}
\caption{The spin-down evolution regimes of a newborn magnetar for the parameter space of mass accretion rate $\dot{M}$ and spin period $P_0$ in different EoSs and $\eta_X$. The solid lines show accretion rates $\dot{M}=\dot{M}_{\rm NS}$ (black), $\dot{M}=\dot{M}_{\rm eq}$ (red), and $\dot{M}=\dot{M}_{\rm lc}$ (blue). The dashed--dotted lines show contours of constant spin-down luminosity $\dot{E}_{\rm sd}$. The red solid circles are our LGRB sample. The sky-blue shaded region and the gray shaded region represent the parameter space of split monopole spin-down evolution and the parameter space of normal dipole spin-down evolution, respectively.}
\label{fig:Mdot-P0}
\end{figure*}

Ultimately, we investigated the spin-down evolution regimes of a newborn magnetar for the parameter space of the mass accretion rate $\dot{M}$ and the spin period $P_0$ in different EoSs and $\eta_X$, and examined the positions of our LGRB sample within the parameter space of $\dot{M}$ and $P_0$, as well as trying to test whether the rotational energy of this newly formed accreting magnetar could sustain the energy required for the propeller model radiation for our LGRB sample in various (EoS, $\eta_{\rm X}$) combinations. Based on equations (\ref{eq:Mdotlc}), (\ref{eq:Mdotns}), (\ref{eq:Peq}), and (\ref{eq:Edotcases}), we can infer the magnetar spin-down power contours in the propeller mechanism and test the energy required for the magnetar spin-down and propeller model radiation for our LGRB sample in various EoSs and $\eta_{\rm X}$. In Figure \ref{fig:Mdot-P0}, we show a series of $\dot{M}-P_{0}$ scatter diagrams in various (EoS, $\eta_{\rm X}$) combinations. As shown in Figure \ref{fig:Mdot-P0}, we find that all GRBs in our sample fall well within the region where the EM dipole radiation and the propeller mechanism jointly modulate the spin evolution of a newborn magnetar and the overall requirement for energy release $\dot E_{\rm sd}\subseteq(10^{49}, 10^{52})~\rm erg ~s^{-1}$. The results are consistent with the magnetar engine hypothesis, namely, the spin energy of the newborn accreting magnetar can sustain the propeller mechanism and EM dipole spin-down emission energy in our LGRB sample.

\section{Conclusions and Discussions}
\label{Sec:discussions}

In Swift's 21 yr of operation, the early-time afterglow emissions of GRBs have been revealed, and a rich trove of X-ray observation data have been collected, which provides valuable information for understanding the GRB central engine. A good fraction of the X-ray afterglows in GRBs exhibit a long-lasting plateau emission feature, which is thought to represent a potentially energy injection into GRB's aftergolw and is commonly taken as evidence of a newborn magnetar as the GRB central engine \citep{Dai1998a,Dai1998b,Zhang2001,Troja2007,Rowlinson2010,Rowlinson2013,Rowlinson2014,Fan2013a,Fan2013b,Gompertz2013,Gompertz2014,Lv2014,Metzger2014,Lv2015,Lv2018,Lv2019,Gao2016,Gao2017a,Gao2017b,Lasky2016,Lasky2017,Ai2018,Ai2020,Lin2019,Zou2019,Zou2021b,Lan2020a,Lan2021,Lan2025,Linjie2020,Sarin2020,Zhao2020,Ai2021,Xie2022a,Xie2022b}. In this work, by considering the $R/I$ evolutionary effects during the magnetar spin-down process, we performed a systematic study in the propeller mechanism properties of a newborn accreting magnetar by reestimating the physical parameters of magnetars and investigating possible relations between the $B_p$ and $P_0$ based on the LGRB sample with X-ray plateau emission, and we tried to constrain the accretion rate of accreting magnetars in different EoSs and X-ray radiative efficiencies, which is crucial in helping us to understand the nature and physical environment of a newborn accreting magnetar and its progenitor. In order to precisely diagnose the propeller properties of the newborn accreting magnetar, we invoked four samples of EoSs with baryonic masses $M_{b}=2.0M_{\odot}$, and two X-ray radiation efficiencies $\eta_{X}=0.1$ and $0.5$ to simultaneously constrain the accretion rate of accreting magnetars. Our main results can be summarized as follows:

\begin{itemize}[leftmargin=*]
\item We found a universal correlation between the surface magnetic field $B_p$ and the initial spin period $P_0$ of newborn magnetar candidates based on 105 LGRBs with X-ray plateau emission sample in \cite{Lan2025}. The overall $B_p-P_0$ correlation can be approximately described as $B_p\propto P_{0}^{1.30\pm0.16}$ for various (EoS, $\eta_{\rm X}$) combination scenarios, which seems to hint that the $P_0$ estimated from plateau data could deviate from that of the magnetar at birth but strongly links to the equilibrium spin period when the accreting magnetar has reached equilibrium in the magnetic propeller mechanism, which follows a positive correlation in agreement with $B_p\propto P_{\rm eq}^{7/6}$. 

\item We found that the accretion rate of the accreting magnetar in our LGRB sample is in the range of $\dot{M}\sim10^{-5}-10^{-2}~M_{\odot}~\rm s^{-1}$ by systematically incorporating $R/I$ evolutionary effects and using the $M_b-M_g$ transition relation in different EoSs and $\eta_{\rm X}$. The accretion rate is one order of magnitude lower compared to the statistical results in \cite{Stratta2018} and \cite{Linweili2020}, which used the constant $R$, $I$ and $M_g$ for their LGRB sample. We suggest that adopting the constant $R/I$ scenario in the propeller model of accreting magnetars will yield a higher mass accretion rate, which may impair our understanding of the physical nature of an accreting magnetar and its surroundings and will further diminish the reliability of inverse inference about the physical properties of its progenitor.

\item We found that the overall $p$-value of the K-S test for the derived accretion rates between different NS EoSs and different $\eta_{\rm X}$ can be approximately expressed as $p_{\rm KS}<7.3\times10^{-1}$ and $p_{\rm KS}<5.3\times10^{-2}$, respectively. The K-S test results indicate that there are no significant differences in the constrained mass accretion rates for different EoSs, and the constraints on magnitude of mass accretion rate seem to be independent of the EoS stiffness. However, there are systematic differences in the constrained mass accretion rates for different $\eta_{\rm X}$, and with higher X-ray radiative efficiency, the constrained mass accretion rates tend to become larger.

\item We found that no matter whether the radiative efficiency $\eta_{\rm X}$ is 0.1 or 0.5, most of the accreting magnetars in our GRB sample can actually survive the equilibrium spin period for two soft SLy and WFF2 EoSs ($M_{\rm TOV}=2.05~M_{\odot}$ and $2.20~M_{\odot}$), and all the accreting magnetars in our GRB sample can actually survive the equilibrium spin period for two stiff EoSs ($M_{\rm TOV}=2.24~M_{\odot}$ and $2.39~M_{\odot}$). The accretion masses of most newborn accreting magnetars in our LGRB sample are less than $0.5M_{\odot}$ for various (EoS, $\eta_{\rm X}$) combination scenarios, which indicates that the majority of the nascent accreting magnetars in our GRB sample are able to survive until they reach the equilibrium spin period and will not collapse into BHs because their total mass after accreting materials to the magnetar surface does not exceed its maximum gravitational mass.

\item We found that the fallback rate of progenitor envelope materials onto the magnetar accretion disk for our LGRB sample in various (EoS, $\eta_{\rm X}$) combination scenarios is compatible with the theoretical mass fallback rate of some low-metallicity massive progenitor stars, such as those with ($Z\lesssim10^{-2}Z_{\odot}$, $M\gtrsim26M_{\odot}$), ($Z\lesssim10^{-2}Z_{\odot}$, $M\gtrsim20M_{\odot}$), ($Z\lesssim10^{-4}Z_{\odot}$, $M\gtrsim15M_{\odot}$) based on the spin profile in \cite{Kumar2008} and density profile in \cite{Liu2018}. These results suggest that the low-metallicity progenitors can provide enough material to satisfy the accretion requirements of the newborn accreting magnetar in our LGRB sample for various (EoS, $\eta_{\rm X}$) combinations. Furthermore, we found that all GRBs in our sample fall well within the region where the EM dipole radiation and the propeller mechanism jointly modulate the spin evolution of a newborn magnetar and that the overall requirement for energy release $\dot E_{\rm sd}\subseteq(10^{49}, 10^{52})~\rm erg ~s^{-1}$, which are consistent with the magnetar engine hypothesis for our LGRB sample.
\end{itemize}

In the end, we would like to point out that several uncertainties exist in the results we have obtained. Our results are based on a sample of LGRBs with X-ray plateau emission. This sample could suffer observational biases from the XRT fluctuation thresholds, and the magnetic dipole emission may be covered by bright jet afterglow emission, leading us to collect an incomplete sample. In other words, the data currently available for LGRBs with plateau emission are still limited, which may affect our results owing to sample incompleteness. In addition, the above results are based on a simple method, with which we estimate the constant mass accretion rates from the $B_p-P_0$ distribution inferred from the magnetar-powered models in given NS EoSs and X-ray radiative efficiency. However, in reality, we know very little about the NS EoS and X-ray radiative efficiency because of the unknown physical properties of matter under extreme conditions and radiation mechanisms of NSs, and in addition to the early-time accretion rate with constant rate, the late-time decay of the accretion rate can also influence the spin period of the magnetar, and hence magnetar outflow can deviate from the magnetic dipole radiation luminosity evolution \citep{Piro2011,Metzger2018}. These uncertainties in NS physics may introduce biases in our results. Furthermore, the physical conditions for a nascent NS in an early stage would be very complicated, and some conditions may significantly alter the dipole radiation light curve, so that the $R/I$ evolution effect we discussed here would be reduced or even completely suppressed, which may impair our estimates of the physical parameters of GRB magnetars and even demolish the $B_p-P_0$ connections between the physical parameters of GRB magnetar. For instance, it has been proposed that the evolution of the inclination angle between the rotation and magnetic axes of the NS could markedly revise the X-ray emission \citep{Cikintoglu2020}. Moreover, the nascent NS is likely to undergo free precession in the early stages of its lifetime when the rotation and magnetic axes of the system are not orthogonal to each other, which can lead to systematic fluctuations in the X-ray light curve \citep{Suvorov2020,Suvorov2021,Zou2021a,Zou2022,Zou2024,Zhang2024}. The impact of these conditions on the constraints for the magnetar physical parameters warrants further investigation in the future to validate the authenticity of the $B_p-P_0$ correlation.

Meanwhile, there are a couple of extra simplifying assumptions that may also introduce uncertainty into our results. One key simplifying assumption in our analysis is the adoption of $M_b=2.0M_{\odot}$ for all newborn magnetars in our LGRBs sample. This is because the primary research purpose of our work is to determine the true mass accretion rate of newborn accreting magnetars after incorporating $R/I$ evolution effects and different gravitational masses. However, the $R/I$ evolution effects depend on the rotational speed of the newborn magnetar. Only when the rotational speed of the newborn magnetar is approaching the breakup limit do its $R/I$ values undergo an obvious evolution as the magnetar spins down, and it was found that only when the baryonic mass is $2.0M_{\odot}$ or greater can the $R/I$ of newborn magnetars maintain an obvious evolution effect over the typical period $P_0\sim1$ ms of the GRB magnetar central engine \citep{Lan2021}. Moreover, \cite{Lan2025} found that adopting a lower baryonic mass yields a smaller $P_0$, particularly when using a baryonic mass below $2.0M_{\odot}$ would lead to some $P_0$ values in their LGRBs sample that break the Keplerian breakup limit, and using a larger baryonic mass ($M_b=2.5M_{\odot}$) would lead to some newborn magnetars rapidly collapsing into BHs in the sample. Taking all these factors into account, we selected only a baryonic mass of $2.0M_{\odot}$ to determine the true mass accretion rate of newborn accreting magnetars after incorporating $R/I$ evolution effects and different gravitational masses. We also reexamined the effect on the paper's final conclusions with a lower mass of $1.4M_{\odot}$, as well as with a distribution of masses. Notably, the evolution effects of $R/I$ can be neglected when using a lower mass. Based on $\dot{M}\propto M_g^{-5/3}$ in Equation (\ref{eq:Mdoteq}), a lower mass will lead to higher mass accretion rate. When using the constant $R/I/M_g$ ($M_g=1.4M_{\odot}$, $R=12~\rm km$, $I=0.35M_gR^2$), we can obtain the range of mass accretion rates $10^{-4}~M_{\odot}~{\rm s^{-1}}<\dot{M}<10^{-1}~M_{\odot}~{\rm s^{-1}}$ for our LGRB sample, which are consistent with the statistical results in \cite{Stratta2018} and \cite{Linweili2020}. This result aligns with the final conclusion of our paper, namely that adopting a constant $R/I/M_g$ scenario for modeling the propeller regime in accreting magnetars results in a higher mass accretion rate.

Another key simplifying assumption in our analysis, as noted in Section \ref{Sec:sample}, is the adoption of $z=1$ for the 62 LGRBs in our sample lacking redshift measurements. While this is a common practice in population studies of LGRBs \citep{Berger2003,Xie2022a}, it is crucial to assess the potential uncertainty this assumption introduces into our primary results. To quantitatively evaluate this, we repeated our entire analysis using only the subsample of 43 LGRBs with measured redshifts. We found that a universal correlation still exists between the surface magnetic field $B_p$ and the initial spin period $P_0$, even when using only the subsample of 43 LGRBs with measured redshifts. The slope of the overall $B_p-P_0$ correlation changes from $B_p\propto P_{0}^{1.30\pm0.16}$ (full sample) to $B_p\propto P_{0}^{1.27\pm0.15}$ ($z$-known sample), which is still approximately consistent with the magnetic propeller model $B_p\propto P_{\rm eq}^{7/6}$. The accretion rate of the accreting magnetar in the $z$-known subsample is also in the range of $\dot{M}\sim10^{-5}-10^{-2}~M_{\odot}~\rm s^{-1}$ by systematically incorporating $R/I$ evolutionary effects and using the $M_b-M_g$ transition relation in different EoSs and $\eta_{\rm X}$. For different NS EoSs and radiative efficiencies, the log-normal distributions in the $z$-known subsample can be described as
$\log\dot{M}^{(\rm SLy,\eta_X=0.1)}/{M_{\odot}~\rm s^{-1}}=-3.01\pm0.62$, $\log\dot{M}^{(\rm SLy,\eta_X=0.5)}/{M_{\odot}~\rm s^{-1}}=-3.07\pm0.63$, $\log\dot{M}^{(\rm WFF2,\eta_X=0.1)}/{M_{\odot}~\rm s^{-1}}=-3.01\pm0.62$, $\log\dot{M}^{(\rm WFF2,\eta_X=0.5)}/{M_{\odot}~\rm s^{-1}}=-3.08\pm0.64$, $\log\dot{M}^{(\rm ENG,\eta_X=0.1)}/{M_{\odot}~\rm s^{-1}}=-2.99\pm0.63$, $\log\dot{M}^{(\rm ENG,\eta_X=0.5)}/{M_{\odot}~\rm s^{-1}}=-3.11\pm0.71$, $\log\dot{M}^{(\rm AP3,\eta_X=0.1)}/{M_{\odot}~\rm s^{-1}}=-2.96\pm0.62$, $\log\dot{M}^{(\rm AP3,\eta_X=0.5)}/{M_{\odot}~\rm s^{-1}}=-3.03\pm0.64$. Most of the accreting magnetars in 43 $z$-known samples can actually survive the equilibrium spin period for two soft SLy and WFF2 EoSs, and all the accreting magnetars in the $z$-known subsample can actually survive the equilibrium spin period for two stiff EoSs. The fallback rate of progenitor envelope materials onto the magnetar accretion disk for the $z$-known subsample in various (EoS, $\eta_{\rm X}$) combination scenarios is also compatible with the theoretical mass fallback rate of some low-metallicity massive progenitor stars. The consistency between the two samples validates our approach of utilizing the larger dataset to maximize statistical precision without compromising accuracy.

From the theoretical point of view, such a universal $B_p-P_0$ correlation for our LGRB sample in various (EoS, $\eta_{\rm X}$) combination scenarios is consistent with the $B\propto P_\mathrm{eq}^{7/6}$ relation for the magnetic propeller model. In an accreting magnetar propeller scenario, the materials at the edge of the accreting disk would flow into the magnetar polar caps along with the magnetic field lines and then form two accreting columns on the polar caps of newborn millisecond magnetars (known as the ``mountain'' ), which could lead to NSs being deformed with an asymmetry in the mass distribution \citep{Frank1992}. In this scenario, the newborn accreting magnetar has a rapid time-varying quadrupole moment, allowing it to become the potential source of continuous GW radiation \citep{Haskell2015,Haskell2017,Zhong2019,Sur2021,Huang2022}. Moreover, accreting NSs should spin up as they gain angular momentum through mass accretion. However, observations reveal a significant lack of submillisecond pulsars in X-ray binaries. This observational evidence implies that accreting NSs may achieve angular momentum balance through GW radiation, thus suppressing infinite acceleration of their spin \citep{Bhattacharyya2017}. But, for an NS in X-ray binaries, the accretion rate is much lower, around the order of $10^{-13}M_{\odot}~\rm yr^{-1}$ \citep{Papitto2015,Jaodand2016}, and their accretion mountain-induced deformation generates too weak GW radiation to be detected by the advanced LIGO/Virgo detector and the next generation ET detector \citep{Konar2016}. In contrast, the accretion columns of millisecond magnetars form under some extreme conditions, such as extremely high accretion rates, as expected for the newborn magnetar in GRBs with the order of $10^{-5}M_{\odot}-10^{-2}M_{\odot}~\rm s^{-1}$ in our sample, and this resultant deformation could also lead to strong GW radiation \citep{Zhong2019}. Unfortunately, until now, we have not detected a single case of GW radiation in the remnants of LGRBs with X-ray plateau emission. Furthermore, because of the interaction between the magnetar and its surrounding accretion disk during the propeller phase, the $P_0$ that we have constrained using the X-ray plateau data may not be the true initial spin period but rather the equilibrium spin period via fallback accretion in the propeller model. In the future, we expect that the GW radiation associated with the X-ray plateau emission can be detected by Advanced LIGO and ET, which not only could offer the first smoking gun that a protomagnetar can serve as the central engine of GRBs but also could also play a crucial role in precisely constraining the true $P_0$ and helping us to understand the nature and physical environment of a newborn accreting magnetar and its progenitor.

\setlength{\tabcolsep}{8mm}
\begin{deluxetable}{cccccccccc}
\tablewidth{0pt} 
\tabletypesize{\footnotesize}
\tablecaption{The center values and the best-fitting correlations between $B_p$ and $P_0$ for various scenarios. \label{table-1}}
\tablenum{1}
\tablehead{ \colhead{EoS}& \colhead{log $B_p$ (G)}&
\colhead{log $P_0$ (s)}& \colhead{log $B_p-P_0$}}
\startdata
$M_{b}=2.0M_{\odot}$,~$\eta_{\rm X}=0.1$\\
\hline
SLy  &($14.24\pm0.37$) &($-3.03\pm0.14$) & $\log B_p=(18.43\pm0.66) + (1.40\pm0.20)\log P_0$ \\
WFF2 &($14.26\pm0.40$) &($-3.04\pm0.14$) &
$\log B_p=(18.37\pm0.64) + (1.37\pm0.21)\log P_0$ \\
ENG  &($14.26\pm0.37$) &($-3.01\pm0.15$) &
$\log B_p=(18.20\pm0.61) + (1.34\pm0.20)\log P_0$ \\
AP3  &($14.22\pm0.38$) &($-3.01\pm0.14$) &
$\log B_p=(18.39\pm0.65) + (1.41\pm0.21)\log P_0$ \\
\hline
$M_{b}=2.0M_{\odot}$,~$\eta_{\rm X}=0.5$\\
\hline
SLy  &($14.58\pm0.42$) &($-2.82\pm0.21$) & $\log B_p=(17.92\pm0.37) + (1.21\pm0.13)\log P_0$ \\
WFF2 &($14.61\pm0.42$) &($-2.83\pm0.21$) & $\log B_p=(17.87\pm0.39) + (1.18\pm0.11)\log P_0$ \\
ENG  &($14.54\pm0.41$) &($-2.78\pm0.22$) & $\log B_p=(17.74\pm0.42) + (1.19\pm0.15)\log P_0$ \\
AP3  &($14.56\pm0.42$) &($-2.79\pm0.21$) & $\log B_p=(18.02\pm0.39) + (1.28\pm0.14)\log P_0$ \\
\enddata
\end{deluxetable}

\setlength{\tabcolsep}{3mm}
\begin{deluxetable}{ccccc|c|c|c|c|c|c|c}
\centering
\tablewidth{0pt} 
\tabletypesize{\footnotesize}
\tablecaption{The characteristic parameters of newborn accreting magnetar in our sample for various EoSs and $\eta_{\rm X}$. \label{table-2}}
\tablenum{2}
\tablehead{ \colhead{EoS} & \colhead{$M_{\rm TOV}$} & \colhead{$P_k$} & \colhead{$\alpha$} & \colhead{$\beta$} & \colhead{($M_b$,$\eta_{\rm X}$)} & \colhead{$P_0/P_k$} & \colhead{$A$} & \colhead{$M_g$} & \colhead{$R$} & \colhead{$\dot{M}$} & \colhead{$t_{\rm ev}$}\\
\colhead{} & \colhead{($M_{\odot}$)} & \colhead{(ms)} & \colhead{($10^{-10}\rm~s^{-\beta}$)} & \colhead{} & \colhead{($M_{\odot}$,--)} & \colhead{} & \colhead{} & \colhead{($M_{\odot}$)} & \colhead{(km)} & \colhead{($M_{\odot}~\rm s^{-1}$)} & \colhead{(s)}}
\startdata
\multirow{2}{*}{SLy}  & \multirow{2}{*}{2.05}  & \multirow{2}{*}{0.55} & \multirow{2}{*}{2.31} & \multirow{2}{*}{-2.73} & (2.0, 0.1) & $P_0=1.7P_k$ & 0.0730 & 1.77 & 14.185 & $7.0\times10^{-5}-2.0\times10^{-2}$ & 26--6149\\
\cline{6-12}
& & & & & (2.0, 0.5) & $P_0=2.8P_k$ & 0.0756 & 1.76 & 11.912 & $4.4\times10^{-5}-1.4\times10^{-2}$ & 13--4127\\
\cline{1-12}
\multirow{2}{*}{WFF2}  & \multirow{2}{*}{2.20}  & \multirow{2}{*}{0.50} & \multirow{2}{*}{2.17} & \multirow{2}{*}{-2.69} & (2.0, 0.1) & $P_0=1.8P_k$ & 0.0790 & 1.76 & 12.948 & $4.2\times10^{-5}-1.2\times10^{-2}$ & 27--6577\\
\cline{6-12}
& & & & & (2.0, 0.5) & $P_0=3.0P_k$ & 0.0830 & 1.75 & 11.495 & $3.7\times10^{-5}-1.4\times10^{-2}$ & 12--4278\\
\cline{1-12}
\multirow{2}{*}{ENG}  & \multirow{2}{*}{2.24}  & \multirow{2}{*}{0.53} & \multirow{2}{*}{3.95} & \multirow{2}{*}{-2.68} & (2.0, 0.1) & $P_0=1.8P_k$ & 0.0780 & 1.76 & 12.966 & $2.9\times10^{-5}-1.2\times10^{-2}$ & 24--6459\\
\cline{6-12}
& & & & & (2.0, 0.5) & $P_0=3.1P_k$ & 0.0821 & 1.75 & 11.701 & $1.7\times10^{-5}-1.5\times10^{-2}$ & 12--5668\\
\cline{1-12}
\multirow{2}{*}{AP3}  & \multirow{2}{*}{2.39}  & \multirow{2}{*}{0.54} & \multirow{2}{*}{1.88} & \multirow{2}{*}{-2.78} & (2.0, 0.1) & $P_0=1.8P_k$ & 0.0740 & 1.77 & 14.957 & $7.9\times10^{-5}-2.2\times10^{-2}$ & 23--5075\\
\cline{6-12}
& & & & & (2.0, 0.5) & $P_0=3.0P_k$ & 0.0780 & 1.76 & 12.586 & $4.2\times10^{-5}-1.5\times10^{-2}$ & 11--4812\\
\enddata
\end{deluxetable}

\begin{acknowledgements}
We thank the anonymous referee for a very thorough analysis of the original version and extremely helpful comments that have helped us to improve significantly the presentation of the paper. This work used data supplied by the UK \emph{Swift} Science Data Centre at the University of Leicester. L.L. is supported by the China Postdoctoral Science Foundation (grant No. GZB20230765). L.Z. is supported by the Hebei Natural Science Foundation (grant No.A2024403004). J.L. is supported by the China Postdoctoral Science Foundation (grant No. GZC20240905). S.A. has received support from the Villum Foundation (Project No.~13164, PI: I. Tamborra). This work is supported by the National Natural Science Foundation of China (Projects 12373040, 12021003, 12303050, 12494573, 12563009, 12403047), the National SKA Program of China (2022SKA0130100) from the Fundamental Research Funds for the Central Universities, the Strategic Priority Research Program of the Chinese Academy of Sciences, Grant No. XDB0550401.

\end{acknowledgements}


\end{document}